\documentclass[article]{aa}
\usepackage{graphicx}
\usepackage{mwe}
\usepackage{longtable}
\usepackage{siunitx}
\usepackage[table]{xcolor}
\usepackage{textcomp}
\usepackage{indentfirst}
\usepackage{chemist}
\usepackage{multirow}
\usepackage{caption}
\usepackage[flushleft]{threeparttable}
\usepackage{subcaption}
\usepackage{float} 
\usepackage{txfonts}
\titlerunning{}

\usepackage{natbib,twoopt}

\usepackage[colorlinks=true, pdfstartview=FitV, linkcolor=blue, citecolor=blue, urlcolor=blue, breaklinks=true]{hyperref} 
\bibpunct{(}{)}{;}{a}{}{,}             
\makeatletter
  \newcommandtwoopt{\citeads}[3][][]{\href{http://adsabs.harvard.edu/abs/#3}%
    {\def\hyper@linkstart##1##2{}%
     \let\hyper@linkend\@empty\citealp[#1][#2]{#3}}}
  \newcommandtwoopt{\citepads}[3][][]{\href{http://adsabs.harvard.edu/abs/#3}%
    {\def\hyper@linkstart##1##2{}%
     \let\hyper@linkend\@empty\citep[#1][#2]{#3}}}
  \newcommandtwoopt{\citetads}[3][][]{\href{http://adsabs.harvard.edu/abs/#3}%
    {\def\hyper@linkstart##1##2{}%
     \let\hyper@linkend\@empty\citet[#1][#2]{#3}}}
  \newcommandtwoopt{\citeyearads}[3][][]%
    {\href{http://adsabs.harvard.edu/abs/#3}
    {\def\hyper@linkstart##1##2{}%
     \let\hyper@linkend\@empty\citeyear[#1][#2]{#3}}}
\makeatother

\usepackage{color}

\usepackage{color}

\usepackage{cleveref}

\usepackage{chemist}
\usepackage{orcidlink}

\begin{document}
\title{Pulsation modelling of the Cepheid Y~Ophiuchi with RSP/MESA}
\subtitle{Impact of the circumstellar envelope and a high projection factor on Baade-Wesselink method}
\titlerunning{Pulsation modelling of the Cepheid Y~Ophiuchus with RSP/MESA}
\authorrunning{Hocd\'e et al. }
\author{V. Hocd\'e \inst{1}\orcidlink{0000-0002-3643-0366}, R. Smolec \inst{1}\orcidlink{0000-0001-7217-4884}, P. Moskalik \inst{1}\orcidlink{0000-0003-3142-0350}, R. Singh Rathour\inst{1}\orcidlink{0000-0002-7448-4285}, O. Zi\'ołkowska \inst{1}\orcidlink{0000-0002-0696-2839}}

\institute{Nicolaus Copernicus Astronomical Centre, Polish Academy of Sciences, Bartycka 18, 00-716 Warszawa, Poland\\
email : \texttt{vhocde@camk.edu.pl}
}

\date{Received ... ; accepted ...}

\abstract{Y~Ophiuchi (Y~Oph) is a classical Cepheid with a pulsation period of $P=$17.12$\,$days. This star is reported to be as dim as a Cepheid of about half its pulsation period, and exhibits a low radial velocity and light-curves amplitude. 
For these reasons, Y~Oph is not used to calibrate period-luminosity (PL) relation and its distance remains uncertain.}
{Our objective is to conduct hydrodynamical pulsation modeling of Y~Oph to derive its distance and provide physical insight to its low amplitude and luminosity, constrained by an extensive set of observations.}{We first perform a linear analysis on a grid of models using hydrodynamical pulsation code \texttt{MESA-RSP} in order to find the combinations of mass, metallicity, effective temperature and luminosity resulting in linear excitation of pulsations with period of about 17$\,$days. Then, for the best combinations of stellar parameters, we perform non-linear computations to obtain the full-amplitude pulsations of these models. Last, we compare the results to a complete set of observations along the pulsation cycle including angular diameter obtained by interferometry, effective temperature and radial velocity obtained by high-resolution spectroscopy, and light curves in $VJHK_SLM$ bands. We adjust simultaneously the distance, the color excess and circumstellar envelope (CSE) model to fit the light curves and the angular diameter.}{We find that all pulsation models at high effective temperature are in remarkable agreement with the observations along the pulsation cycle. This result suggests that the low amplitude of Y~Oph can be explained by its location close to the blue edge of the instability strip. We also find that a pulsational mass of about 7-8$\,\mathrm{M}_\odot$ is consistent with a non-canonical evolutionary model with moderate overshooting, PL relation and \textit{Gaia} parallax. However, a much lower mass below 5$\,$M$_\odot$ is required to match Baade-Wesselink (BW) distance measurements from the literature. We show that the combination of the impact of the CSE on the photometry together with a projection factor of about 1.5 explains the discrepant distance and luminosity values obtained from BW methods.
}{Our findings indicate that small pulsation amplitude of Y~Oph can be attributed to its close proximity to the blue edge of the instability strip. Additionally, our analysis reveals that the distances obtained using the BW method are biased compared to \textit{Gaia}, mainly due to the impact of circumstellar envelope on the photometries and a high $p$-factor close to 1.5. Despite these unique characteristics, Y Oph is a long-period classical Cepheid that holds potential for calibration of the PL relation in the Galaxy.}
\maketitle


\section{Introduction}\label{s_Introduction}
The discovery of the Y~Ophiuchi (Y~Oph) variability, which has a pulsation period of about 17$\,$days, dates back to 130$\,$years ago \citep{Sawyer1890,Sawyer1892,Luizet1905}. At the same epoch, the period-luminosity relation  (hereafter PL relation) of Cepheids was discovered \citep{leavitt08,leavitt1912}. Y~Oph, however, is generally not taken into account in the calibration of the PL relation because of its unusual behaviour. Indeed, using Baade-Wesselink (BW) method, \cite{Abt1954} found that this star is 1$\,$mag fainter than what would be derived using PL relation. In other words, the luminosity derived by the BW method corresponds to a Cepheid with a pulsation period of 10$\,$days or less.

Several variants of BW analysis inferred the distance of Y~Oph, but they are scattered between about 400 and 650$\,$pc \citep{Gieren1993,kervella04a,merand07,Groenewegen2008,storm11a,Groenewegen2013}. Most of these methods used the Surface Brigthness Color Relation (SBCR). However, Y~Oph is close to the galactic plane and exhibits an important color excess of about $E(B-V)=0.650$ \citep{fernie95,Laney2007,Kovtyukh2008} which might bias the distance measurements based on photometric relations if the reddening is not well known. Among these BW variants, \cite{merand07} observed directly the angular diameter along the pulsation cycle with infrared interferometry. The advantage is to not use SBCRs which are sensitive to empirical calibrations and interstellar extinction. The derived distance from this method is $d=491\pm$18$\,$pc. On the other hand, the distance obtained from \textit{Gaia} DR3 parallax \citep{gaia2016,Gaia2022} is strongly discrepant with $d=$742$\pm 21\,$pc \citep[corrected from zero-point,][]{Lindegren2021}. The renormalized weight error (RUWE) of Y~Oph indicates a good quality of the parallax (RUWE = $1.04<1.4$), and, thus, can be used for astrometry. Nevertheless, Y~Oph slightly saturates the detector with $G=5.6\,$mag. Indeed, \cite{Lindegren2018} warns that stars with $G<6\,$mag generally have inferior astrometry due to calibration issues. Despite small saturations we do not expect such large errors of measurements that would explain the difference obtained with BW methods. As an example, T Vul has both comparable RUWE and magnitude in $G$-band with Y~Oph, but the distance inferred from its parallax $d$=581$\pm$20$\,$pc is consistent with HST parallax and BW distance from the literature \citep{fouque07}. Moreover, the color variation of Y~Oph is smaller than others Cepheids due to its small light curve amplitude, thus the chromaticity effects on \textit{Gaia} detector are likely mitigated.





Because of its low amplitude, Y Oph is sometimes supposed to be a first-overtone pulsator \citep[see e.g.][]{Owens2022}. However, it is questionable to assume that Y Oph pulsates in the first overtone since this mode is usually excited for Cepheids of much shorter pulsation period. From observations, the longest pulsation period for a securely identified first-overtone Cepheid is 7.57$\,$day in the case of V440 Per \citep{Baranowski2009}. In the OGLE Milky Way survey, the longest overtone period claimed is 9.437$\,$day \citep{Udalski2018,Pietru2021}. Another explanation for the small amplitudes is proposed by \cite{Luck2008}. The small amplitudes are not a direct result of the first-overtone mode, but are associated with the proximity to the blue edge of the fundamental mode instability strip. This might be the case for Y Oph which exhibits a mean effective temperature of about 5800$\,$K  as measured by spectroscopic observations \citep{Luck2018,Proxauf2018,daSilva2022}. This is a high temperature compared to the stars of the same pulsation period which have a mean temperature of about 5500$\,$K as in the the case for example for CD~Cyg ($P=17.07\,$day) or SZ~Aql ($P=17.14\,$day) \citep{Trahin2021}.

    Several studies have shown spectroscopic binarity evidences of Y Oph, although its effect on photometric and radial velocity measurements remains negligible. \cite{Abt1978} suggested that Y~Oph is a spectroscopic binary system having an orbital period of 2612$\,$days and concluded that its effect on photometry and radial velocity are likely marginal. \cite{szabados89} extended the analysis of the mean radial velocity variation and derived an orbital period of 1222.5$\,$days. However, the companion was not detected in \textit{IUE} spectra \citep{Evans1992} and NACO lucky imaging which yields a magnitude difference of at least 2.5$\,$mag in the $K_s$ band \citep{gallenne14}. More recently, a third companion of much longer orbital period (about 28000$\,$days) is suspected from light-travel time effect observed in $O-C$ diagram \citep{Csornyei2022}.






In order to give physical insights to this intriguing star, in particular to explain its low radial velocity and light curve amplitude, brightness and distance discrepancy, we aim at constraining Y Oph properties with hydrodynamical pulsation models. Non-linear pulsation modeling offers a valuable approach for exploring the underlying physics of variable stars, as demonstrated for the first time by \cite{Christy1964,Christy1966WVir,Christy1966RRLyrae,Christy1966Cep}. Additionally, it provides a tool to accurately determine the mass of Cepheids, a crucial factor when considering the mass discrepancy associated with these stars \citep{Caputo2005,bono06,keller08}. Pulsation models are also useful to determine the distance of Cepheids \citep{Marconi2013_binary,Marconi2013,Ragosta2019}.
In the case of Y Oph, a non-linear model was proposed to fit radial velocity measurements and $V$-band apparent magnitude \citep{Ruoppo2004}. These authors found that a 7$\,$M$_\odot$ model is consistent with short distance and low absolute luminosity in $V$-band (respectively 423$\,$pc and $M_V$ = $-3.996\,$mag). However, Y~Oph is modeled with an average effective temperature by about 1000$\,$K lower than recent high-resolution spectroscopic temperature ($T_\mathrm{eff}=4720\,$K$<5800\,$K).

 In this paper, we propose for the first time to constrain hydrodynamical pulsation models of Y Oph with an extensive set of observations. On the modelling side, the recently released Radial Stellar Pulsation (RSP) 
tool in Modules for Experiments in Stellar Astrophysics \citep[MESA,][]{Smolec2008,paxton13,Paxton2015,Paxton2018,Paxton2019,MESA2023} can be used to model non-linear pulsation of Y~Oph.  \texttt{MESA-RSP} is a one-dimensional Lagrangian convective code that was already used for modelling RR Lyrae, BL Her and classical Cepheids \citep[see, e.g.,][]{Paxton2019,Susmita2020,Susmita2021,Kurbah2023}. On the observational side, we benefit from a complete set of observations to constrain the models. To this end, we gathered from literature effective temperature and radial velocity curves, angular diameter from interferometric observations and light curves in the visible and the infrared. 





The paper is organized as follows. We first construct a grid of equilibrium static models and we perform a linear non-adiabatic analysis with \texttt{MESA-RSP} in Sect.~\ref{sect:linear}. Then, we use the solutions that have a positive growth rate for pulsation period of $\approx$17$\,$days to perform a nonlinear calculations for this star in Sect.~\ref{sect:full}. We then present our fitting strategy to a full set of observations in Sect.~\ref{sect:fit_strategy}. We finally present the results in Sect.~\ref{sect:results} and we discuss the discrepancy between \textit{Gaia} and BW distances in Sect.~\ref{sect:discussion}. We conclude in Sect.~\ref{sect:conclusion}.


\section{Linear analysis with RSP}\label{sect:linear}
Our method consists first of computing a linear non-adiabatic (LNA) stability analysis of a grid of models with \texttt{MESA-RSP} as described e.g., in \cite{Smolec2016,Paxton2019}. For this study we used \texttt{MESA} version \texttt{r15140} and OPAL opacity tables from \cite{Iglesias1996,Ferguson2005}, using solar abundance mixture from \cite{Asplund2009}. We note that while subsequent \texttt{MESA} versions were released after this study, the changes introduced does not affect pulsation calculation with the \texttt{RSP} module \citep{Jermyn2023}.  The objective is to determine the metallicity, mass, luminosity and temperature at which the pulsations are linearly unstable for a pulsation period of approximately 17$\,$days. To this end, we constructed equilibrium static models with a fixed chemical composition from the literature and a grid of mass $M$, luminosity, $L$, and effective temperature $T_\mathrm{eff}$. The results of this computation provide the linear periods of the different radial pulsation modes and their growth rates $\gamma$. The growth rate of a mode $\gamma$, is defined by the fractional growth of the kinetic
energy per pulsation period.

In the following we first present our choice for the convective parameters and the chemical composition adopted, and we then describe the construction of the grid and the LNA computation.

\subsection{Convective parameters adopted}\label{sect:convective}
\texttt{MESA-RSP} uses the time-dependent model of turbulent convection described in \cite{Kuhfuss1986} and follows the implementation of \cite{Smolec2008}. We used Set A of convective parameters as given in Table 4 in \cite{Paxton2019}. Set~A represents a simple convection model without radiative cooling, turbulent pressure and flux. We note that in the particular case of a \textit{time-independent} version of Set A, this model would reduced to to the standard mixing-length theory (MLT) \citep{Kuhfuss1986,Wuchterl1998,Paxton2019}. \cite{Paxton2019} have shown that the set of convective parameters proposed are reasonably reproducing the Fourier parameters of both the radial velocity and $I$-band light curves of Cepheids \citep[see Figure 12 in ][]{Paxton2019}. On the other hand, the choice of convective parameters also impacts directly the linear period and associated growth rate of the different modes. The introduction of the turbulent pressure, which involves slight inflation of the star, tends to shift the blue edge of the instability strip red-ward \citep[see Figure 13 in ][]{Paxton2019}. Therefore, this will affect the temperature of Y Oph to decrease and modify the growth rates and linear periods associated to fundamental and first-overtone modes. The convective parameters and their value respected to each other have to be properly calibrated and used with caution \citep{Kovacs2023}. However, we do not expect a significant change of the convective parameters used in this study, because our modeling is constrained to the observed effective temperature of Y Oph (in average 5800$\,$K), which prevents the introduction of too large turbulent pressure, as for example introduced by the Set D \citep{Paxton2019}. For this same reason, the growth rate associated to different pulsation modes will not change significantly as compared to Set A. While we are confident that the Set A of convective parameters is robust to obtain reasonable models of Cepheids, we stress that calibrating these parameters are needed to provide a more precise modeling of Y Oph in the future.

\subsection{Chemical composition adopted}\label{sect:chemical}
Several authors derived the metallicity of Y~Oph from spectroscopic observations. We summarized the most recent measurements of the iron-to-hydrogen ratio [Fe/H] in Table \ref{Tab:Z}. From this table, we see that the measurements are in good agreement with solar metallicity. However, the most recent [Fe/H] determination given by \cite{daSilva2022} is not in agreement with preceding estimates.

As noted by \cite{daSilva2022}, their [Fe/H] determination for several Cepheids is about 0.1$\,$dex below others measurements. In particular, their study estimated Y~Oph metallicity from the same spectra as \cite{Proxauf2018} who found [Fe/H] that is shifted by $+0.13\,$dex. According to \cite{daSilva2022}, this difference is due to their very careful selection of the lines adopted to estimate the iron abundance. Although more studies are needed to confirm this trend, the recent \cite{daSilva2022} measurement is likely the most accurate [Fe/H] value. Since Y~Oph metallicity is well constrained to solar abundance, we choose to adopt only a standard solar abundance value for the metallicity of Y~Oph that is $\mathrm{[Fe/H]}=0.0\,$dex.

From [Fe/H] we can now estimate the mass fraction of hydrogen $X$, helium $Y$, and metals $Z$ related by $X+Y+Z=1$. By definition, the metallicity [Fe/H] is given by :

\begin{equation}
    [\mathrm{Fe}/\mathrm{H}]=\mathrm{log}\left(\frac{Z}{X}\right)-\mathrm{log}\left(\frac{Z}{X}\right)_\odot
\end{equation}
where $X_\odot$ and $Z_\odot$ stand for the solar mass fraction of hydrogen and metals respectively. For these values we adopted the solar mixture $Z_\odot/X_\odot=0.134$ from \cite{Asplund2009}. To derive the helium abundance $Y$ in the precedent equation we assumed the linear relation between $Y$ and $Z$ \citep{Peimbert1974}:
\begin{equation}
    Y(Z)=Y_p + \frac{\Delta Y}{\Delta Z} Z
\end{equation}
where $Y_p$ is the primordial helium abundance and $\Delta Y / \Delta Z$ is helium-to-metal enrichment ratio. We adopted the primordial helium abundance $Y_p=0.2484\pm0.0005$ \citep{Cyburt2004}. Several estimates for $\Delta Y/ \Delta Z$ can be found in the literature \citep{Tognelli2021}. These values are generally centered on $\Delta Y/ \Delta Z$ = 2 with large uncertainties. Hence, we choose to explore different $\Delta Y/ \Delta Z$ values for the adopted solar abundance in the LNA analysis. We choose to extend our grid to $\Delta Y/ \Delta Z$=1.5, 2, and 2.5. Several studies investigated the dependence of helium content at fixed metallicity on the edge of the instability strips derived with non-linear pulsation models \citep[see e.g.][]{Fiorentino2002,Marconi2005}. 




\begin{table}[]
\caption{Spectroscopic metallicities gathered from the literature for Y~Oph. \label{Tab:Z}}
\begin{center}
\begin{tabular}{l|c}
\hline
\hline
Reference& [Fe/H] (dex)\\

\hline
\cite{daSilva2022}& $-0.05\pm$0.03\\ 
\cite{Proxauf2018}&  $+0.08\pm$0.04 \\ 
\cite{Genovali2014} & $+0.12\pm$0.04\\ 
\cite{Luck2008} & $+0.06\pm$0.04\\ 

\hline
\end{tabular}
\normalsize
\end{center}
    \begin{tablenotes}
    \item \textbf{Notes :} These values are scaled to the solar iron abundance value of 7.50 \citep{Asplund2009}. 
    \end{tablenotes}
\end{table}

\subsection{Results of the linear non-adiabatic analysis}
We choose to perform LNA analysis for 6 different stellar masses: 3, 4, 5, 6, 7 and 8$\,\mathrm{M}_\odot$. Our grid is also evenly spaced for the effective temperature by 50$\,$K between 5500 and 6000$\,$K and for the luminosity by 250$\,\mathrm{L}_\odot$ between 3000 and 10000$\,\mathrm{L}_\odot$. The temperature grid is chosen in order to cover the average effective temperature measured by \cite{Luck2018} that is 5819$\,$K. On the other hand, the extension of the luminosity grid is chosen to explore the different distance scenarios in the context of the discrepancy between BW methods and \textit{Gaia} parallax. This ensures that the physical properties of Y~Oph are within the limits of the grid. However, we emphasize that this is beyond the scope of this paper to provide a finer resolution of the grid to constrain precisely the physical parameters.

We used a standard grid structure in RSP which consists of 150 Lagrangian mass shells with variable and constant shell mass, below and above the anchor temperature respectively \citep[see Figure 2 in ][]{Paxton2019}. This zoning is defined to ensure a good spatial resolution of the ionized hydrogen and helium driving regions around the anchor temperature of $11000\,$K. The effective temperature is attributed to the outer shell, while the inner boundary temperature is set to 2.10$^{6}\,$K (see Appendix \ref{appendix:mesa}). This setting is customary in the use of RSP \citep[see e.g. ][]{Paxton2019,Kovacs2023}.

We computed linear periods of the fundamental mode together with the corresponding linear growth rates. We also derived the linear periods and growth rates for the 11 consecutive radial overtones, to investigate the possibility of overtone pulsations for this star. High radial overtones in Cepheids are often trapped in the outer layer of the star. Such trapped modes, called by Buchler the "strange modes", are characterized by very small light amplitudes \citep{Buchler1997}. These modes can be at the origin of non-linear pulsation of Cepheids outside the instability strip on the blue side, which might be interesting in the case of Y~Oph. The \texttt{MESA-RSP} script used for the computation is presented in appendix \ref{appendix:mesa}.

\begin{figure*}[] 

\begin{subfigure}{0.50\textwidth}
\includegraphics[width=\linewidth]{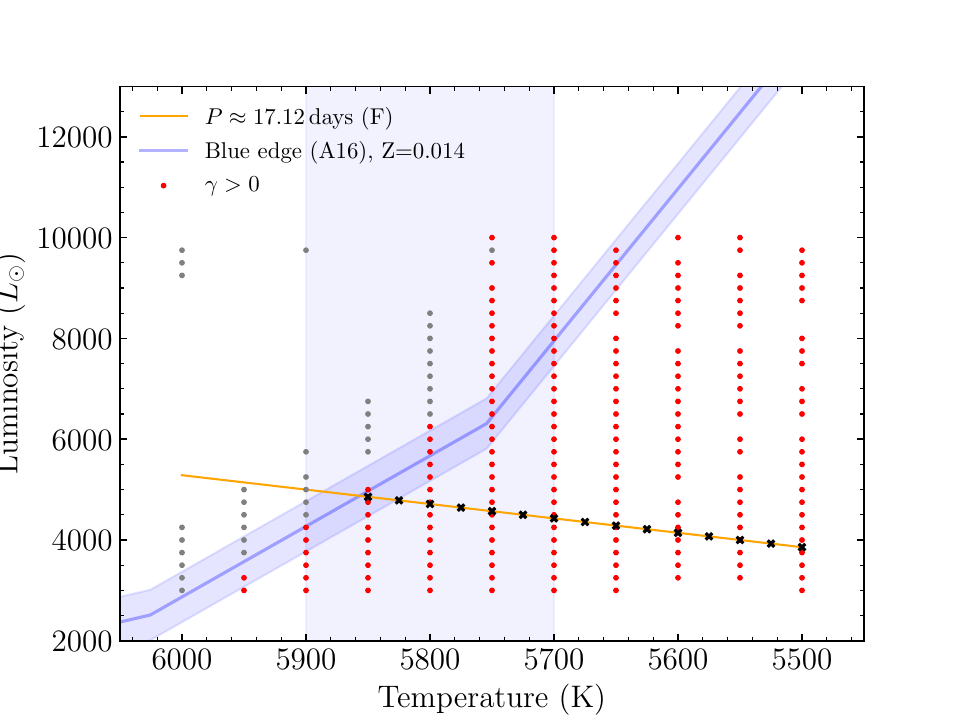}
\caption{3$\,\mathrm{M}_\odot$} \label{fig:grid_3M}
\end{subfigure}\hspace*{\fill}
\begin{subfigure}{0.50\textwidth}
\includegraphics[width=\linewidth]{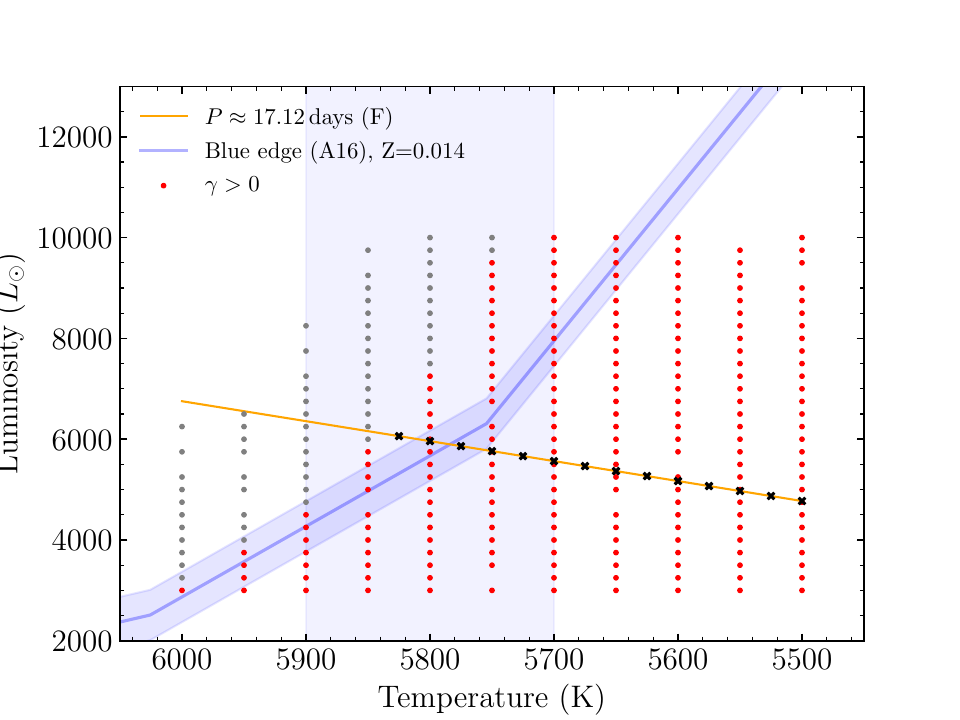}
\caption{4$\,\mathrm{M}_\odot$} \label{fig:grid_4M}
\end{subfigure}
\medskip
\begin{subfigure}{0.50\textwidth}
\includegraphics[width=\linewidth]{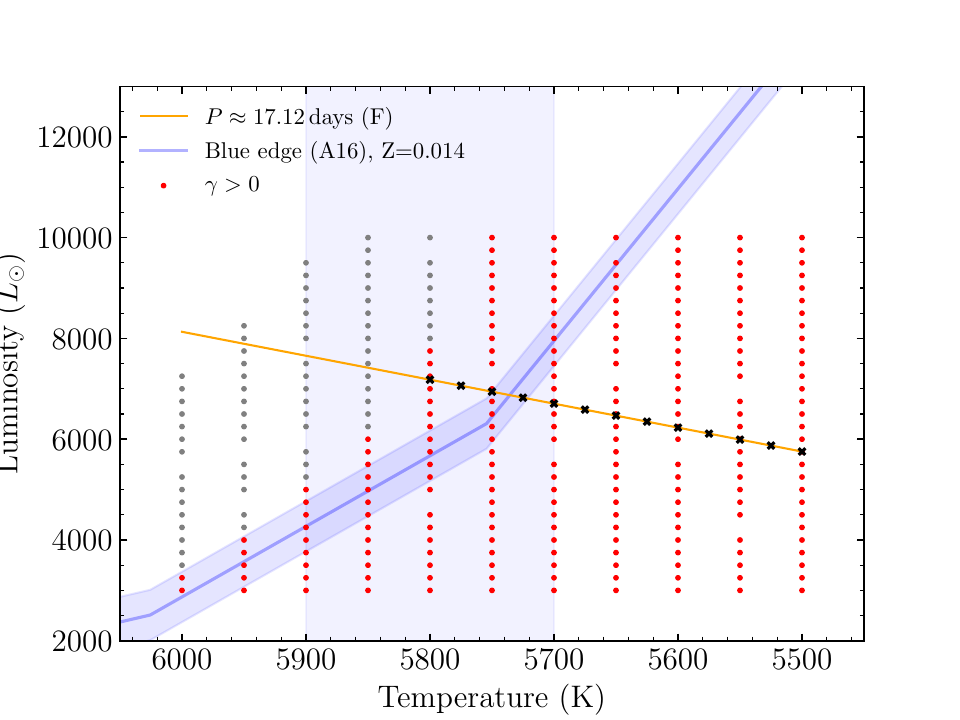}
\caption{5$\,\mathrm{M}_\odot$} \label{fig:grid_5M}
\end{subfigure}\hspace*{\fill}
\begin{subfigure}{0.50\textwidth}
\includegraphics[width=\linewidth]{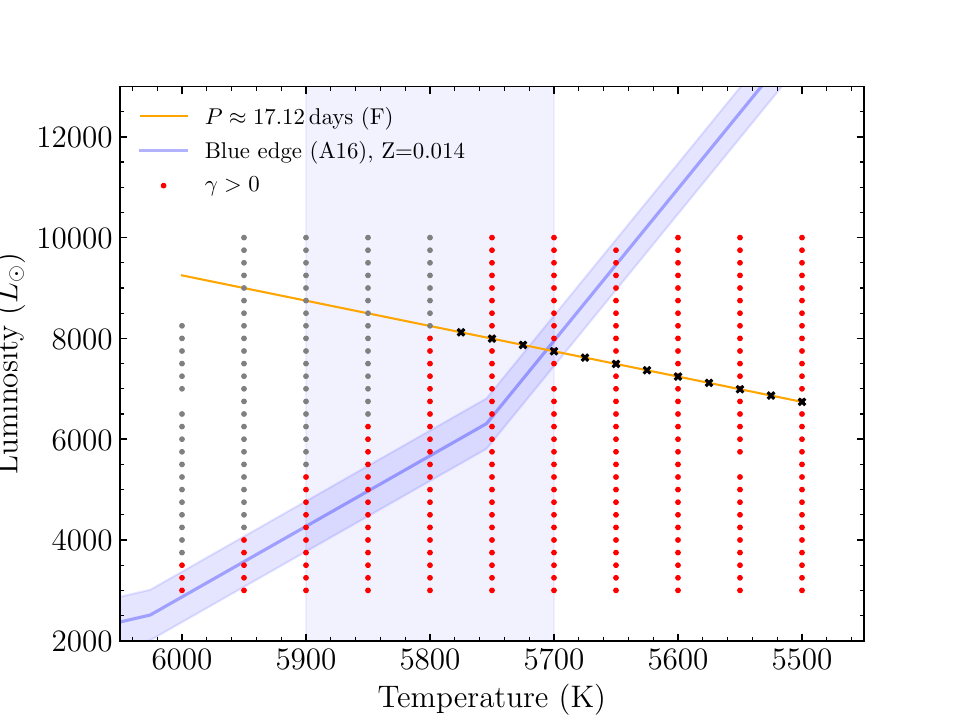}
\caption{6$\,\mathrm{M}_\odot$} \label{fig:grid_6M}
\end{subfigure}
\medskip
\begin{subfigure}{0.50\textwidth}
\includegraphics[width=\linewidth]{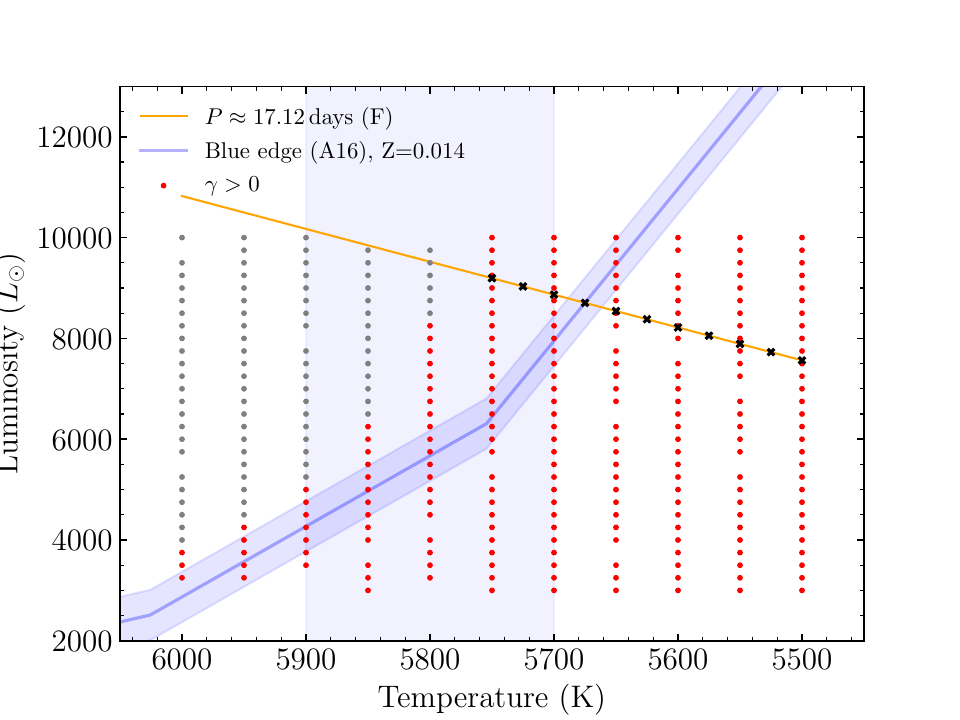}
\caption{7$\,\mathrm{M}_\odot$} \label{fig:grid_7M}
\end{subfigure}\hspace*{\fill}
\begin{subfigure}{0.50\textwidth}
\includegraphics[width=\linewidth]{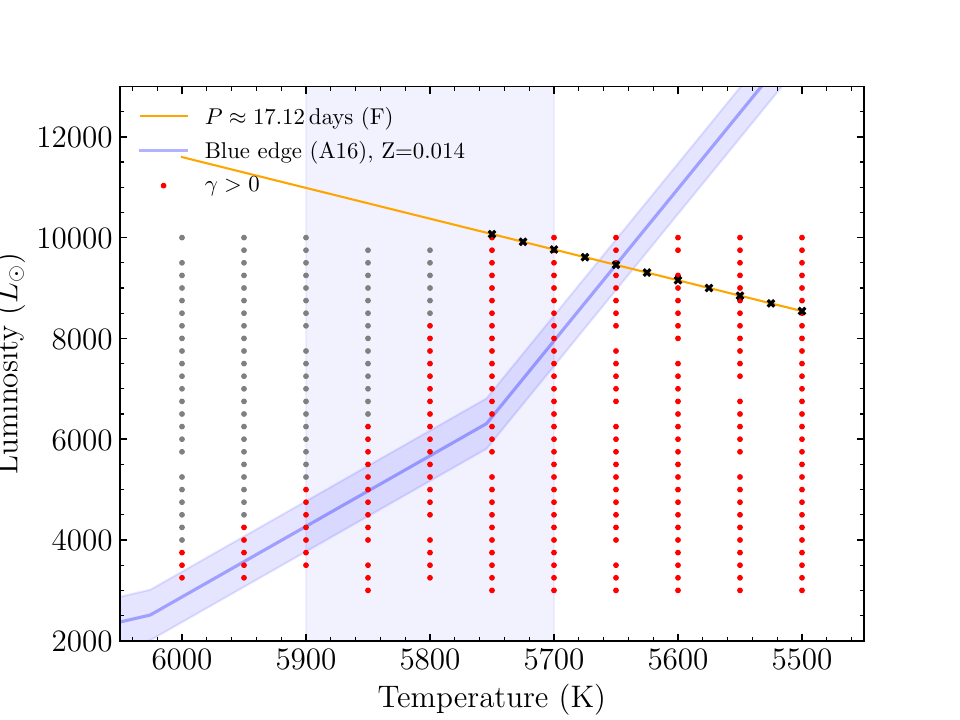}
\caption{8$\,\mathrm{M}_\odot$} \label{fig:grid_8M}
\end{subfigure}
\caption{Results of the linear non-adiabatic (LNA) analysis with \texttt{MESA-RSP}. Positive and negative growth rates $\gamma$ for the fundamental mode are indicated in red and grey points respectively. The black crosses are the selected models for non-linear analysis as summarized in Table \ref{Tab:models}. The vertical blue strip corresponds to the expected value of the mean effective temperature for Y~Oph i.e. $T_\mathrm{eff}=5800\pm100\,$K \citep{Luck2018,Proxauf2018,daSilva2022}. The fundamental blue edge from \cite{Anderson2016evol} is shown for comparison ($Z=0.014$) with $\Delta L\pm500\,$L$_\odot$}.\label{fig:linear}
\end{figure*}

As a result of the computation, we found positive growth rates for the first overtone mode but only for linear pulsation period below about 5$\,$days. All higher order overtones are linearly damped in our grid. Finally, we can rule out the possibility that Y~Oph is a first overtone pulsator or pulsates in even higher overtone modes.
The results of the LNA analysis for the fundamental mode are displayed in Fig.~\ref{fig:linear}.  From this Figure, we can see that a few of these calculated models are missing because the computation failed for numerical reasons, especially at very high temperature off from the instability strip. This has no impact for the following of the study.

In order to find models with positive growth rates for the fundamental mode with period of $P\simeq 17.12$ days, we first interpolated growth rates along the luminosity axis. Then, we interpolated the pulsation period along $L$ and $T_\mathrm{eff}$, in order to determine an iso-period line at which models pulsate with the period close to $P=17.12\,$days, as is observed in the star (see orange lines in Fig.~\ref{fig:linear}). The intersection of the iso-period line and the positive growth rates region provides the combination of mass, metallicity, luminosity and effective temperature we can consider for non-linear modeling of Y~Oph pulsations. We note that we adopted the non-linear pulsation period to constrain the linear period from the grid. In principle, these two values are not equal for a same model, the non-linear value being slightly longer than the linear period. For the sake of simplicity, we have chosen to neglect this difference which is at the percent level. For each set of mass, we selected models on iso-periods evenly spaced by 25$\,$K (see black crosses in Fig.~\ref{fig:linear}). For each mass, these combinations of effective temperature and luminosity will be then used for a non-linear analysis in the next section (see Table~ \ref{Tab:models}). From our computations, we do not observe any significant difference of luminosity and temperature from the different adopted values of $\Delta Y / \Delta Z$. This is in agreement with non-linear computation results from \cite{Marconi2005} who found no clear trend on the fundamental blue edge with helium content change. Hence, we fix $\Delta Y / \Delta Z=2$ and we use the grid computed in Table~\ref{Tab:models} in the following section.

\begin{table}[]
\caption{Grid of models selected for non-linear computations, based on
   results of the LNA analysis for a linear period of $P\simeq 17\,$days (see Fig.~\ref{fig:linear}). \label{Tab:models}}
\begin{center}
\begin{tabular}{l|cccccc}
\hline
\hline
$T_\mathrm{eff}$(K)  & $L_{3\mathrm{M}_\odot}$ & $L_{4\mathrm{M}_\odot}$ & $L_{5\mathrm{M}_\odot}$ & $L_{6\mathrm{M}_\odot}$& $L_{7\mathrm{M}_\odot}$ & $L_{8\mathrm{M}_\odot}$\\
\hline
5850 &  4857 &  -   &   -  &   - &  -   & -       \\
5825 &  4785 & 6060 &  -   &   -  &  -  & -    \\
5800 &  4714 & 5961 & 7180 &  -   &  -  & -     \\
5775 &  4642 & 5862 & 7061 & 8120 &   - & -     \\
5750 & 4571 & 5763 & 6942 & 7995 & 9194 & 10069\\
5725 &  4499 & 5664 & 6828 & 7870 & 9031 & 9916\\
5700 &  4428 & 5565 & 6705 & 7744 & 8868 & 9764\\
5675 &  4357 & 5466 & 6586 & 7619 & 8705 & 9611\\
5650 &  4285 & 5367 & 6467 & 7493 & 8542 & 9458\\
5625 &  4214 & 5268 & 6348 & 7368 & 8379 & 9305\\
5600 & 4142 & 5169 & 6229 & 7242 & 8216 & 9153\\
5575 &  4071 & 5070 & 6111 & 7117 & 8053 & 9000\\
5550 &  3999 & 4971 & 5992 & 6991 & 7890 & 8847\\
5525 &  3928 & 4872 & 5873 & 6866 & 7727 & 8694\\
5500 &  3857 & 4773 & 5754 & 6740 & 7564 & 8542\\
\hline
\hline
\end{tabular}
\normalsize
\end{center}
    \begin{tablenotes}
    \item \textbf{Notes :} The effective temperature of each model $T_\mathrm{eff}$(K), and the luminosity given for different masses, $L_{X\mathrm{M}_\odot}$, are indicated.
    \end{tablenotes}
\end{table}

\section{Non-linear analysis with RSP}\label{sect:full}

\subsection{Non-linear computations}
Full-amplitude stable pulsations are reached when
kinetic energy per pulsation period becomes constant. In other
words, a fractional growth of the kinetic energy per pulsation period, $\Gamma$, is equal to 0. Given a running window, our models verify the following conditions:

\begin{itemize}
    \item Period stability: The ratio of the standard deviation of
      the pulsation period to the average pulsation period is less
      than $10^{-5}$.
    \begin{equation}
        \frac{\sigma(P)}{P}<10^{-5}.
    \end{equation}
    \item Amplitude stability: standard deviation of the radius amplitude to the average radius is less than $10^{-4}$.
    \begin{equation}
        \frac{\sigma(\Delta R)}{R}<10^{-4}.
    \end{equation}
    \item Kinetic energy stability: The change of the absolute kinetic energy per pulsation period is below $10^{-5}$.
    \begin{equation}
        \left| \Delta \mathcal{E}_k \right| < 10^{-5}.
    \end{equation}

\end{itemize}
Once these criteria are verified altogether, we let the computations run for 200 additional pulsation cycles. For each model we retrieved the 2 last pulsation cycles computed. Given our time resolution, it corresponds to about 2400 data points. The results along the pulsation cycle are then phased to $\phi=0$ for maximum light of the $V$-band light curve.

\subsection{Atmosphere models along the pulsation cycle}\label{sect:atmosph}
 
 From the non-linear computation, \texttt{MESA/RSP} provides bolometric corrections to obtain the photometry in specific bands of the Johnson-Cousins system \citep{Lejeune1998}. However, since we retrieved light curves observations in different filters (Sect.~\ref{sect:obs}) we rather apply the specific synthetic filters to a grid of atmosphere models derived for every phase of non-linear model. The radius and effective temperature are defined by RSP following the Stefan-Boltzman law $L_\mathrm{bol}\propto R_\mathrm{RSP}^2T_\mathrm{eff}^4$ \citep{Smolec2008,Paxton2019}, and thus are convenient for computing static atmosphere models. To this end, for each effective temperature $T_\mathrm{eff}(\phi)$, radius $R_\mathrm{RSP}(\phi)$ and log$\,g(\phi)$ computed by \texttt{MESA-RSP} along the pulsation phase $\phi$, we interpolated over ATLAS9 model grid to obtain the spectral energy distribution of these models. We used ATLAS9 models\footnote{\url{https://wwwuser.oats.inaf.it/castelli/grids.html}} \citep{castelli2003} with solar metallicity and a standard
turbulent velocity of 2$\,$km/s.  We note that the quasi-static assumption of the atmosphere of Y Oph is particularly well adapted as a result of both small amplitude and long pulsation period. Then, we apply the synthetic filters corresponding to the observations in $VJHK_sLM$-bands presented in the next section. These synthetic filters (Johnson, 2MASS  and \textit{Spitzer} IRAC) are publicly available from the Spanish Virtual Observatory\footnote{\url{http://svo2.cab.inta-csic.es/theory/fps3/}} (SVO).


\section{Fitting strategy}\label{sect:fit_strategy}
\subsection{Set of observations}\label{sect:obs}
In order to assess which models are the best to reproduce Y~Oph characteristics we compare the results with a full set of observations. We gathered the following set of observations along the pulsation cycle:

\begin{itemize}

    \item Uniform disk angular diameters along the pulsation cycle are provided by near-infrared interferometric observations in the $K$-band from the CHARA/FLUOR instrument \citep{merand07,Gallenne2011PhD}. Angular diameter measurements are displayed in Fig.~\ref{fig:R_8M}.

    \item  Radial velocities measurements are retrieved from \cite{Petterson2005,Borgniet2019,Eaton2020} and phased together and corrected for zero-point offsets. The RV measurements are displayed in Fig.~\ref{fig:RV_8M}. 

    \item The effective temperature measurements were obtained from high-resolution spectroscopy by \cite{Luck2018} and \cite{daSilva2022}. The measurements obtained by \cite{Luck2018} cover most of the pulsation cycle whereas those obtained by \cite{daSilva2022} are close to the minimum temperature. The combination of these two independent measurements is in excellent agreement as we can see from the temperature plot in Fig.~\ref{fig:Teff_8M}.

    \item The light curve in $V$-band was obtained from the Johnson filter \citep{Berdnikov2008}. These observations are a compilation of many observations between $\mathrm{MJD}=46000$ and $53000$. To mitigate the effet of phase mismatch, we selected $V$-band photometry of MJD before 48000 to match the observation epoch of the near-infrared observations. We retrieved the light curves in $J$, $H$ and $K$ from \cite{laney1992}. We transformed these photometries from the South African Astronomical Observatory \citep[SAAO,][]{Carter1990} system into the Two Micron All Sky Survey \citep[2MASS,][]{2MASS2006} system using transformations from \cite{Koen2007}. This transformation allows to use the interstellar extinction model calibrated with this same system (see Sect.~\ref{sect:ISM}). We complemented these data with mid-infrared light curves obtained from Infrared Array Camera (IRAC) onboard the \textit{Spitzer} telescope at 3.5 and 4.5$\,\mu$m (namely $L$ and $M$ bands) \citep{Monson2012}.
    These light-curves are displayed in Figs.~\ref{fig:V_8M}, \ref{fig:J_8M}, \ref{fig:H_8M}, \ref{fig:K_8M}, \ref{fig:L_8M} and \ref{fig:M_8M}.
\end{itemize}

The rate of period change of Y Oph is known to be about $+8\,$s/yr \citep{Fernie1995YOph,Csornyei2022}. However, in-depth analysis of  $O-C$ diagram reveals a wave-like signal superimposed to the evolutionary parabolic trend \citep{Csornyei2022} as presented in Fig.~\ref{fig:OC}.  Unfortunately, our sequence of observations used in this paper falls exactly in the wave-like signal observed in the $O-C$ diagram (see blue strip in Fig.~\ref{fig:OC}). To correct for these effects, we took into account the $O-C$ correction modeled by a parabola and we also applied a second order correction corresponding to a linear fit of the $O-C$ residuals in our observation range (see Fig.~\ref{fig:OC}). In order to phase the observations to maximum light in $V$-band we used reference epoch $T_0$=39853.30 and pulsation period $P$=17.12413$\,$days from \cite{GCVS2017}. We authorize a slight phase shift ($\delta \phi=$0.02) to perfectly align at the phase of maximum light.

\subsection{Modeling interstellar extinction}\label{sect:ISM}
Since we used photometric bands in multiple filters from the visible to the infrared, we must define a consistent model to correct for interstellar extinction in the line of sight of Y~Oph. We choose a standard total-to-selective extinction ratio $R_V=A_V/E(B-V)=3.1$ corresponding to an average extinction law of diffuse interstellar medium along the line of sight \citep{Savage1979,Wang2017}. In the infrared regime, the extinction law is approximated by a power-law $A_\lambda\propto\lambda^{-\alpha}$. However, as highlighted by \cite{Wang2017,Lara2019} there is a diversity of extinction curves in the diffuse interstellar medium and towards the Galactic Center.
 The determination of the extinction coefficient $\alpha$ increased significantly in the past decades from about 1.5 \citep{Rieke1985,cardelli89} up to a steeper slope with $\alpha>2.0$ \citep[see][and references therein]{Lara2019}. 
 Among the variety of Milky Way extinction curves available in the near- and mid-IR \citep[e.g.][]{cardelli89,indebetouw2005,Nishiyama2009,Wang2017,Chen2018}, we choose to adopt interstellar extinction law in the near- and mid-IR from \cite{Nishiyama2008,Nishiyama2009} which are calibrated from 2MASS and \textit{Spitzer}/IRAC observations consistently with our photometric bands in Sect.~\ref{sect:obs}.

\subsection{Modeling the circumstellar envelope}\label{sect:ir_excess}
\subsubsection{Impact on the photometry}
In the case of classical Cepheids it is often difficult to fit all the photometry in different bands at once from a simple atmosphere model because of the degeneracy between distance, interstellar extinction, and circumstellar envelope (CSE) emission and absorption. Unfortunately, the impact of CSE on the Cepheid photometry is often omitted and might be a source of systematic error on individual distance and color excess measurement.
 CSEs were resolved from visible and infrared interferometry around several Galactic Cepheids \citep{kervella06a,merand06,merand07,kervella09,gallenne11,gallenne13b,nardetto16,Hocde2021}. In the case of Y~Oph, \cite{merand07} have shown that a CSE emission of about $-0.05\,$mag in the $K$-band can explain the bias between CHARA/FLUOR and VLTI/VINCI interferometric observations \citep{kervella04c}. \cite{Gallenne2012} also detected IR emission from photometry analysis and modeled a significantly hotter CSE than others Cepheids.
  The infrared emission could be caused by a shell of ionized gas around the Cepheids \citep{Hocde2020a,Hocde2020b} or a dust envelope \citep{gallenne13b,Gro2020dust}. However, an excess in the near-IR is likely only caused by breaking radiation from a hot ionized gas envelope \citep{Hocde2020a}, as the dust is too cold to produce significant emission. In the case of ionized hydrogen opacity, absorption can be caused by bound-free processes in the optical, while free-free emission dominates in the infrared. The IR excess exhibits an asymptotic behaviour with longer wavelength because the ionized gas becomes optically thick \citep{Hocde2020a}. Previous studies modeled the IR excess with a parametric power law, assuming that there is no excess, nor deficit in the visible range \citep{merand15,Trahin2021,Gallenne2021}.
  In order to model analytically this process in a convenient way, we introduce the following logistic function:
\begin{equation}\label{eq:ir_excess}
    \Delta \mathrm{mag}_\lambda(\alpha,\beta,\lambda_0)=\alpha \left(\frac{1}{1+e^{\beta(\lambda-\lambda_0)}}-\frac{1}{2}\right)
\end{equation}
where $\alpha$ and $\beta$ represent the intensity and the slope of the logistic function respectively, and $\lambda_0$ is the pivot wavelength for which $\Delta \mathrm{mag}_{\lambda_0}=0$. Although this parametric model certainly provides more physical justification compared to a simple power law, we note that it cannot reproduce complex features from bound-free absorption, a task that can only be accomplished by radiative transfer models.

\subsubsection{Impact on angular diameter measurement}
The CSE model has also an impact on the measured uniform disk (UD) angular diameter $\theta_\mathrm{UD}$ via interferometry. \cite{merand07} have shown that $\theta_\mathrm{UD}$ measurements are slightly biased toward larger diameter in the $K$-band because the CSE emission is partially resolved. More precisely, the bias is $k=\theta_\mathrm{UD}/\theta_\star\approx1.023$ in the case of Y~Oph \citep{merand07}, where $\theta_\star=\theta_\mathrm{LD}$ is the limb-darkened angular diameter. In principle, it is necessary to assume a CSE geometry and opacity to derive the bias on $\theta_\mathrm{UD}$ on the CHARA/FLUOR band. For the sake of simplicity, we assume in the following a bias of $k=1.023$ to take into account the CSE \citep{merand07}.

\subsection{Fitting process}\label{sect:fit}
Our fitting approach consists of adjusting simultaneously the distance, color excess and CSE parametric model to fit the angular diameter and the photometries.

Indeed, the photosphere radius $R_\mathrm{RSP}$ derived by \texttt{MESA-RSP} allows to determine unambiguously the distance via the angular diameter observations following:
\begin{align}\label{eq:distance}
&\theta_\mathrm{LD}(\phi) \propto \frac{R_\mathrm{RSP}(\phi)}{\mathrm{d}_\mathrm{pc}}
\end{align}

Simultaneously to the angular diameter, we fit the entire set of observed light curves in $VJHK_SLM$ bands. To this end, we converted the absolute magnitudes $M_\lambda(\phi)$ in each photometric bands along the pulsation cycle into apparent magnitudes following:
\begin{align}\label{eq:mag}
&m_\lambda(\phi)=M_\lambda(\phi)+5\,\mathrm{log\left(\frac{d_\mathrm{pc}}{10}\right)}+A_\lambda+\Delta\mathrm{mag}_\lambda(\alpha,\beta,\lambda_0)
\end{align}
where $A_\lambda$ is the interstellar absorption in each band and $\Delta\mathrm{mag}_\lambda(\alpha,\beta,\lambda_0)$ is the absorption or extinction produced by the parametric CSE model as a function of the effective wavelength $\lambda$ of each synthetic filter (see Eq.~\ref{eq:ir_excess}).

To summarize, we fit simultaneously Eqs.~\ref{eq:distance} and \ref{eq:mag}, adjusting the free parameters $d_\mathrm{pc}$, $E(B-V)$ and the CSE model parameters, $\alpha$, $\beta$ and $\lambda_0$.
We performed a reduced $\chi^2$ statistic from python Kapteyn package \citep{KapteynPackage} which makes use of the Marquardt-Levenberg algorithm \citep{Levenberg1944,Marquardt} to solve the least-squares problem. For every quantity fitted in our study, we derived statistical errors using the bootstrap method. The fitting of each light curves is displayed in Fig.~\ref{fig:non_linear_8M} and in the appendix (Figs.~\ref{fig:non_linear_3M},  
 \ref{fig:non_linear_4M},  
  \ref{fig:non_linear_5M}, \ref{fig:non_linear_6M}, \ref{fig:non_linear_7M}). The resulting CSE model is displayed in Fig.~\ref{fig:IR_excess} and the quantitative results are provided in Table \ref{Tab:results}.

Despite small statistical uncertainties of the order of a few parsecs, as presented in Table~\ref{Tab:results}, our distance calculation might be biased by larger systematic errors due to the assumptions of our modelling. For example, the derived radius might be sensitive to the various choices of hydrodynamical modeling parameters. It is difficult to estimate the systematic uncertainty of the derived distance without in-depth pulsation modeling. 
Hence, for the rest of the study we assumed arbitrarily an uncertainty of $\pm$15$\,$pc.

In this method, we note that we did not take into account the RV curves because it requires to fit the $p$-factor to transform RSP pulsation velocity ($V_p$) into RV measurements ($V_\mathrm{RV}$) following $V_{p}=pV_\mathrm{RV}$. This is possible in principle, as it was done for example by \cite{Marconi2013}. However, this method is relevant only if the light amplitude modeled from \texttt{MESA/RSP} is adjusted in order to fit the observed light amplitude. Since we do not perform a fit but simply compare the observations to a grid of models, we cannot include the RV curves in the calculation. For each model however, we simply assume $p=1.27$ and present the result for consistency only.

   \begin{figure*}[]
    \centering
    \begin{subfigure}{0.33\textwidth}
        \includegraphics[width=\linewidth]{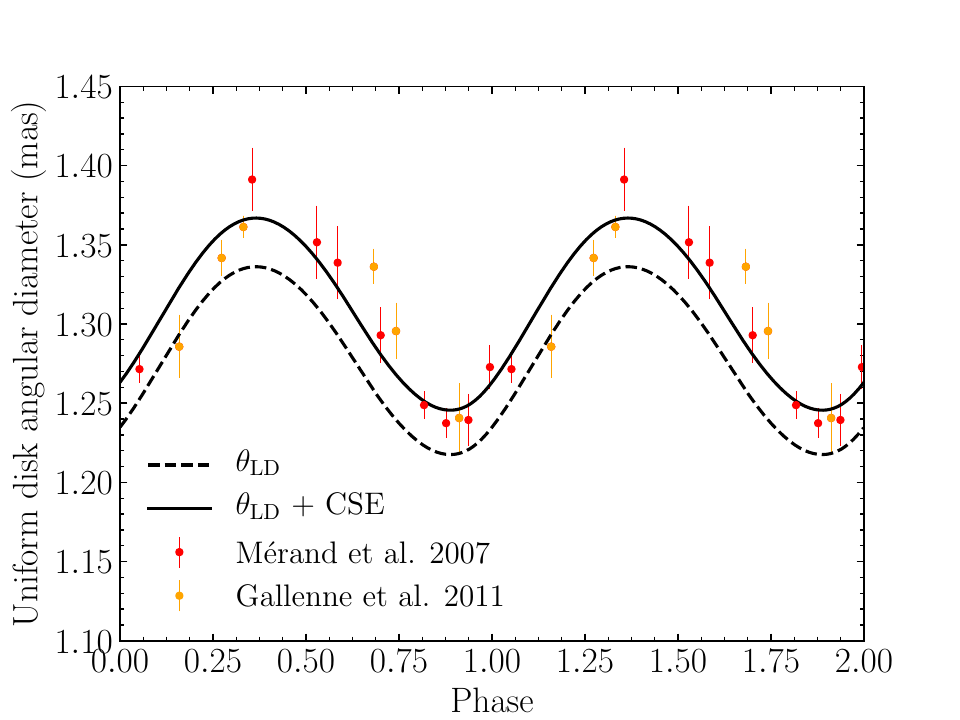}
        \caption{}
        \label{fig:R_8M}
    \end{subfigure}
    \begin{subfigure}{0.33\textwidth}
        \includegraphics[width=\linewidth]{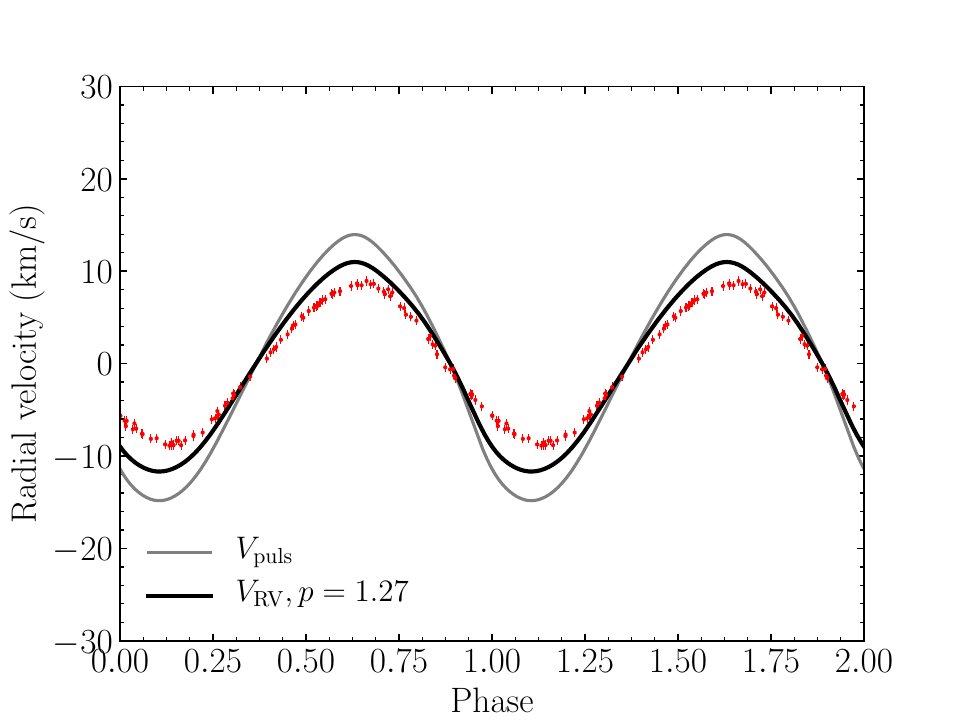}
        \caption{}
        \label{fig:RV_8M}
    \end{subfigure}
    \begin{subfigure}{0.33\textwidth}
        \includegraphics[width=\linewidth]{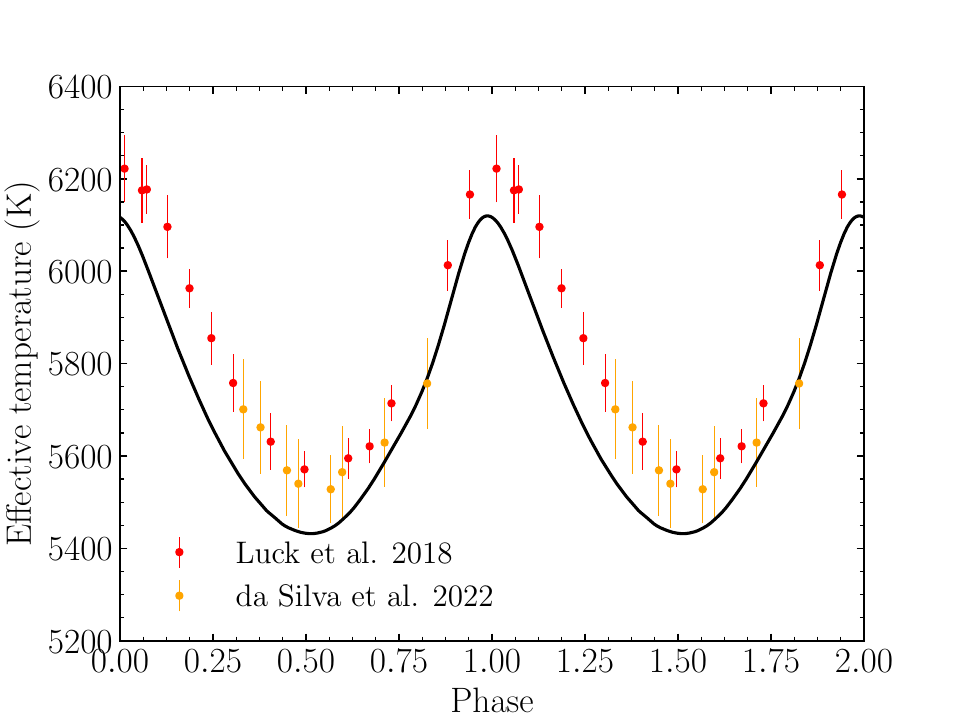}
        \caption{}
        \label{fig:Teff_8M}
    \end{subfigure}

    \medskip
        \begin{subfigure}{0.33\textwidth}
        \includegraphics[width=\linewidth]{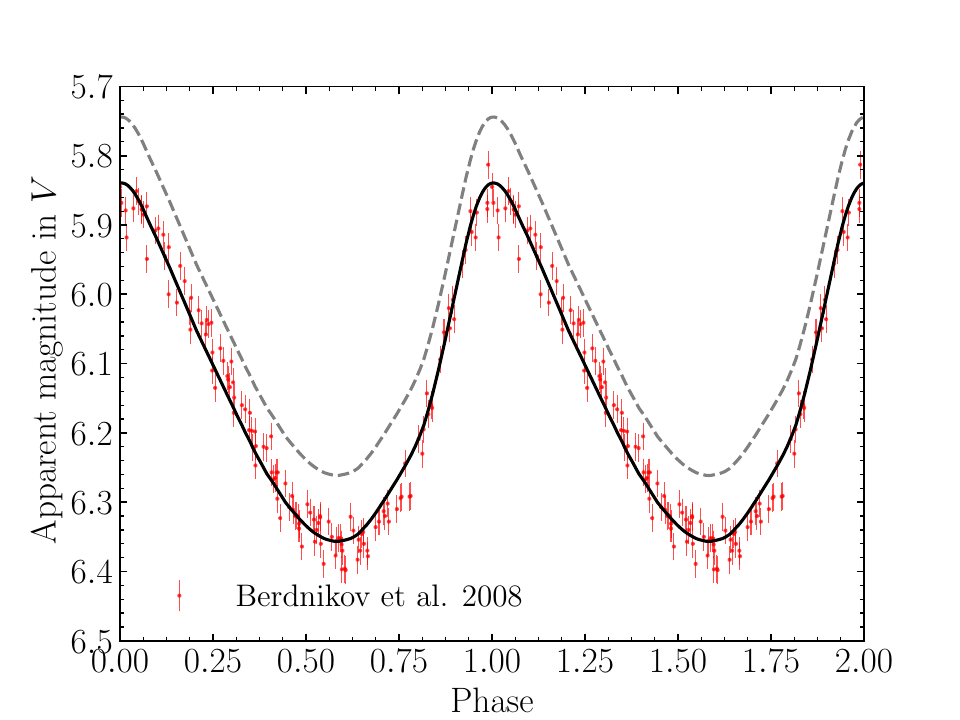}
        \caption{}
        \label{fig:V_8M}
    \end{subfigure}
    \begin{subfigure}{0.33\textwidth}
        \includegraphics[width=\linewidth]{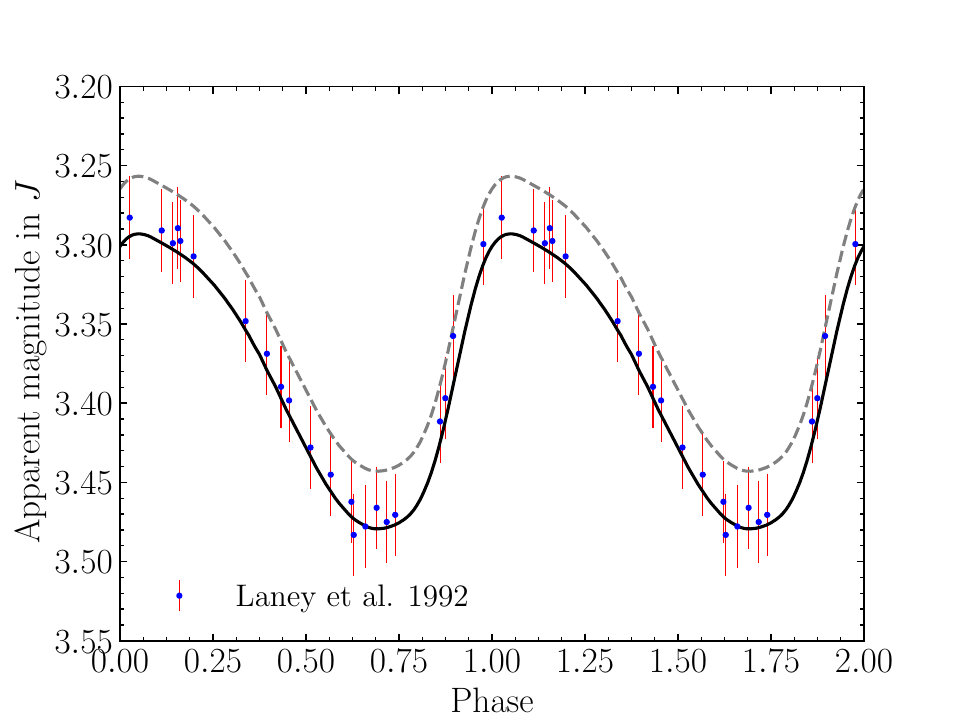}
        \caption{}
        \label{fig:J_8M}
    \end{subfigure}
    \begin{subfigure}{0.33\textwidth}
        \includegraphics[width=\linewidth]{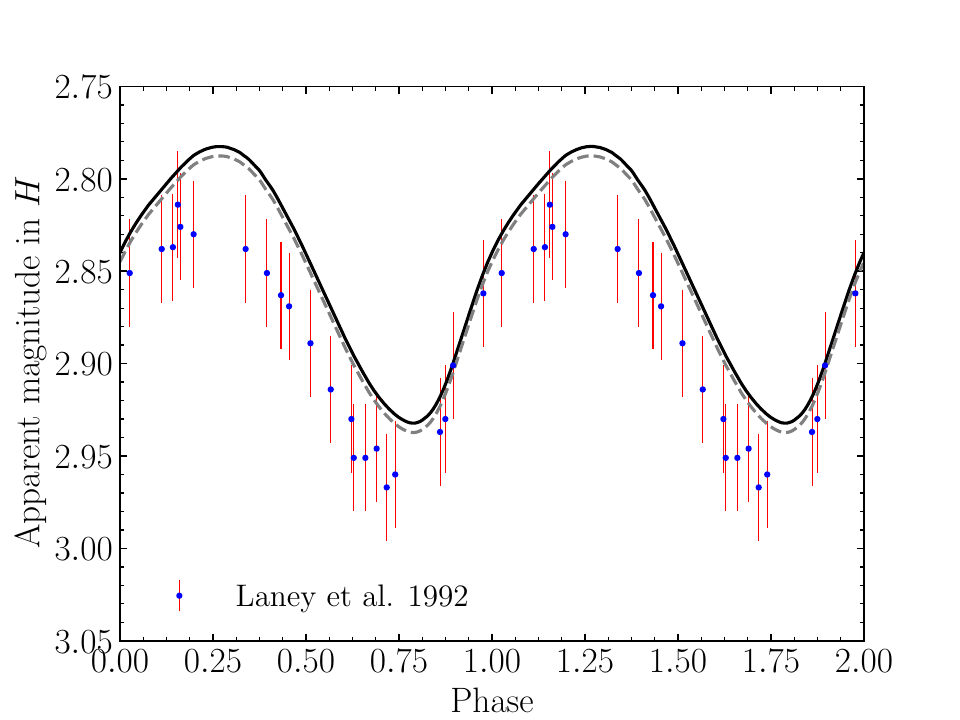}
        \caption{}
        \label{fig:H_8M}
    \end{subfigure}

    \medskip

    \begin{subfigure}{0.33\textwidth}
        \includegraphics[width=\linewidth]{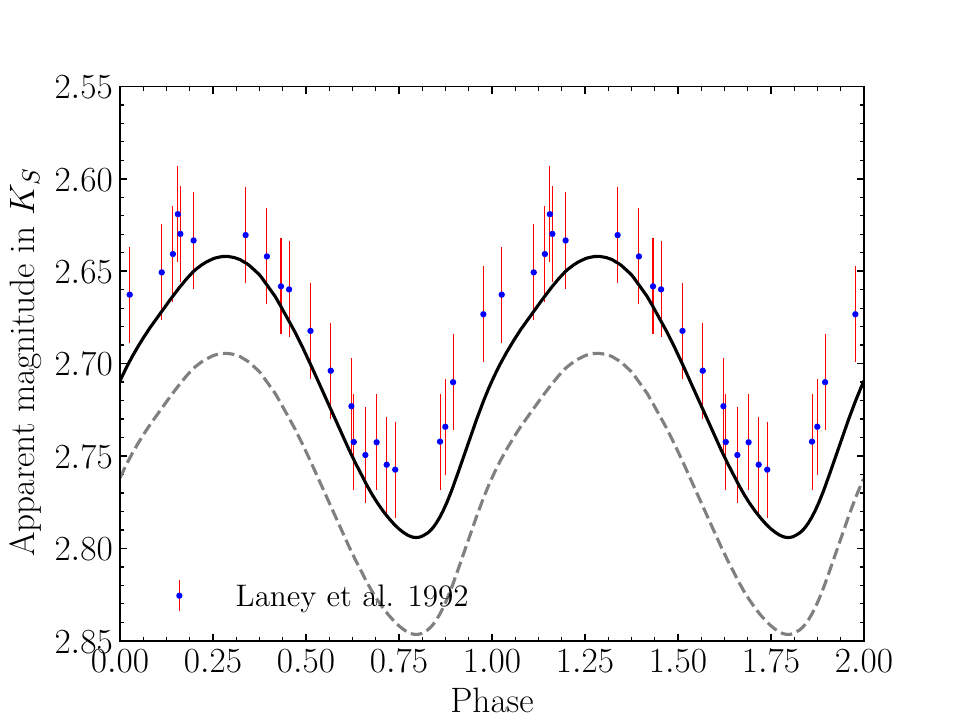}
        \caption{}
        \label{fig:K_8M}
    \end{subfigure}
    \begin{subfigure}{0.33\textwidth}
        \includegraphics[width=\linewidth]{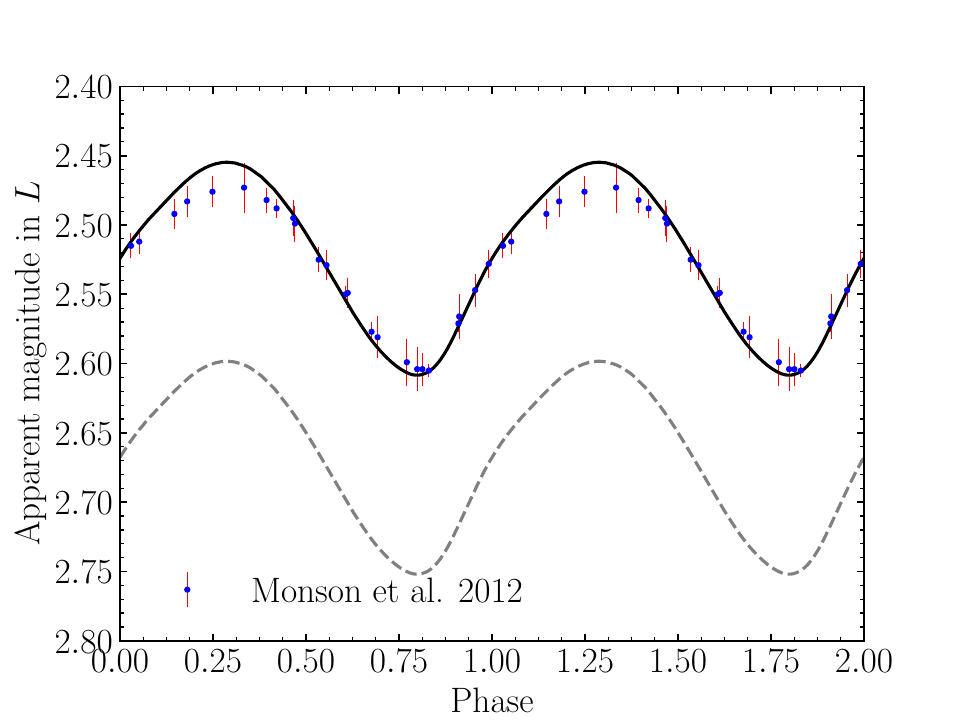}
        \caption{}
        \label{fig:L_8M}
    \end{subfigure}
    \begin{subfigure}{0.33\textwidth}
        \includegraphics[width=\linewidth]{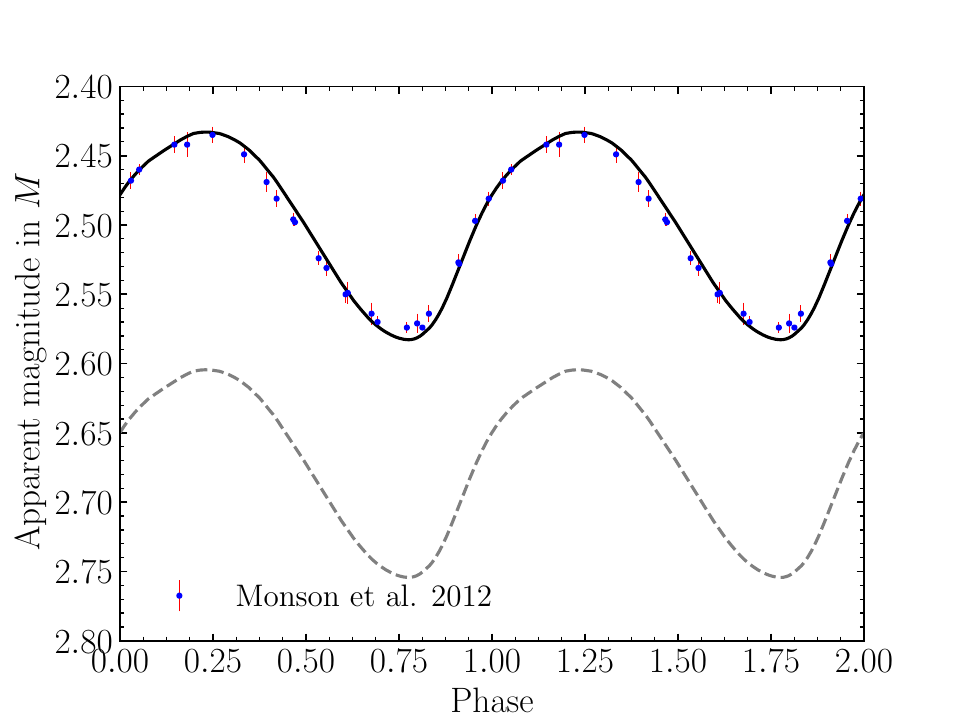}
        \caption{}
        \label{fig:M_8M}
    \end{subfigure}

\caption{Best result of the non-linear analysis with \texttt{RSP/MESA} for 8$\,M_\odot$ model. Uniform disk angular diameter, radial velocity curve and effective temperature are displayed in (a), (b) and (c) respectively. The photometric panels indicate the apparent magnitudes in (d) $V$-band, (e) $J$-band (f) $H$-band (g) $K_S$-band (h) $L$-band (i) $M$-band. In the angular diameter and photometric panels, thick black line and dashed grey lines are RSP models with and without CSE models respectively.}\label{fig:non_linear_8M}
\end{figure*}

\begin{figure}
\includegraphics[width={0.52\textwidth}]{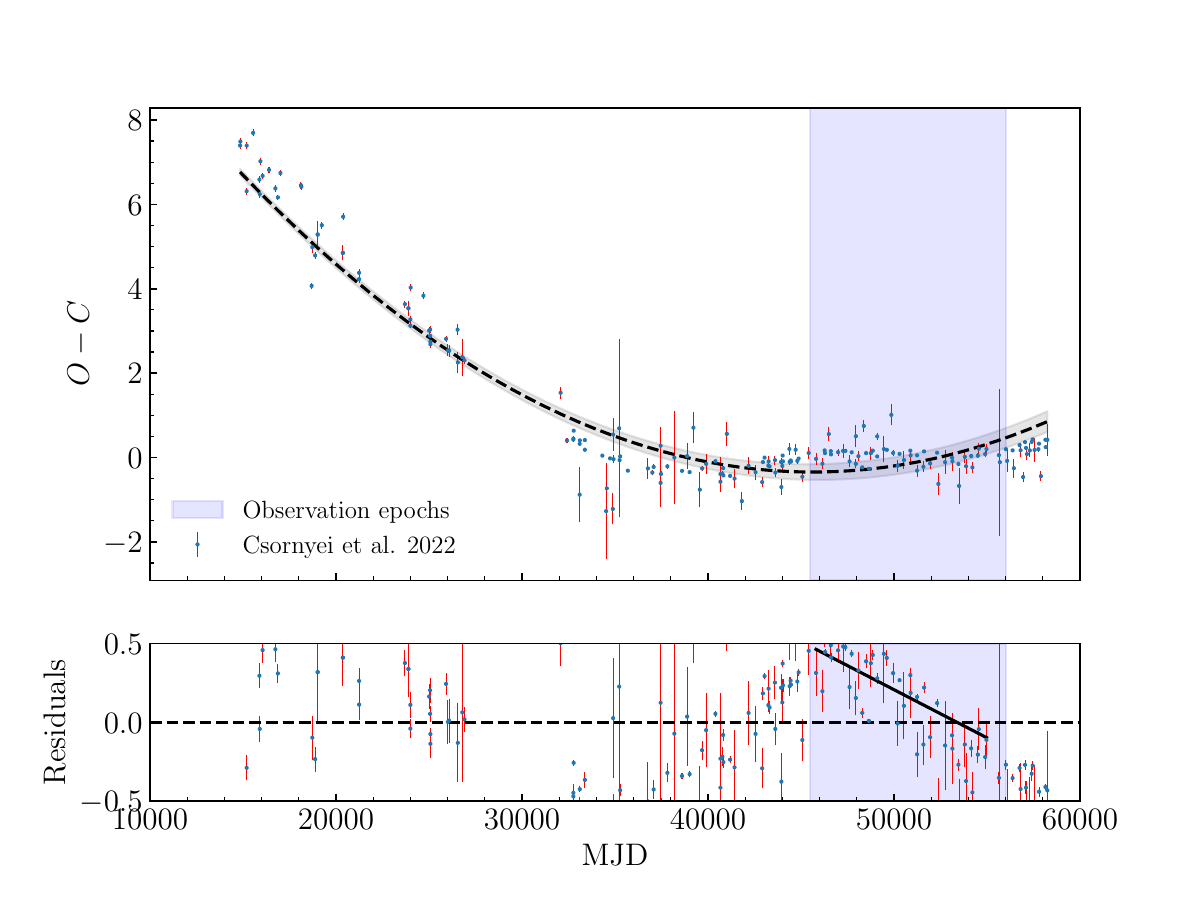}
\caption{\small $O-C$ diagram of Y Oph. $O-C$ calculations were performed by \cite{Csornyei2022}. The dashed line is a parabola fit over the entire range of data. The continuous line on the bottom plot is the linear fit of the $O-C$ residuals over the observation period used in this paper. \label{fig:OC}}
\end{figure}

\begin{figure}[h]
\includegraphics[width={0.52\textwidth}]{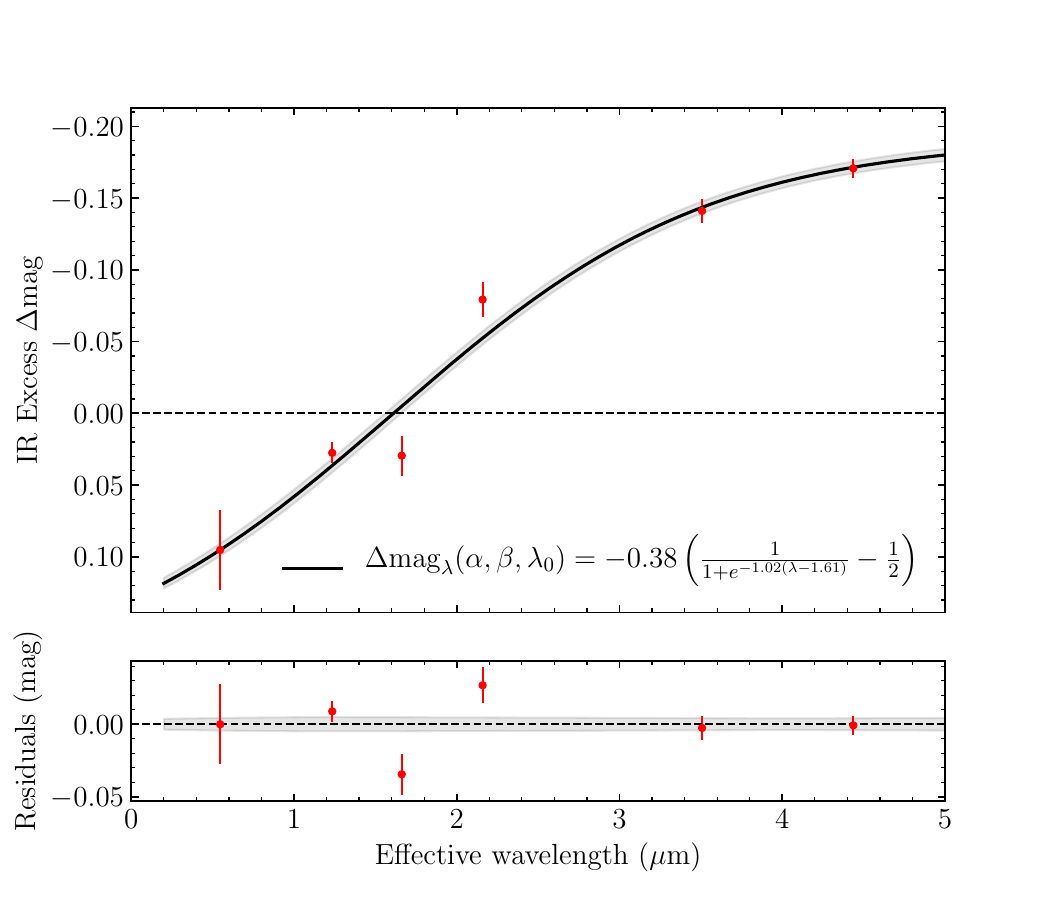}
\caption{\small Fit of the parametric CSE model derived from Eq.~\ref{eq:ir_excess} in the case of the best fit at 8$\,$M$_\odot$. Errors bars represent the standard deviation of photometric measurements compared to the model.  The grey region is the error
on the magnitude obtained using the covariance matrix of the fitting result. \label{fig:IR_excess}}
\end{figure}

\begin{figure}
\includegraphics[width={0.5\textwidth}]{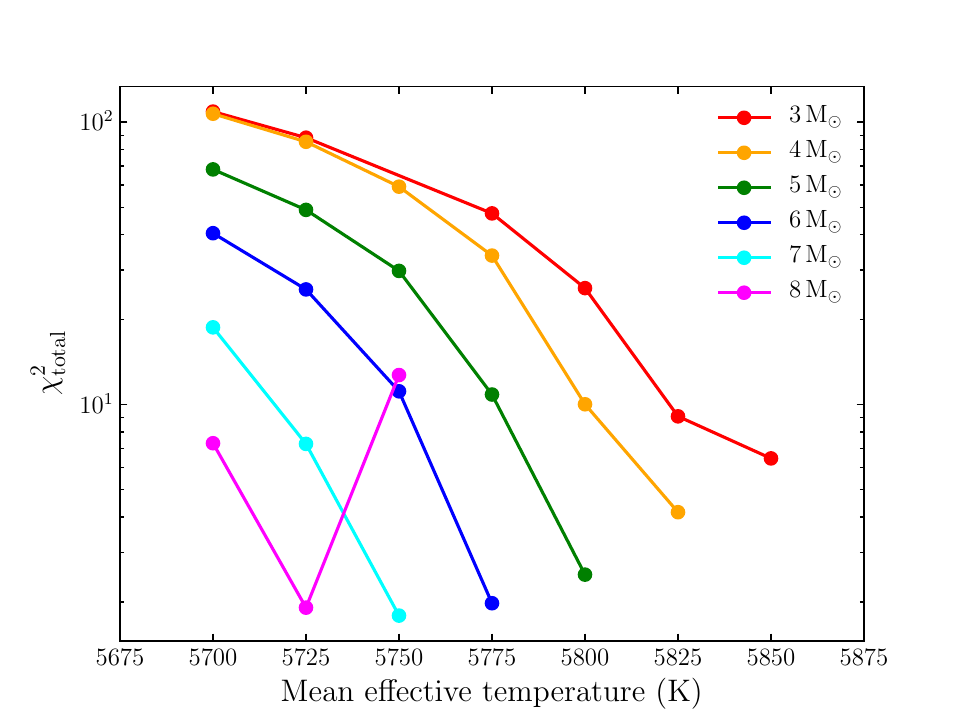}
\caption{\small Mean $\chi_r^2$ {\it vs.} the mean effective temperature for models of different masses.\label{fig:chi2}}
\end{figure}

Finally, in order to assess which models are the best to fit the observations, we computed the total $\chi^2_\mathrm{total}$ to take into account the following observations along the pulsation cycle: light curves, angular diameter and effective temperature. To this end, the total $\chi^2_\mathrm{total}$ is derived as the average $\chi^2$ for all observable. Hence, each observable contributes equally to the final likelihood estimation. Results are presented in Table~\ref{Tab:results}. For each mass, we displayed $\chi^2_\mathrm{total}$ as a function of the effective temperature in Fig.~\ref{fig:chi2}.

\begin{table*}[h]
\centering
\caption{Best fits of the non-linear models. \label{Tab:results}}
\begin{tabular}{ccr|ccc|ccccc|S}
\hline
\multicolumn{3}{c|}{Grid of models (Table \ref{Tab:models})}&\multicolumn{3}{c|}{Non-linear calculation}&\multicolumn{6}{c}{Fitting results}\\
\hline
$M_\star$ & $T_\mathrm{eff}$(K) & $L_\star(L_\odot)$ & $R_\star(R_\odot)$ & $P\,$(days) & $M_{K_S} \,$(mag) & $d\,$(pc)  & $E(B-V)$ & $\alpha$ & $\beta$ &$\lambda_0(\mu$m)& $\chi^{2}_\mathrm{tot}$\\

\hline

\hline
                     & 5850 & 4857 &68.84 & 17.05 & $-5.89$ & $504.0_{1.6}$ & $0.664_{0.006}$& $-0.27_{0.03}$ & $-1.79_{0.32}$& 1.82$_{0.14}$ & 6.45\\
3$\,\mathrm{M}_\odot$ & 5825 & 4785 &69.52 & 17.17 & $-5.90$ & $504.9_{1.3}$ & $0.653_{0.004}$& $-0.32_{0.03}$ & $-1.30_{0.17}$& 1.84$_{0.13}$ & 9.09\\
                     & 5800 & 4714  &70.09 & 17.26 & $-5.91$ & $505.9_{1.9}$ & $0.643_{0.007}$& $-0.37_{0.05}$ & $-1.06_{0.23}$& 1.83$_{0.23}$ & 25.86\\

             \hline
                        & 5825 & 6060  &77.59 & 17.06 & $-6.14$ & $566.7_{1.7}$ & $0.655_{0.004}$& $-0.29_{0.02}$ & $-1.55_{0.20}$& 1.79$_{0.12}$ & 4.16\\
4$\,\mathrm{M}_\odot$   & 5800 & 5961 &78.37 & 17.13 & $-6.15$ & $567.2_{1.7}$ & $0.643_{0.004}$& $-0.35_{0.03}$ & $-1.14_{0.15}$& 1.80$_{0.14}$ & 10.03\\
                        & 5775 & 5862 &79.10 & 17.23 & $-6.16$ & $568.9_{2.7}$ & $0.631_{0.008}$& $-0.43_{0.07}$ & $-0.87_{0.24}$& 1.77$_{0.28}$ & 33.72\\

             \hline

& 5800 & 7180   &85.45 & 17.15 & $-6.34$ & $622.0_{1.6}$ & $0.645_{0.003}$& $-0.32_{0.02}$ & $-1.34_{0.12}$& 1.76$_{0.09}$ & 2.50\\
5$\,\mathrm{M}_\odot$ & 5775 & 7061 &86.25 & 17.21 & $-6.35$ & $622.6_{2.0}$ & $0.633_{0.004}$& $-0.38_{0.03}$ & $-1.01_{0.14}$& 1.75$_{0.15}$ & 10.86\\
& 5750 & 6942 &86.81 & 17.27 & $-6.35$ & $623.8_{2.9}$ & $0.622_{0.008}$& $-0.46_{0.08}$ & $-0.80_{0.22}$& 1.72$_{0.27}$ & 29.76\\

             \hline

& 5775 & 8120  &91.83 & 17.02 & $-6.49$ & $666.5_{1.7}$ & $0.636_{0.002}$& $-0.34_{0.01}$ & $-1.19_{0.08}$& 1.72$_{0.07}$ & 1.98\\
6$\,\mathrm{M}_\odot$  & 5750 & 7994 &92.61 & 17.09 & $-6.50$ & $667.6_{2.1}$ & $0.625_{0.005}$& $-0.41_{0.04}$ & $-0.93_{0.14}$& 1.70$_{0.16}$ & 11.15\\
& 5725 & 7870 &93.14 & 17.16 & $-6.50$ & $669.2_{2.9}$ & $0.615_{0.008}$& $-0.48_{0.08}$ & $-0.76_{0.21}$& 1.65$_{0.26}$ & 25.59\\

\hline

& 5750 & 9194 &98.38 & 17.24 & $-6.64$ & $714.5_{1.9}$ & $0.630_{0.002}$& $-0.35_{0.01}$ & $-1.16_{0.08}$& 1.67$_{0.07}$ & 1.79\\
7$\,\mathrm{M}_\odot$  & 5725 & 9031 &99.07 & 17.28 & $-6.64$ & $715.0_{2.2}$ & $0.619_{0.004}$& $-0.42_{0.03}$ & $-0.90_{0.11}$& 1.63$_{0.13}$ & 7.26\\
& 5700 & 8868 &99.56 & 17.33 & $-6.64$ & $716.3_{3.0}$ & $0.608_{0.007}$& $-0.49_{0.07}$ & $-0.73_{0.18}$& 1.56$_{0.24}$ & 18.78\\
\hline

                     & 5750 & 10069 &102.14 & 16.97 & $-6.73$ & $748.4_{3.7}$ & $0.634_{0.007}$& $-0.32_{0.04}$ & $-1.37_{0.29}$& 1.64$_{0.20}$ & 12.73\\
8$\,\mathrm{M}_\odot$ & 5725 & 9915&103.36 & 17.04 & $-6.74$ & $748.6_{2.1}$ & $0.620_{0.002}$& $-0.38_{0.01}$ & $-1.02_{0.06}$& 1.61$_{0.06}$ & 1.91\\
& 5700 & 9764  &104.02 & 17.11 & $-6.74$ & $750.4_{2.6}$ & $0.610_{0.004}$& $-0.45_{0.04}$ & $-0.83_{0.12}$& 1.54$_{0.14}$ & 7.30\\
\hline
\hline

\hline
\end{tabular}

    \begin{tablenotes}
    \item \textbf{Notes : }For each model computed ($M_\star$,$T_\mathrm{eff}$,$L_\star$), we derive the average stellar radius $\mathrm{R}_\star(R_\odot)$, the pulsation period $P$ in days, and the mean absolute magnitude in $K_S$-band $M_{K_S}$. We fitted simultaneously the distance $d(pc)$, color excess $E(B-V)$ and the parameters $\alpha$, $\beta$ and $\lambda_0$ of the CSE model (see Eq.~\ref{eq:ir_excess}). The total $\chi^2_\mathrm{tot}$  is the mean reduced $\chi^2$ from the observations (see Sect~\ref{sect:fit_strategy}).
    \end{tablenotes}
\end{table*}

 \section{Results}\label{sect:results}
 \subsection{Proximity with the blue edge of the IS}
The first striking feature of the non-linear computations is that there is a remarkable agreement between the measurements along the pulsation cycle and almost all models at the highest effective temperature. This is summarized in Fig.~\ref{fig:chi2}, where we plot variations of the total $\chi^2$ {\it vs.} the effective
temperature. We also emphasize the overall success of these pulsation models to remarkably reproduce  simultaneously many different types and complex observations along the pulsation cycle. Indeed, we observe that the best RSP models follow closely all together the variations of angular diameter measured by interferometry, radial velocity and effective temperature obtained by high-resolution spectroscopy and light-curves in $VJHK_sLM$ bands. Hence, our first conclusion is that the quasi-symmetric and low amplitude of the different curves are attributed to the extreme proximity of Y~Oph with the blue edge of the IS, independently of the assumed stellar mass adopted.

\subsection{Distance discrepancy}
From the results of our non-linear computations it is not possible to determine what is the best mass for Y~Oph based only on the shape of the different curves. Indeed, even if the best $\chi_\mathrm{total}^2$ is obtained for the 7$\,$M$_\odot$ models at $T_\mathrm{eff}$=5750$\,$K, all the other best models are essentially equivalent regarding the various modeling assumptions and observational uncertainties. 

\begin{figure*}[] 
\begin{subfigure}{0.50\textwidth}
\includegraphics[width=\linewidth]{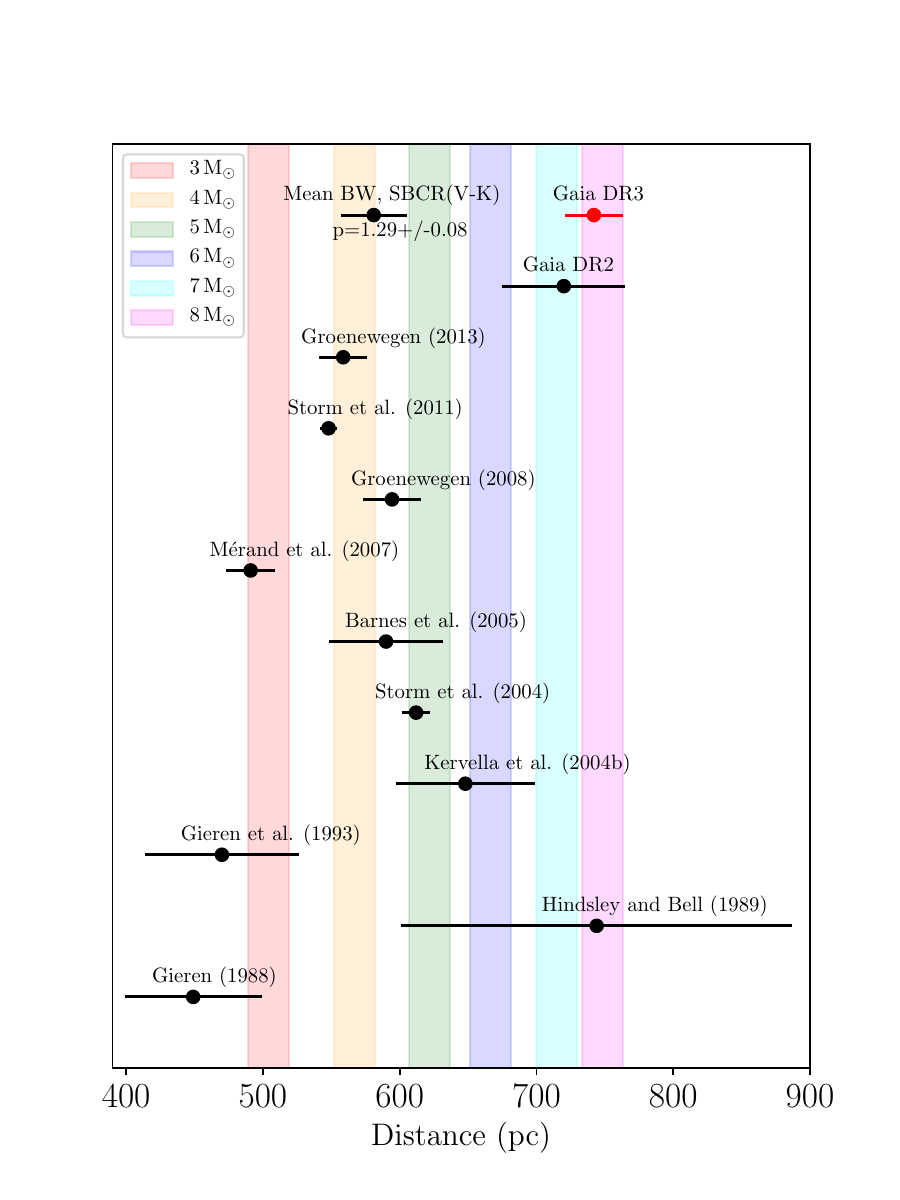}
\caption{} \label{fig:HR_distance}
\end{subfigure}\hspace*{\fill}
\begin{subfigure}{0.50\textwidth}
\includegraphics[width=\linewidth]{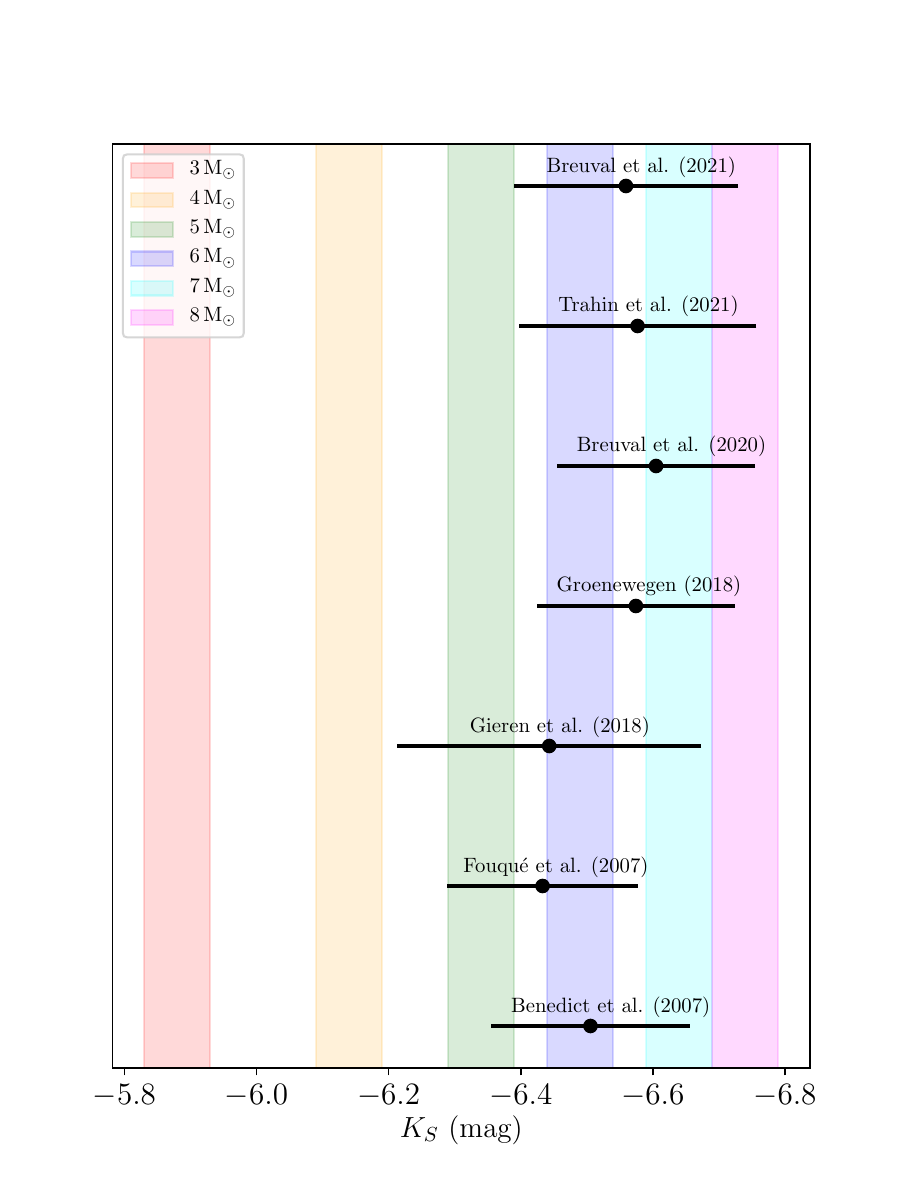}
\caption{} \label{fig:comparison_MK}
\end{subfigure}

\caption{\small \textbf{(a)} Comparison of distances obtained using variant of SBCR Baade-Wesselink methods based on $(V-R)$ \citep{Gieren1988,Hindsley1989,Gieren1993}; $(V-K)$ \citep{storm04,barnes05,storm11a,Groenewegen2008,Groenewegen2013}; Interferometric Baade-Wesselink method \citep{merand07,kervella04a}; distance obtained from the parallax measured by \textit{Gaia} DR2 and DR3 \citep{Gaia2022}; and distance inferred from RSP models  (see Table \ref{Tab:results}) as vertical strips for each stellar mass assuming uncertainty of $\pm$15$\,$pc. The average of both BW distances based on $(V-K)$ color and their $p$-factor are  indicated. \textbf{(b)} Comparison of $K_S$-band absolute magnitude derived from different PL relations \citep{benedict07,fouque07,Gieren2018,Groenewegen2018,Breuval2020,Trahin2021,Breuval2021} established in the 2MASS filter (see \cite{Breuval2020}), and the magnitude as we derived in the 2MASS filter from \texttt{MESA-RSP} non-linear models.}
\end{figure*}

Nevertheless, the derived distance for each mass can be used to discriminate which model is the best. In Table \ref{Tab:BW_literature} and Fig.~\ref{fig:HR_distance}, we provide a comparison of the different distances from the literature, together with the distance we derived from RSP models for each mass. As we can see, the different variants of the BW method yield distances in agreement with the lower stellar mass models for Y~Oph. In particular, the interferometric BW distance (hereafter IBW) from \cite{merand07} is in excellent agreement with the RSP distance obtained for a stellar mass of about 3$\,$M$_\odot$. On the other hand, the \textit{Gaia} distance yields a stellar mass of at least 7$\,$M$_\odot$ according to our modeling. This result is in agreement with masses inferred from pulsation models constrained by \textit{Gaia} parallax of Cepheids with similar pulsation period \citep{Marconi2020,DeSomma2020}, although a slightly higher mass was derived in the case of Y~Oph ($10.4\pm1.4\,$M$_\odot$). Last, the luminosity as predicted by higher stellar mass models are in good agreement with PL relations in the $K_S$ band (see Fig.~\ref{fig:comparison_MK}) while this is discrepant for lower stellar mass in particular for 3$\,$M$_\odot$.

\begin{table*}
\begin{center}
\caption{Summary of Baade-Wesselink distances for Y~Oph in the literature.  \label{Tab:BW_literature}}
\begin{tabular}{lllllc}

\hline
\hline
 	Reference & $d$(pc) &   	$p$-factor	 &   RV data        	&			Angular diameter	&	 	$E(B-V)$\\
\hline
\cite{Groenewegen2013} &	559$\pm$18		&		1.20 &	Multiple					&		SBCR($V-K$)	     &		0.645\\
\cite{storm11a}	        &	548$\pm$6		&		1.32   &   \cite{Petterson2005}    &		SBCR($V-K$)	     &		0.647\\
			                 &				   &				&	\cite{nardetto06a}     &						 &	 \\
\cite{Groenewegen2008} &	594$\pm$21		&		1.20	&	Multiple					&		SBCR($V-K$)		 &		0.645\\
\cite{merand07}         &	491$\pm$18		&		1.27	&   \cite{Gorynya1998}		&		CHARA/FLUOR	 	 &			\\
\cite{barnes05}          &	590$\pm$42		&		1.35 &	\cite{Gorynya1998}		&		SBCR($V-K$)		 &		0.655\\
\cite{storm04}          &	612$\pm$10		&		1.39	&	\cite{Gorynya1998}	&		SBCR($V-K$)		 &		0.655\\
\cite{kervella04a}	  &	648$\pm$51	&		1.36	&	\cite{Evans1986}		&		VLTI/VINCI &			\\
\cite{Gieren1993}       &	470$\pm$56		&	1.35	&	\cite{Coulson1985}		&		SBCR($V-R$)		 &		0.655\\
\cite{Hindsley1989}  &	744$\pm$141		&		1.39	&	\cite{Coulson1985}		&		SBCR($V-R$)		 &		0.660\\
			                 &				   &				&	\cite{LloydEvans1980}    &						 &	 \\
\cite{Gieren1988}  &	449$\pm$50		&		1.39	&	\cite{Coulson1985}		&		SBCR($V-R$)		 &		0.630\\
\hline
Mean BW SBCR($V-K$) &   581$\pm$24          & 1.29$\pm$0.08 &                           &                            &  \\
\hline

\hline
\end{tabular}
\end{center}
    \begin{tablenotes}
    \item \textbf{Notes :} For each reference of BW measurement in the literature, we indicate the derived distance $d$(pc), the projection factor used, the set of radial velocity data, the source of angular diameter which is either derived using SBCR from $(V-K)$ or $(V-R)$ colors, or measured by the mean of interferometric measurements. The last column indicates the color excess adopted. The last line presents the average distance and $p$-factor as derived from BW methods using $(V-K)$ color. The distance measurements are summarized in Fig.~\ref{fig:HR_distance}.
    \end{tablenotes}
\end{table*}

\subsection{Color excess and CSE model}
\begin{figure}
\includegraphics[width={0.50\textwidth}]{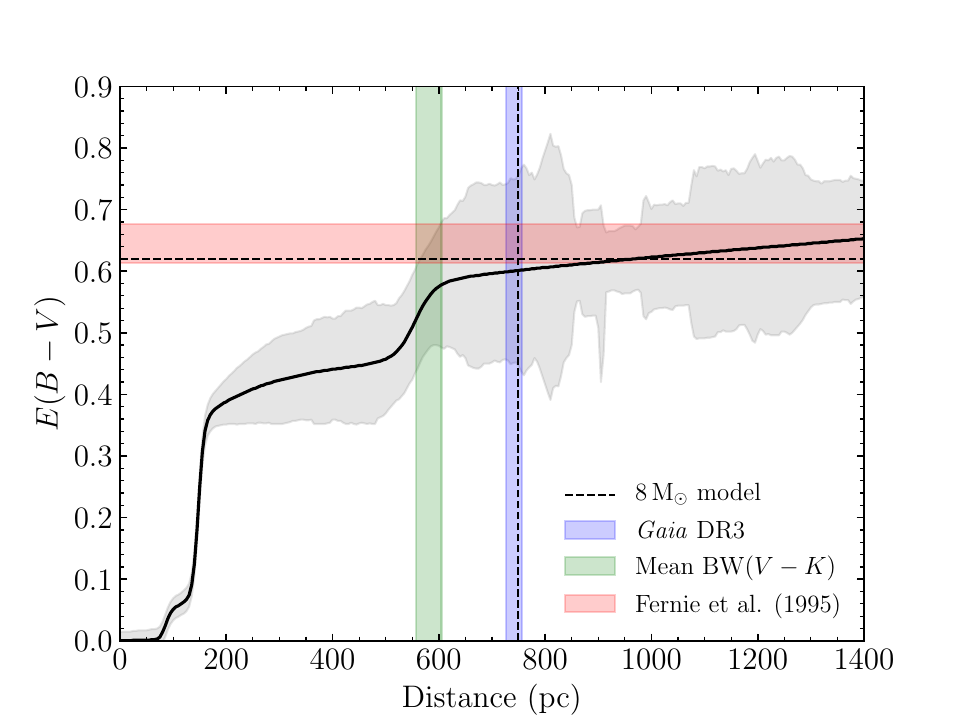}
\caption{\small Color excess $E(B-V)$ in the direction of Y Oph {\it vs.} the distance, from 3D extinction map \citep{Lallement2014,2017stilism}. The horizontal red strip indicates mean $E(B-V)$ and standard deviation of the mean as compiled by \cite{fernie95}. Green and blue strips indicate mean BW$(V-K)$ distance and $d_\textit{Gaia}$ distance \textit{Gaia} DR3 \citep{Gaia2022} respectively.\label{fig:EBV}}
\end{figure}

 For every fit we computed, we observe a range of $E(B-V)$ from about 0.620 and 0.670. These values are in excellent agreement with color excess measurements from the literature: $0.645\pm0.032\,$mag \citep{fernie95}\footnote{\url{http://www.astro.utoronto.ca/DDO/research/cepheids/}}; $0.660\pm0.02\,$mag \citep{Laney2007} and $0.683\pm0.01\,$mag \citep{Kovtyukh2008}.
 In order to check the consistency of the color excess with the derived distance, we compared our result with 3D extinction map in the Galaxy from \cite{Lallement2014,2017stilism} in Fig.~\ref{fig:EBV}. In this Figure, we observe that the \textit{Gaia} DR3 distance is in agreement with $E(B-V)=0.600\pm0.15\,$mag. Our best 8$\,$M$_\odot$ model is also consistent with both \textit{Gaia} DR3 distance and color excess measurements (see dashed lines in Fig.~\ref{fig:EBV}).
 On the contrary, shorter distances derived from BW measurements are only in marginal agreement with 3D extinction map. In particular, distances below 500$\,$pc \citep{Gieren1988,Gieren1993,merand07} are not consistent with a high color excess of 0.600$\,$mag or more. This result suggests that the color excess derived in our calculation is consistent only for distance higher than about 600$\,$pc which excludes almost all distances derived from BW variants (see Fig.~\ref{fig:HR_distance}). The choice of near-IR and mid-IR extinction law in Sect.~\ref{sect:ISM} does not change our main result which is the derived distance for each model. Indeed, the distance depends almost solely on the radius and angular diameter measurement. However, it has an impact on the derived shape of the CSE model adopted. 
 
 As expected, the CSE absorption in the visible range is an additional source of extinction. We derive a CSE absorption in the visible of about $0.10\,$mag (see Fig~\ref{fig:IR_excess}). This level of absorption is in agreement with bound-free absorptions derived by \cite{Hocde2020a} from radiative transfer models. If we recalculate the fit, but omitting the modeling of the CSE in Eq.~\ref{eq:mag}, we still obtain the same distance, but the extinction compensates for the absence of CSE in the visible. For example, in the case of the 8$\,$M$_\odot$ model we find $E(B-V)$=0.666$\pm0.003\,$mag which is 0.033$\,$mag larger. We obtained a  pivot wavelength $\lambda_0$ between 1.5 and 2$\,\mu$m which is also consistent with radiative transfer of ionized gas \citep{Hocde2020a}. In the infrared, we obtain a small excess in the $K$-band close to the CSE emission derived from FLUOR/CHARA observations \citep{merand07}. At longer wavelength, the IR excess stabilizes at about $-0.15\,$mag which is in agreement of the results of \cite{Gallenne2012} who found an excess of 15\% compared to the stellar photosphere at 8.6$\,\mu$m for Y~Oph. This confirms the importance of modeling the CSE in the framework of adjusting atmosphere and pulsation models to visible and infrared observations (see Figs.~\ref{fig:V_8M} to \ref{fig:M_8M}).

\section{Discussion}\label{sect:discussion}

\begin{figure*}[] 
\begin{subfigure}{0.50\textwidth}
\includegraphics[width=\linewidth]{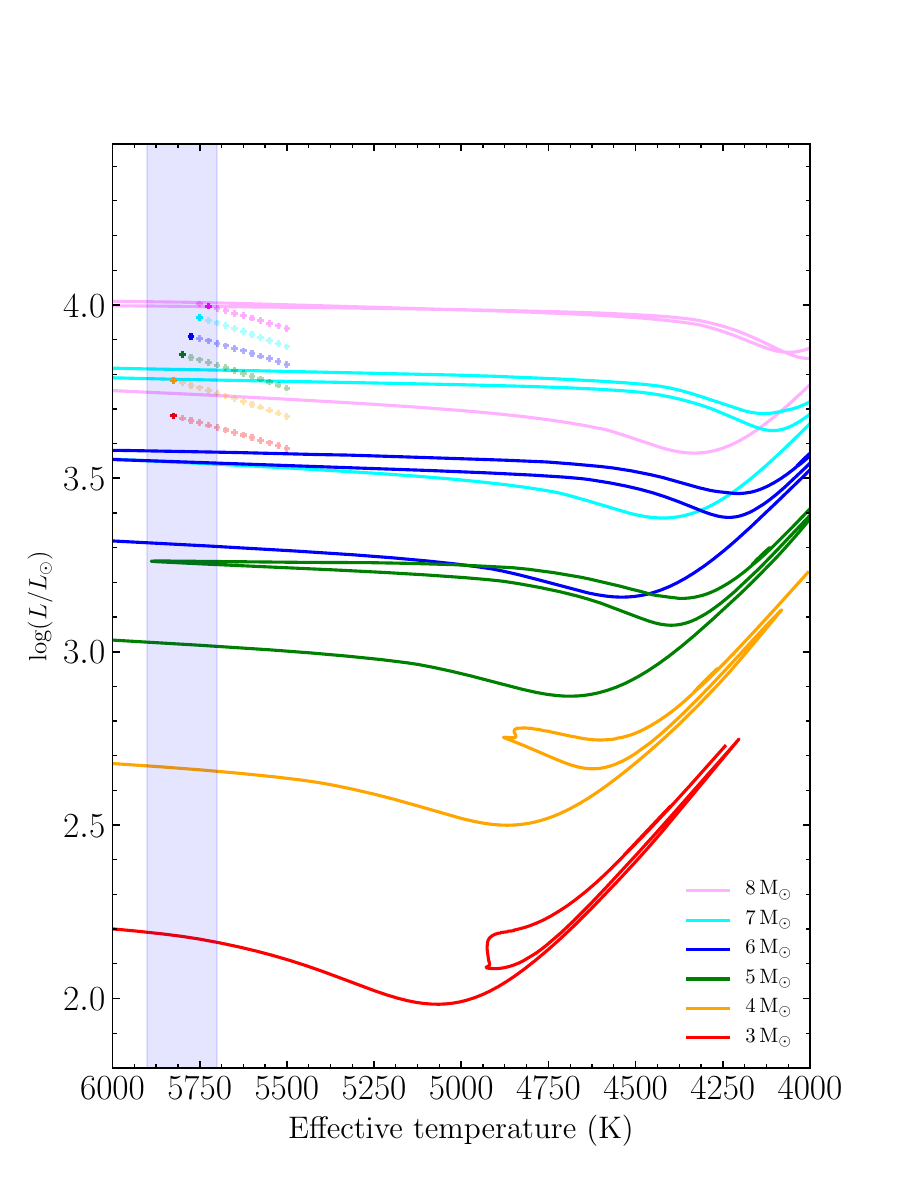}
\caption{} \label{fig:comparison}
\end{subfigure}\hspace*{\fill}
\begin{subfigure}{0.50\textwidth}
\includegraphics[width=\linewidth]{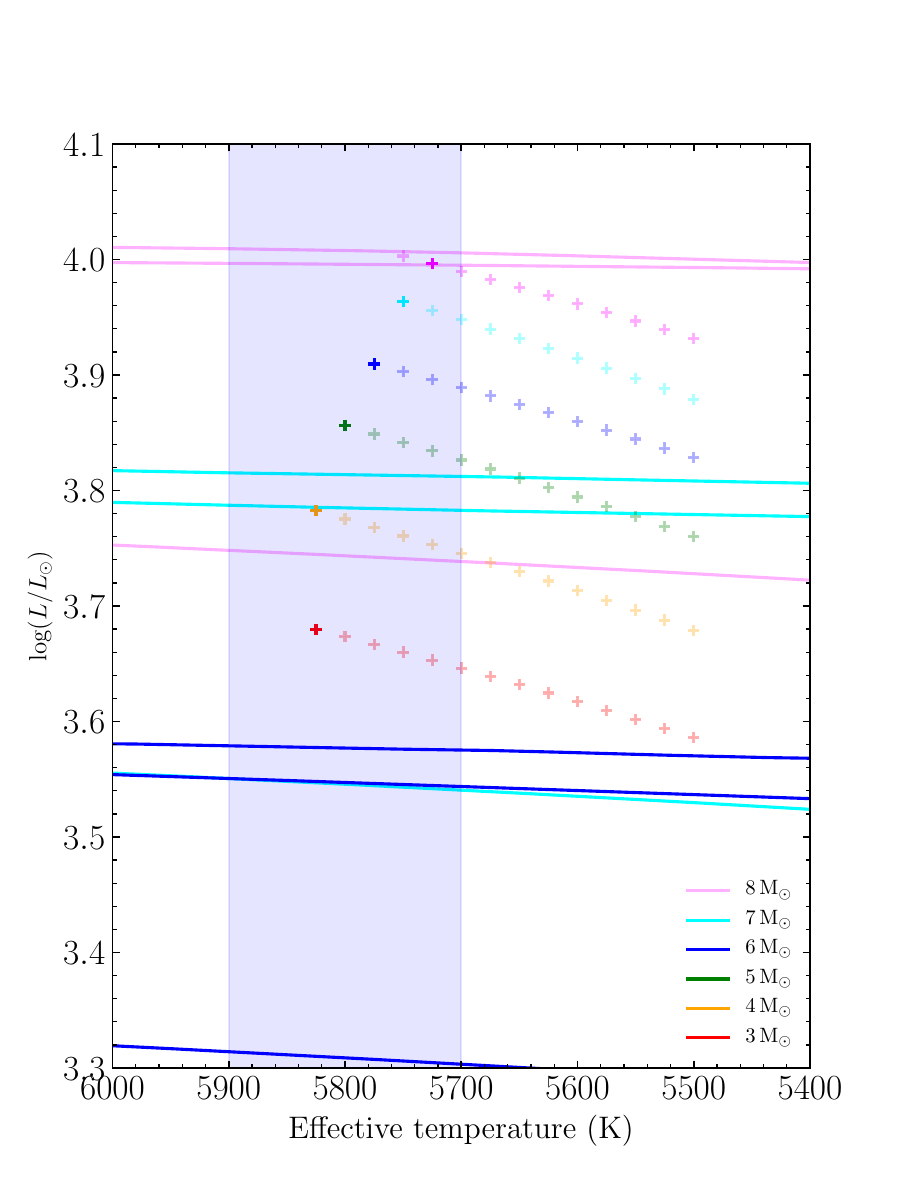}
\caption{} \label{fig:comparison_zoom}
\end{subfigure}

\caption{\small Hertzsprung-Russell diagram comparing parameters of \texttt{MESA-RSP} models of Y~Oph  given in Table \ref{Tab:models} (crosses) with evolutionary tracks computed with MESA for different stellar masses (MESA version r-21.12.1 used by Ziółkowska et al. in prep). These models assume an overshooting parameter $f_\mathrm{ov}=0.02$. The vertical blue strip corresponds to the measured value of the mean effective temperature for Y~Oph, i.e. $T_\mathrm{eff}=5800\pm100\,$K.}
\end{figure*}\label{fig:mesa}

\subsection{Comparison with evolutionary masses}\label{sect:evol}
In order to compare our results with evolutionary models, we computed evolutionary tracks at solar metallicity for the same helium abundance and sample of stellar masses with \texttt{MESA}. The use of MESA for this comparison is most appropriate, since both \texttt{MESA} and \texttt{MESA-RSP} are based on the same implementation of microphysics and use the same numerical algorithms. 
We considered non-canonical models with an overshooting parameter $f_\mathrm{ov}=0.02$, without rotation and no mass-loss. We choose this model to simply illustrate the consistency of evolutionary models with our pulsation modeling. More detailed investigation are necessary but beyond the scope of this paper that is dedicated to the pulsation. Other details of this model will be given in paper in preparation (Ziółkowska et al. in prep).
Comparing \texttt{MESA} evolutionary tracks with \texttt{MESA-RSP} pulsation models (see Fig.~\ref{fig:comparison} and \ref{fig:comparison_zoom}), we observe that only the blue loops of the higher mass models at 8$\,$M$_\odot$ can reach the luminosity of the RSP models. In other words, a simple non-canonical evolutionary model with moderate convective overshooting is able to explain simultaneously the pulsational mass derived for Y~Oph and the \textit{Gaia} distance. The rate of period change of Y Oph $+8.14\pm0.25\,$s/yr \citep{Csornyei2022} is consistent with typical values of long-period Cepheids in the 3rd crossing of the instability strip \citep{turner2006}. The rate of period change for stars of similar pulsation period in the first crossing is at least one order of magnitude larger. We can thus rule out that the star is on its first crossing.

On the contrary, Fig.~\ref{fig:mesa} also shows that the pulsation models with masses of 3 and 5$\,$M$_\odot$, i.e. masses derived for the BW distances, are overluminous as compared to the corresponding evolution tracks. In fact, the blue loops of 3 to 4$\,\mathrm{M}_\odot$ evolutionary tracks do not even penetrate the $T_\mathrm{eff}$ range of Y Oph, as it is also shown by other evolutionary computations \citep{anderson14,DeSomma2021}.

We note that the presence of CSE could indicate significant prior mass-loss although we did not take this phenomenon into account in our evolutionary models. In this case, Y~Oph could be produced by a higher prior mass Cepheid. This is not surprising, as mass-loss is one of the ingredient to solve mass discrepancy of Cepheids. The range of mass-loss rates generally attributed to Cepheids are between $10^{-10}$ and $10^{-6}\mathrm{M}_\odot/\mathrm{yr}$ \citep[see e.g. ][]{Deasy1988,neilson2008,gallenne13b}. In the case of Y~Oph, \cite{Gallenne2012} derived a minimum mass-loss of $2\times 10^{-10}\mathrm{M}_\odot/\mathrm{yr}$ assuming that the mass loss is driven by radiation pressure on the dust envelope. Although there is no consensus on the mass-loss mechanism, it is expected that pulsation-driven mass loss might be the most efficient \citep{neilson2008,neilson12}. The effect of mass-loss is likely not significant on our pulsation modeling which represents a too short time interval for the Cepheid, but the structure of the blue loop as computed by evolutionary models might be affected. However, evolutionary codes implemented simple empirical models such as \cite{Reimers1977,deJager1988} that are used by default to study Cepheid mass-loss. Unfortunately, these models are not only inefficient, but also not relevant for Cepheids for which pulsation likely plays a crucial role in driving the mass-loss. Therefore, the effect of mass-loss on Cepheid evolution remains to be understood to compare consistently together pulsation and evolution models. More detailed evolutionary calculations, will be necessary to better understand the evolutionary state of Y~Oph.

\subsection{BW distances versus \textit{Gaia} distance}
If we assume that the \textit{Gaia} parallax is accurate for Y~Oph, the question is now to understand why \textit{all} BW distances are discrepant (see Fig.~\ref{fig:HR_distance}).
For every BW distance presented in this paper, we note the heterogeneity of these analyses in terms of fitting procedure, source of radial velocity, angular diameter measured by interferometry or deduced by SBCRs, photometry used and color extinction as presented in Table~\ref{Tab:BW_literature}. All these elements impact the distance measurements differently \citep[see e.g. ][]{Nardetto2023} and might explain why BW distances are scattered. Overall, the mean BW distance derived from these methods is $581\pm$24$\,$pc for an average $p$-factor $p=1.29\pm0.08$ (see Fig.~\ref{fig:HR_distance}). This average distance is clearly in disagreement with our computed models at 7 and 8$\,$M$_\odot$ and \textit{Gaia} distances (DR2 and DR3) and would require a $p$-factor above 1.5 (formally (742/581)$\times$1.29=1.65). Before any conclusion about causes of discrepancy between BW and \textit{Gaia} distances is drawn, we must first scrutinize the SBCR used for Y Oph distance determination.

\subsection{Impact of the CSE on the SBCR}
\begin{figure}
\includegraphics[width={0.5\textwidth}]{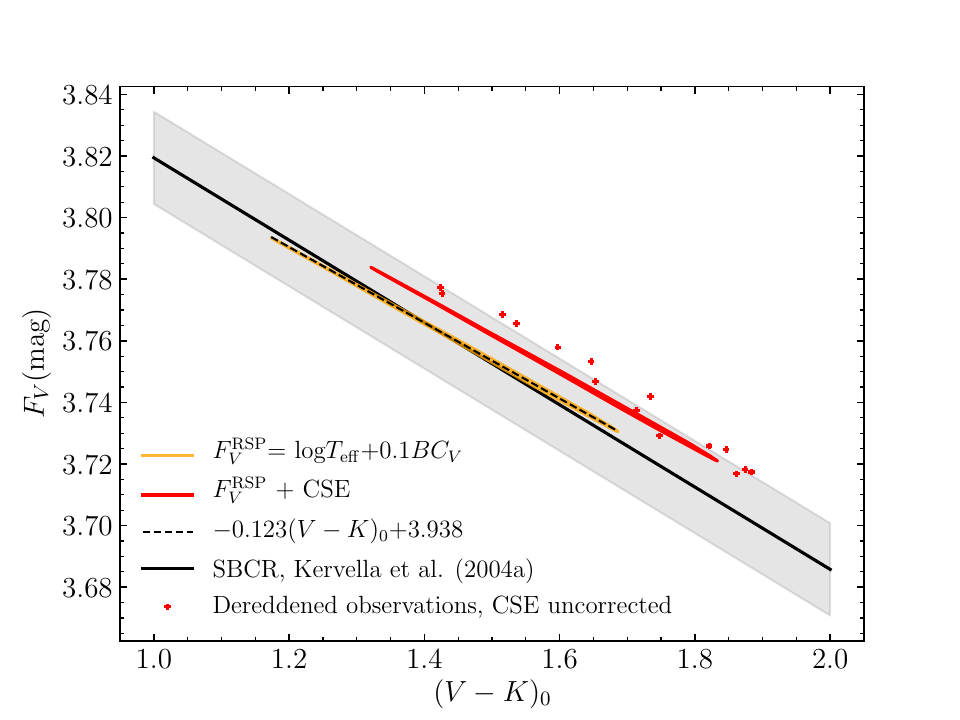}
\caption{\small Comparison of SBCR derived for Y Oph with RSP along the pulsation cycle and empirical SBCR from \cite{kervella04c}.\label{fig:SBCR}}
\end{figure}

The Surface Brightness Color Relation (SBCR) is critical in BW methods to determine the distance of Y~Oph. We investigate the SBCR of Y~Oph using a theoretical SBCR based on the 8$\,$M$_\odot$ RSP non-linear model previously derived:
\begin{equation}
    F^\mathrm{RSP}_V=\mathrm{log}T_\mathrm{eff}+0.1\,\mathrm{BC}_V
\end{equation}
where $T_\mathrm{eff}$ and the bolometric correction in the visible $\mathrm{BC}_V$ are directly given by the output of the RSP model, without any hypothesis needed on the distance, nor the extinction, nor CSE model. We then plot this value against $(V-K)_0$ where $V$ and $K$ are the absolute magnitude derived from atmosphere models interpolated from RSP stellar parameters (see Sect.~\ref{sect:atmosph}). As we can see from Fig.~\ref{fig:SBCR}, we obtain an excellent agreement with empirical SBCR relation of Cepheids derived by \cite{kervella04c}. A linear fit based on the SBCR derived with RSP gives $F^\mathrm{RSP}_V=-0.123_{{0.001}}(V-K)_0+3.939_{{0.001}}$. 

At first sight, there is no significant difference between theoretical SBCR for Y Oph as derived from RSP and the fiducial SBCR of Cepheids. However, \cite{Groenewegen2007,Groenewegen2013} has shown that the \textit{empirical} SBCR derived for Y~Oph is discrepant with the relation calibrated by \cite{kervella04c}. In particular, this star appears too red at a given surface brightness or conversely, the surface brightness is too high at a given color. 
Until now the origin of this discrepancy was attributed to uncertainty of the interstellar extinction. \cite{Groenewegen2013} argued that a value of $E(B-V)\approx1$ would be enough to bring Y Oph onto the relation. We show hereafter that the CSE model derived in this paper for Y~Oph is able to give an interesting alternative explanation without assuming exceptionally large $E(B-V)$ (see also Fig.~\ref{fig:EBV} for comparison). In order to check for the CSE impact we reproduced the SBCR \textit{observed} for Y~Oph, by omitting the existence of the CSE:
\begin{equation}\label{eq:SBCR}
    F_V=4.2196-0.5\,\mathrm{log}\theta_{\mathrm{LD}}-0.1\,{V_0}
\end{equation}
where $\theta_\mathrm{LD}$ is the limb-darkened angular diameter obtained from $\theta_\mathrm{UD}$ observed by interferometry (corrected for LD effect), and ${V_0}$ is the dereddened apparent magnitude in the visual band. Dereddened apparent magnitudes in $V$ and $K$ bands, interpolated to the phase of angular diameter measurements, are also used to compute ($V-K$)$_0$. We stress that we deliberately do not correct for the influence of the CSE on the apparent $V$ and $K$ magnitudes and the angular diameter measurements, to compare with previous studies \citep[see, e.g.,][]{Groenewegen2013}. 

As a result, the empirical SBCR of Y Oph appears to be significantly higher than the fiducial SBCR, although within the uncertainty (see red points in Fig.~\ref{fig:SBCR}). This result is in agreement with the discrepancy previously observed by \cite{Groenewegen2007,Groenewegen2013}.  Finally, if now we add the CSE model presented in Fig.~\ref{fig:IR_excess} as derived in Sect.~\ref{sect:fit}, on the theoretical SBCR, we obtain a result which is in agreement with observations (see red line in Fig.~\ref{fig:SBCR}).

Now the question is to quantify the impact of the CSE on the derived distance when using SBCR relation. We define the CSE effect on the visible and the infrared magnitudes as $\Delta V$ and $\Delta K$ respectively. The observed magnitudes of the star, dereddened, but not corrected for the presence of CSE, can be expressed as $V^\mathrm{cse}_0=V_0+\Delta V$ and  $K^\mathrm{cse}_0=K_0+\Delta K$, where $V_0$ and $K_0$
are dereddened magnitudes of the star itself. We can then substitute these magnitudes into the following equations that are used to derive the angular diameter:
\begin{align}
    F_V &= a(V-K)^\mathrm{cse}_0 + b \\
    \mathrm{log}\theta_{\mathrm{LD}} &= 8.4392 - 2\,F_V - 0.2\,V^\mathrm{cse}_0.
\end{align}
The combination of these equation allows to derive the CSE impact on the distance:
\begin{equation}\label{eq:cse_effect}
    d \propto 10^{(2a+0.2) \Delta V}10^{-2a \Delta K}.
\end{equation}
In the case of absence of CSE, $\Delta V=\Delta K=0$, and there is obviously no bias on the derived distance.
If there is a CSE absorption and emission in the visible and infrared respectively, then $\Delta V > 0$ and $\Delta K < $0. Thus, both terms of the above equation are contributing to lower the derived distance. In the case of our CSE model, we have $\Delta V = 0.10\,$mag and $\Delta K=-0.05\,$mag, which translates into a distance 5\% lower (We used $a=-0.133$ consistently to \cite{kervella04c}). We find that the infrared emission from the CSE has the most significant impact on the SBCR, in agreement with the findings of \cite{Nardetto2023}. As a conclusion, we have shown that the CSE of Y~Oph might be a significant bias on the SBCR. Therefore, studies which made use of any calibrated SBCRs to derive the distance of Y~Oph very likely underestimated the surface brightness of Y~Oph. The unbiased average BW distance, i.e. corrected from CSE effect, is thus closer to 610$\,\pm24$pc. Although the tension with \textit{Gaia} distance is relaxed, the discrepancy is still at the 4$\,\sigma$ level. We discuss the possibility of a high $p$-factor in Sect.~\ref{sect:pfactor} and \ref{sect:comparison}.

\subsection{Comment on SBCR based on $V-R$}

Using Eq.~\ref{eq:cse_effect} we can estimate the systematic error on the distance in SBCR using $V-R$ color. We find that the derived BW distance must be overestimated by a few percent because of the presence of the CSE. Therefore, it is difficult to explain why \cite{Gieren1988,Gieren1993} found much smaller distance than subsequent BW methods. Several independent reasons may explain this difference. First, \cite{Gieren1993} noticed that Y~Oph is biased compared to their calibrated SBCR relation, which simply invalidates their distance for Y Oph. Second, the latter studies use RV data from \cite{Coulson1985} which is not accurate enough and suffers from an amplitude 20\% too small as noted by \cite{Hindsley1989}. Last, they calibrate $(V-R)$ on the basis of atmosphere models which has been later proved biased by \cite{fouque97}. Curiously, despite large uncertainties, \cite{Hindsley1989} found a mean distance in excellent agreement with \textit{Gaia} observations. We think this result is mostly accidental, as most of their distance determinations of Cepheids are strongly biased towards larger distance.

\subsection{Discussion on high $p$-factor}\label{sect:pfactor}
Although BW distance measurements presented in Fig.~\ref{fig:HR_distance} were probably underestimated, the BW methods (corrected by CSE) can only be matched with \textit{Gaia} DR3 distance if in turn the $p$-factor is about $p=1.5$.
Qualitatively, a high $p$-factor (but below 1.5) is associated to lower velocity gradient in the atmosphere or a weaker limb-darkening effect \citep{nardetto06a}. This is not usually the case for long-period Cepheids which have larger velocity gradient and large limb-darkening effect \citep{nardetto06a}, however this could be a good guess for Y~Oph regarding its small light and radial velocity amplitude.
Moreover, large values slightly above 1.5 were also occasionally reported in the literature for different Cepheids \citep{storm11a,Trahin2021}, but they can be in general attributed to observational uncertainties. It is physically conjectured that a reverse atmospheric velocity gradient or a limb-brightening effect can bring the $p$-factor above 1.5 \citep{albrow1994,sabbey95, nardetto06a,storm11a,ngeow12,neilson12}. A limb-brightening effect might be caused for instance by the presence of a chromosphere \citep{neilson12}.  In addition, \cite{sabbey95} found $p$=1.6 from radiative hydrodynamical models during the expansion of the photosphere which suggests a limb-brightening of the spectral lines. However, the absence of Ca II K emission for Y~Oph should in principle exclude any strong chromospheric activity \citep{kraft57}. 

Paradoxically, \cite{Wallerstein1992} concluded that Y~Oph would be more reliable to use for BW analysis because of the smooth and synchronized variation of H$\alpha$ and metallic lines observed for this star. 
In contrast, \cite{albrow1994} argued that in the case of low pulsation velocity in combination of radial macroturbulence, the central intensity profiles are broadened, and the major contribution of the radial velocity comes actually from region closer to the limb of the stellar disk. As a result, the lines that have an opposite asymmetry are able to produce a larger $p$-factor. As a conclusion, we suggest that Y~Oph has peculiar photospheric characteristics that makes it extremely interesting for studying the physics behind the $p$-factor.
 

\subsection{Interferometric BW distance versus Gaia distance}\label{sect:comparison}
\begin{figure}
\includegraphics[width={0.5\textwidth}]{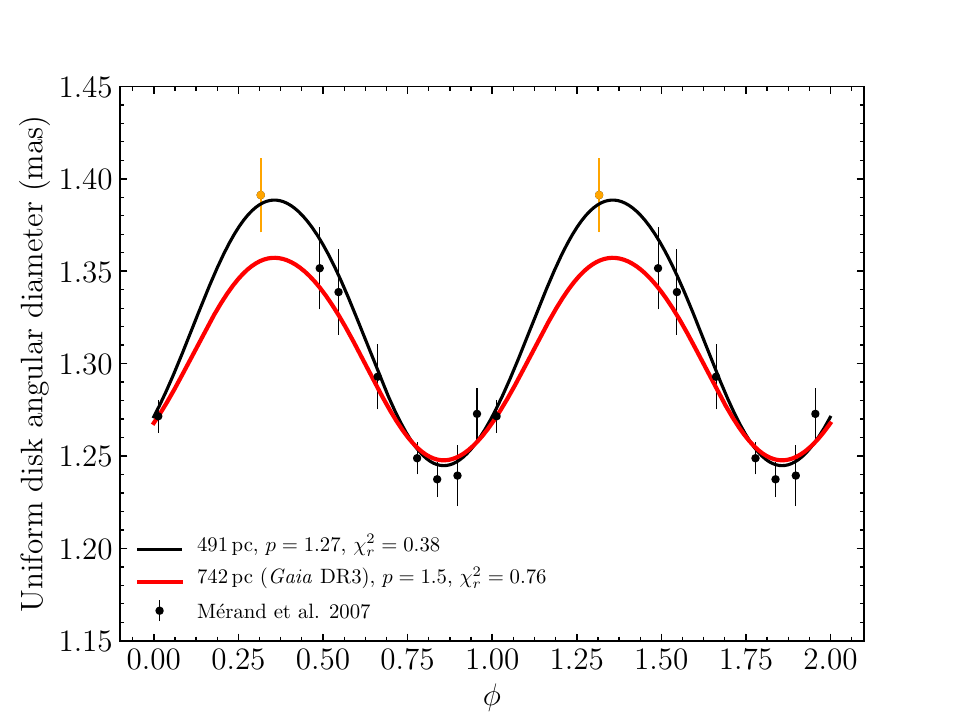}
\caption{\small Comparison of uniform disk angular diameter measurements in $K$-band from \cite{merand07} with a combination of the \textit{Gaia} DR3 distance and a $p$-factor of 1.5. The orange bar is a discrepant measurement compared to \textit{Gaia} distance model (See Sect.~\ref{sect:comparison}). \label{fig:comparison_merand}}
\end{figure}

    We might argue that the SBCR is not considered in the interferometric Baade-Wesselink distance determination from \cite{merand07} (hereafter M07), and thus an even larger discrepancy remains with the \textit{Gaia} distance, since the IBW distance is particularly small compared to other methods (see Fig.~\ref{fig:HR_distance}). However, a simple test is able to show that angular diameter distance measurements from M07 can be also explained by a high distance in combination of $p$-factor limited by the geometry as we can see in Fig.~\ref{fig:comparison_merand}. To this end, we performed a RV curve fitting using observations presented in Sect.~\ref{sect:obs}. We derived the angular diameter considering two cases, the first one taking into account $p=1.27$ similarly to M07, and the second one where we assumed the \textit{Gaia} distance together with a $p$-factor limited by the geometry ($p=1.5$). We used the following equation which transforms radial velocity into angular diameter variation:
\begin{equation}
    \theta_\mathrm{UD}(t) = -2\frac{kp}{d}\int^t_0 V_\mathrm{RV}(t)\mathrm{d}t + \theta_\mathrm{UD}(0) 
\end{equation}
where $\theta_\mathrm{UD}$ is the interferometric uniform disk diameter, and $k$ is
defined as previously $k=\theta_\mathrm{UD}/\theta_\star=1.023$ consistently with M07. In the first case, we chose $p=1.27$ as assumed in M07 and we fit the distance and the mean angular diameter $\theta_\mathrm{UD}(0)$. We find a distance of $d=480\,$pc with a reduced chi-square $\chi^2_r=0.38$, in close agreement with the results obtained by M07 and \cite{Gallenne2011PhD}. In the second case we fixed the $p$-factor to $p=1.5$ and the distance to \textit{Gaia} DR3 $d=742\,$pc, and we adjust the mean angular diameter only. We obtain $\chi^2_r=0.76$. Although the distance from M07 has the smallest $\chi^2$, the goodness of fit obtained from the \textit{Gaia} distance together with a high $p$-factor has also an excellent likelihood. Moreover, if we remove the most discrepant measurement (see orange error bar in Fig.~\ref{fig:comparison_merand}) we obtained essentially the same likelihood for both models with $\chi^2_r=0.41$ vs. 0.48 for M07 and \textit{Gaia} distance respectively. 

Interestingly, \cite{kervella04a} derived a distance of $d=648\pm50\,$pc which is the closest BW distance to the \textit{Gaia} observation. This later result was established using an hybrid version of the BW method where a value of the linear radius was assumed to be about 100$\,\mathrm{R}_\odot$ using a canonical Period-Radius relation. Therefore, it is not surprising that this determination is closest to the \textit{Gaia} distance. However, we think that two elements prevent this method to perfectly match the \textit{Gaia} distance. First of all, the $p$-factor used in this method is probably too low ($p$=1.36) as we argued in the previous section. Secondly, the VINCI/VLTI uniform disk angular diameter are likely appearing too large because it resolves the CSE emission in the $K$-band as shown by \cite{merand07}. This effect tends also to derive a lower distance. Therefore, this test supports our result that all BW distances of Y Oph can be explained by the combination of CSE impact and high $p$-factor.

\subsection{Systematic errors of temperature measurements}
The effective temperature is a critical parameter of our modeling since the high average temperature of Y~Oph of about 5819$\,$K \citep{Luck2018} places the star close to the blue edge of the instability strip.  Therefore, it is important to discuss the possibility of systematic errors of the effective temperature measurements. The effective temperature used in this study was determined using the Line Depth Ratio (LDR) of several pair of absorption lines from \cite{Luck2018,daSilva2022}.  According to \cite{Kovtyukh2008}, although $T_\mathrm{eff}$ are provided with a small internal error (typically 100$\,$K or less), a systematic error could exist. \cite{daSilva2022} re-analyzed HARPS spectra used by \cite{Proxauf2018}. As explained in Sect~\ref{sect:chemical}, \cite{daSilva2022} focused on possible systematics affecting the estimation of atmospheric parameters and they developed a careful analysis of the adopted lines. Although they derived a systematically smaller [Fe/H], the mean $T_\mathrm{eff}$ derived for Y~Oph is in agreement with the determination of \cite{Proxauf2018} at minimum temperature of the pulsation cycle (5609$\pm33\,$K vs 5612$\pm33\,$K). Moreover, as we noted previously, temperature measurements of \cite{Luck2018} and \cite{daSilva2022}, which are based on different sets of spectra, are also in agreement along the pulsation cycle. Therefore, we are confident that $T_\mathrm{eff}$ determination of Y~Oph is accurate. 

The presence of systematic errors in temperature determinations remains however a topic of discussion \citep{Mancino2021}. To investigate possible systematic effect, \cite{Gro2020dust,Groenewegen2020fluxgrav} compared the average $T_\mathrm{eff}$ determination of Cepheids from \cite{Luck2018} (for 52 stars with more than 5 spectra) to photometric temperature using SED fitting. Interestingly, \cite{Groenewegen2020fluxgrav} found an agreement within the uncertainties for all stars except Y~Oph and S Vul. The photometric temperature of Y~Oph is found to be 570$\,$K lower than spectroscopic temperature. However, photometric temperatures based on SED fitting is mostly sensitive to the optical data (Wien's law).  Hence, this method can be strongly biased because of CSE visual absorption and the color excess of this star. If the CSE absorption is ignored during the fit, then its contribution is hidden in the fitted $E(B-V)$ term. On the contrary, if $E(B-V)$ is fixed to literature values, which do not take into account the CSE, then the SED of the star will appear fainter in the optical, and as consequence, colder than it really is. \cite{Gro2020dust} also notes that photometric temperatures of Cepheids are found in average 200$\,$K cooler than spectroscopic determination. This might be a hint for a common photometric biased caused by a CSE.

\subsection{Evolutionary stage of Y Oph}
As mentioned previously, Y Oph most likely undergoes the 3rd crossing of the instability strip. This identification follows from its rate of period change. However, Y~Oph is the second Cepheid known after Polaris to exhibit a decreasing light curve amplitude \citep{Fernie1990,fernie95} with a slope of about $-1\,$mmag/yr similar to Polaris \citep{Fernie1993}.  An examination of the $V$-band light curve over the past century suggests that its amplitude might have been around 0.7$\,$mag at the beginning of the 20th century, compared to its current value of 0.5$\,$mag. Therefore, the location close to the blue edge together with the decreasing light amplitude of Y~Oph lead us to consider this star as being on the 2nd crossing, leaving the instability strip. However, in-depth statistical analysis of the light amplitude decline performed by \cite{Pop2010} shows that drawing a firm conclusion on an amplitude change in Y Oph is not possible with the available data.

\subsection{Absolute magnitude of Y Oph}
\cite{Kovtyukh2010} empirically determined the absolute visual magnitude of Y Oph to be $M_V=-3.90\pm0.15\,$mag at the pulsation phase $\phi=0.454$. This is puzzling, since such a magnitude would require a low mass of about 3$\,$M$_\odot$ and is also discrepant with \textit{Gaia} distance. Their empirical relation is calibrated from FeII/FeI line depth ratios of F and G supergiants and is shown to be accurate for application of Cepheids. They applied their method from Cepheid spectra observed around maximum radius ($\phi \approx0.40$) to mitigate dynamical effects. However, they scrutinized the application of this method along the pulsation cycle in the particular case of $\delta$ Cep. They find absolute magnitudes up to 0.8$\,$mag fainter at phases between minimum and maximum light ($\phi \approx 0.75-1.00$) compared to expected values. Therefore, we think that systematic errors cannot be excluded for Y Oph that has a different dynamic than the bulk of Cepheids.

\section{Conclusion}\label{sect:conclusion}
Y~Oph is a long-period Cepheid known in the literature for its small light amplitude and low luminosity among other peculiarities discussed in the introduction. We provided hydrodynamical modeling with \texttt{MESA-RSP} in combination with a full set of observations in order to constrain the physical characteristics of this star. This comparison with the observations allows to constrain the mass and the distance of the star. We thus provide the following conclusions:

\begin{enumerate}
\item On the basis of the linear nonadiabatic analysis, we conclude that Y Oph is a fundamental-mode Cepheid.

\item From the non-linear analysis, we find that the low radial velocity and light curve amplitude of Y~Oph is caused by the proximity of the star to the blue edge of the instability strip.

\item We find that the \textit{Gaia} DR2 and DR3 distances of Y Oph are in agreement with models of pulsational mass of about 7-8$\,$M$_\odot$. These masses are also in agreement with masses inferred by non-canonical evolutionary models assuming a mild convective core overshoot.  The luminosity is also consistent with PL relations calibrated in the $K_s$ band.

\item On the other hand, BW distances determination from the literature are discrepant with \textit{Gaia} distance, and are consistent with a pulsation mass of 5$\,$M$_\odot$ or less. Such combination of mass and distance cannot be explained by evolutionary models. Moreover, the luminosity derived from pulsation models is much lower than expected from PL relation.

\item We find that the distance of Y~Oph derived by the BW method based on SBCR is biased if the impact of the CSE on the photometry is not taken into account. In addition, our result suggests that the $p$-factor of Y~Oph might be close to the geometric limit of 1.5.

\end{enumerate}
 As a conclusion, pulsation modeling with the \texttt{MESA/RSP }code is a valuable tool to provide constraints on the physical parameters of Cepheids. However, exploring the physics of the CSE is crucial to refine the photometric modeling of Cepheids and make it more realistic. Lastly, we stress that Y Oph is an important Cepheid for better understanding of the physics behind the $p$-factor. Nevertheless, Y~Oph is a long-period Cepheid with reliable \textit{Gaia} parallax that can be useful to calibrate the PL relation. The development of pulsation models with higher grid resolution will be essential for a more precise determination of the $p$-factor and the physical parameters of this star. On the other hand, thorough investigation of evolutionary state of Y~Oph will be important to conclude about its nature.

\begin{acknowledgements}
We thank the referee for their helpful comments and suggestions.
VH, RS, OZ, RSR are supported by the National Science Center, Poland,
Sonata BIS project 2018/30/E/ST9/00598. This research made use of the SIMBAD and VIZIER databases at CDS, Strasbourg (France) and the electronic bibliography maintained by the NASA/ADS system. This research also made use of Astropy, a community-developed core Python package for Astronomy \citep{astropy2018}. This research has made use of the Spanish Virtual Observatory (https://svo.cab.inta-csic.es) project funded by MCIN/AEI/10.13039/501100011033/ through grant PID2020-112949GB-I00.
\end{acknowledgements}
\bibliographystyle{aa}  
\bibliography{bibtex_vh} 

\begin{appendix} 



\section{MESA/RSP inlist}

\subsection{MESA inlist}\label{appendix:mesa}
\begin{verbatim}
&star_job

    show_log_description_at_start = .false.

    create_RSP_model = .true.

    save_model_when_terminate = .true.
    save_model_filename = 'final.mod'

    initial_zfracs = 6

    ! History and profile columns 
    history_columns_file = 'history_columns.list'

    set_initial_age = .true.
    initial_age = 0

    set_initial_model_number = .true.
    initial_model_number = 0
      
    set_initial_cumulative_energy_error = .true.
    new_cumulative_energy_error = 0d0
      
    pgstar_flag = .false.
   
/ ! end of star_job namelist

&eos
/ ! end of eos namelist

&kap
    Zbase = 0.0134
    kap_file_prefix = 'a09'
    kap_lowT_prefix = 'lowT_fa05_a09p'
    kap_CO_prefix = 'a09_co'
/ ! end of kap namelist

&controls
!max_model_number = 2 ->  to uncomment for LNA
! limit max_model_number as part of test_suite
     RSP_max_num_periods = 6000 
! True convergence criteria defined in run_star_extras

! RSP controls

    RSP_do_check_omega = .true. 
  
    ! Mass, Teff, luminosity and hydrogen abundance
    ! Set by the next script provided in A.2
    RSP_mass = mmmm
    RSP_Teff = tttt
    RSP_L    = llll
    RSP_X    = xxxx
    RSP_Z    = 0.0134 

    !Uncomment for LNA
    !RSP_nmodes = 12 ! number of modes analyzed
    RSP_kick_vsurf_km_per_sec = 10d0
    RSP_fraction_1st_overtone = 0d0
    RSP_fraction_2nd_overtone = 0d0

! :: SET A OF CONVECTIVE PARAMETERS
    RSP_alfa   = 1.5d0
    RSP_alfac  = 1.0d0
    RSP_alfas  = 1.0d0
    RSP_alfad  = 1.0d0
    RSP_alfap  = 0.0d0
    RSP_alfat  = 0.0d0
    RSP_alfam  = 0.25d0
    RSP_gammar = 0.0d0

! controls for building the initial model
    RSP_nz = 150
    RSP_nz_outer = 40
    RSP_T_anchor = 11d3
    RSP_T_inner = 2d6
    RSP_target_steps_per_cycle = 600

! output controls
    terminal_show_age_units = 'days'
    trace_history_value_name(1) = 'rel_E_err'
    trace_history_value_name(2) = 'log_rel_run_E_err'
    photo_interval    = 1000 
    profile_interval  = -1 
    history_interval  = 1 
    terminal_interval = 4000
/ ! end of controls namelist
\end{verbatim}
\subsection{Script}
Below we provide a useful bash script that can be used or adapted for conveniently prepare and run several models for the previous inlist, with parameters defined in a text file.
\begin{verbatim}
file="parameters_list.dat"

# number of models to be computed
n=$(wc -l $file |awk '{print $1}')

for i in $(seq 1 $n);
do
    # l, t, m & x in a single variable
    arg=$(head -n $i $file |tail -n 1)

    # separate variables for l, t, m & z
    l=$(head -n $i $file |tail -n 1 | awk '{print $1}')
    t=$(head -n $i $file |tail -n 1 | awk '{print $2}')
    m=$(head -n $i $file |tail -n 1 | awk '{print $3}')
    x=$(head -n $i $file |tail -n 1 | awk '{print $4}')
    echo $l $t $m $x

    cp inlist_project_ltz inlist_project
    sed -i "s/llll/$l/g" inlist_project
    sed -i "s/tttt/$t/g" inlist_project
    sed -i "s/mmmm/$m/g" inlist_project
    sed -i "s/xxxx/$x/g" inlist_project
done
\end{verbatim}

\clearpage
\section{Results of the non-linear analysis}
   \begin{figure*}[htbp]
    \centering
    \begin{subfigure}{0.33\textwidth}
        \includegraphics[width=\linewidth]{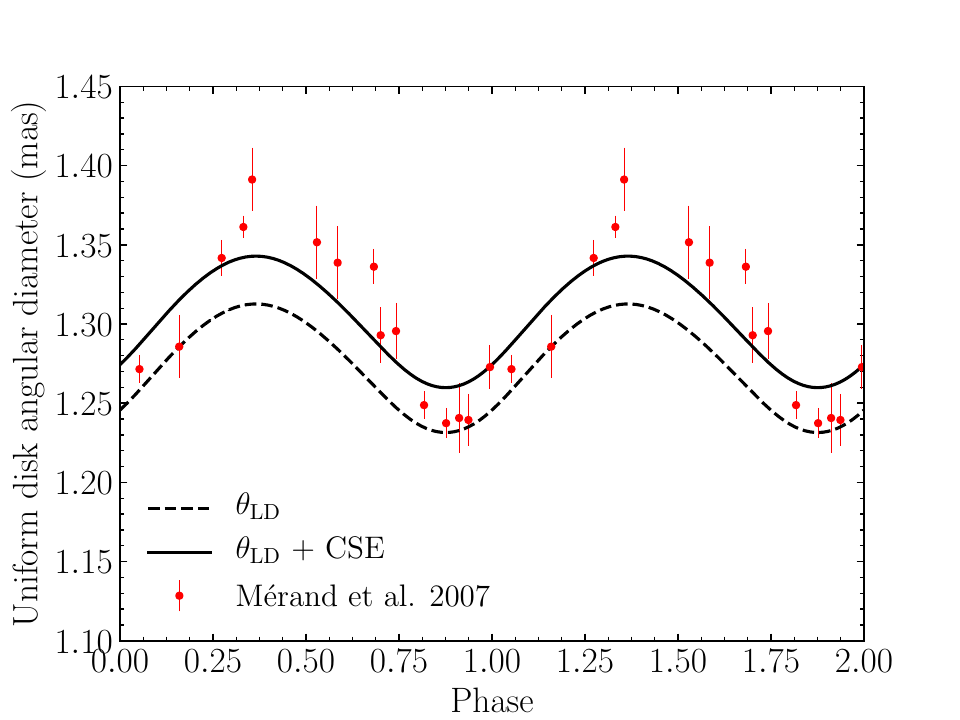}
        \caption{}
        \label{fig:R_3M}
    \end{subfigure}
    \begin{subfigure}{0.33\textwidth}
        \includegraphics[width=\linewidth]{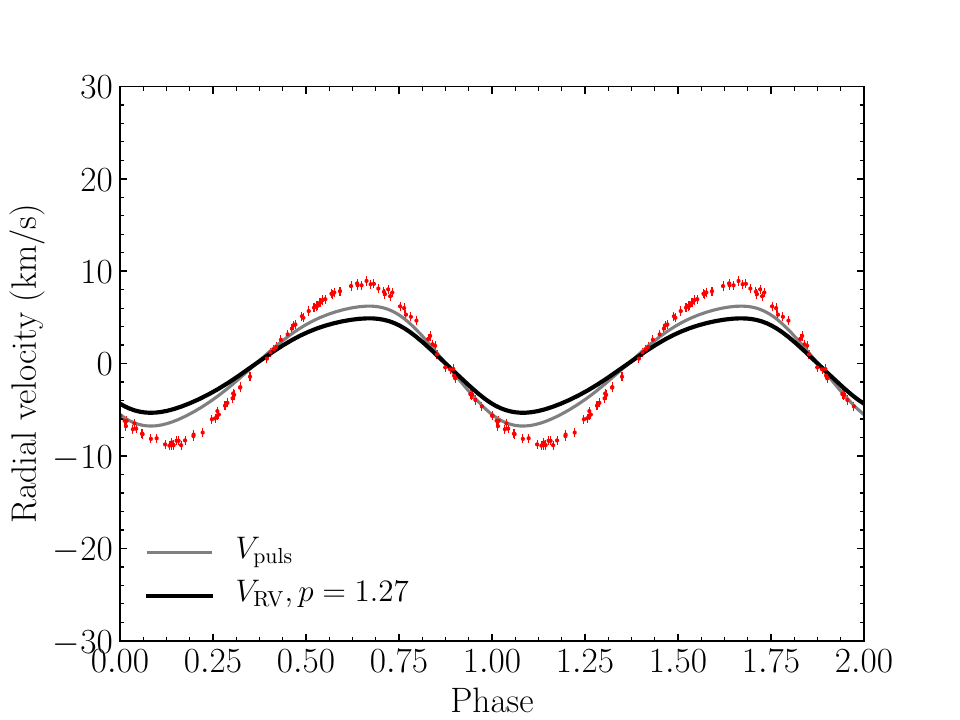}
        \caption{}
        \label{fig:RV_3M}
    \end{subfigure}
    \begin{subfigure}{0.33\textwidth}
        \includegraphics[width=\linewidth]{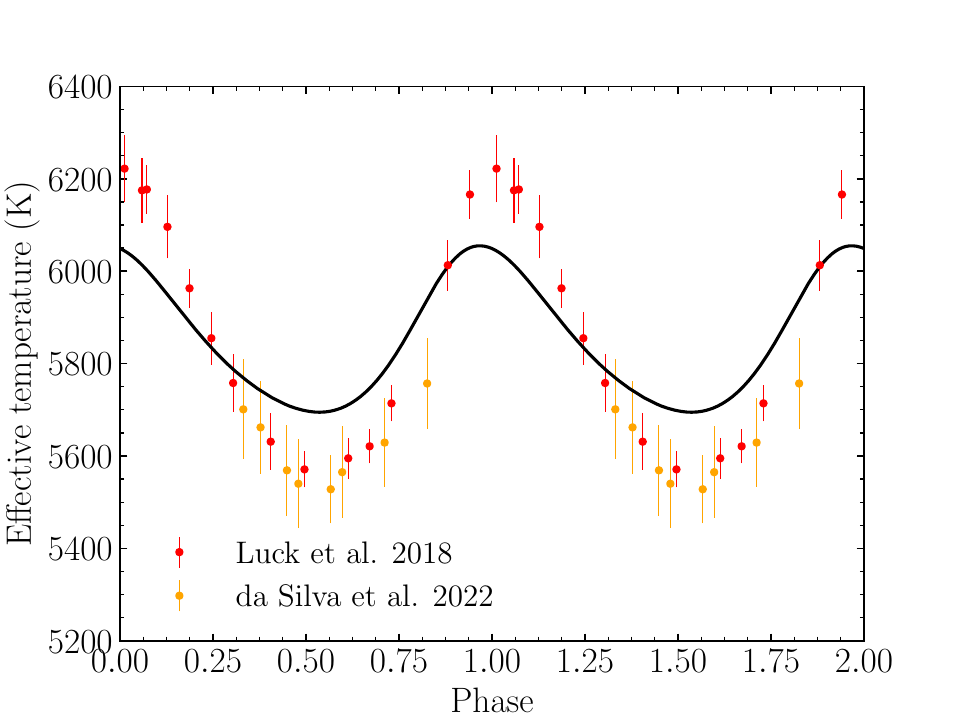}
        \caption{}
        \label{fig:Teff_3M}
    \end{subfigure}

    \medskip
        \begin{subfigure}{0.33\textwidth}
        \includegraphics[width=\linewidth]{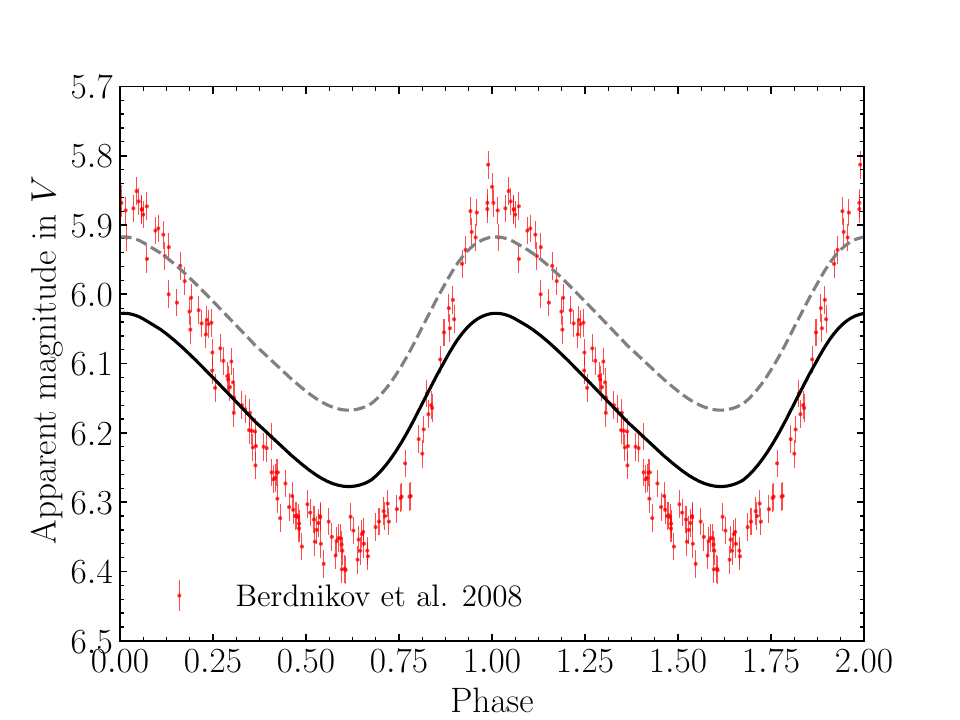}
        \caption{}
        \label{fig:V_3M}
    \end{subfigure}
    \begin{subfigure}{0.33\textwidth}
        \includegraphics[width=\linewidth]{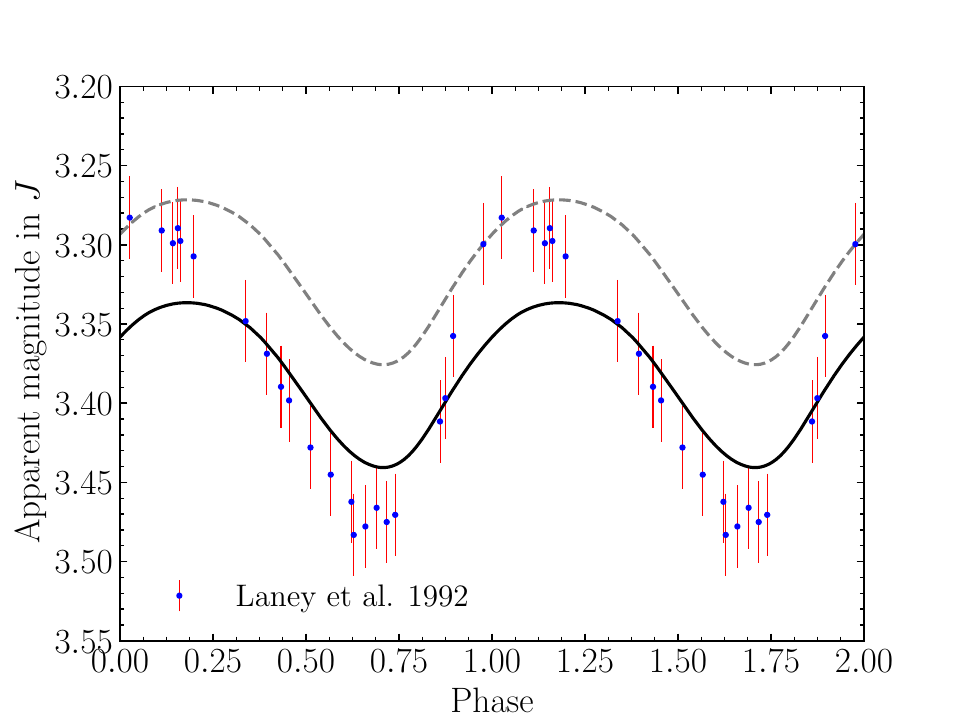}
        \caption{}
        \label{fig:J_3M}
    \end{subfigure}
    \begin{subfigure}{0.33\textwidth}
        \includegraphics[width=\linewidth]{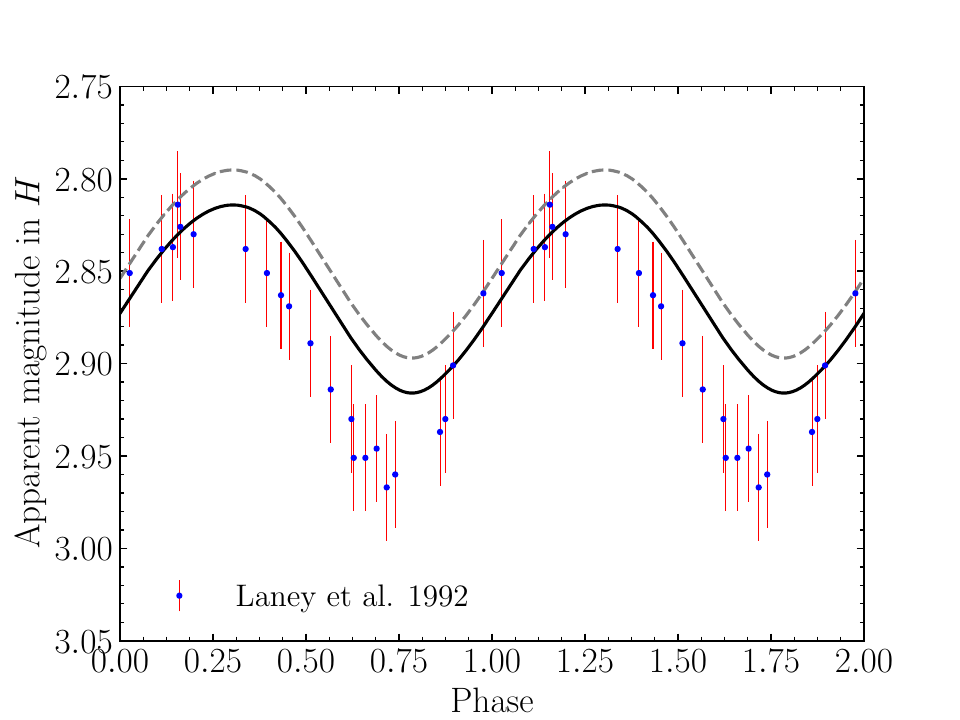}
        \caption{}
        \label{fig:H_3M}
    \end{subfigure}

    \medskip

    \begin{subfigure}{0.33\textwidth}
        \includegraphics[width=\linewidth]{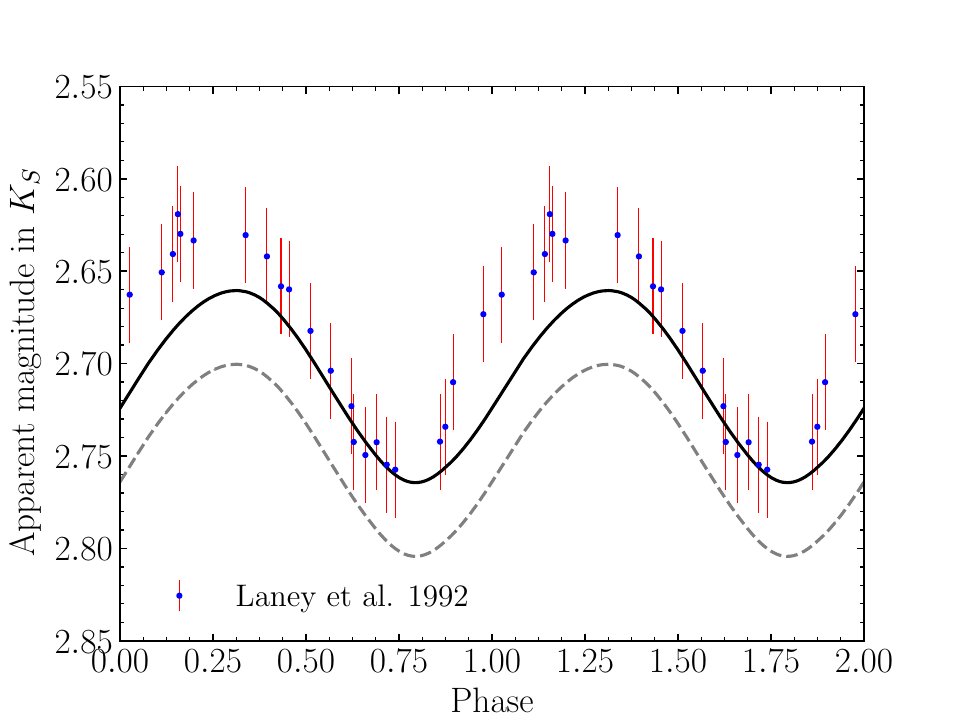}
        \caption{}
        \label{fig:K_3M}
    \end{subfigure}
    \begin{subfigure}{0.33\textwidth}
        \includegraphics[width=\linewidth]{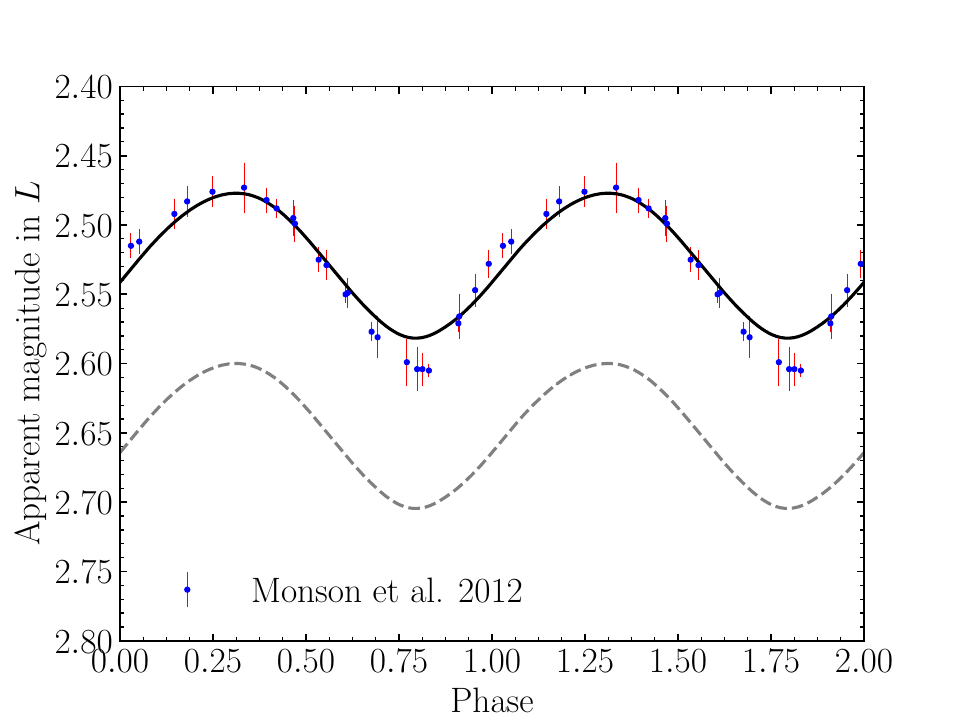}
        \caption{}
        \label{fig:L_3M}
    \end{subfigure}
    \begin{subfigure}{0.33\textwidth}
        \includegraphics[width=\linewidth]{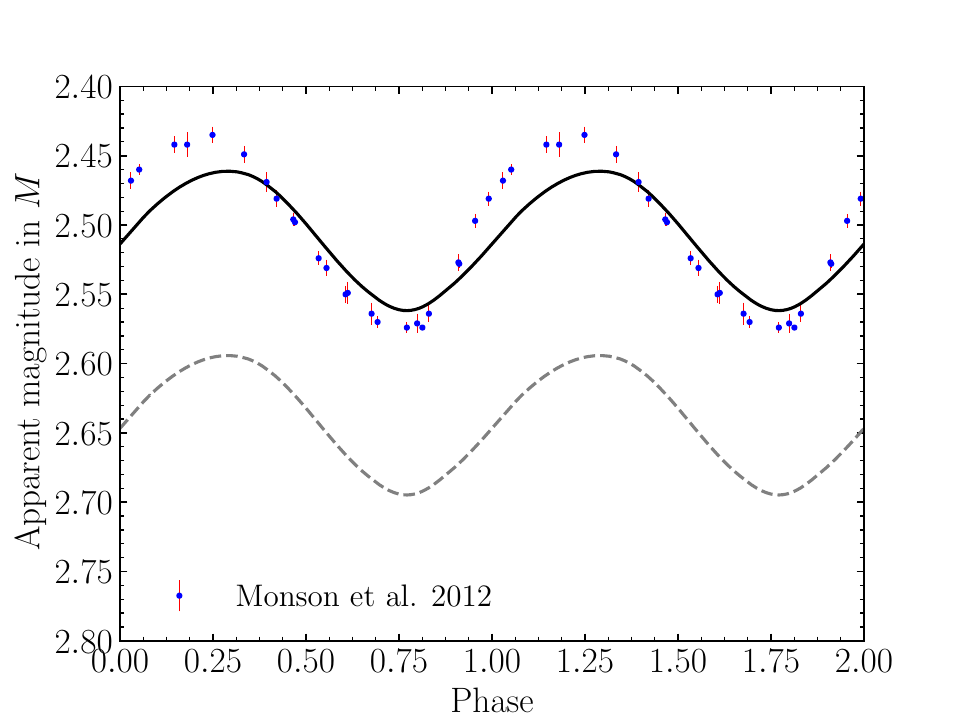}
        \caption{}
        \label{fig:M_3M}
    \end{subfigure}
\caption{Best result of the non-linear analysis with RSP for 3$\,M_\odot$.}\label{fig:non_linear_3M}
\end{figure*}

   \begin{figure*}[]
    \centering
    \begin{subfigure}{0.33\textwidth}
        \includegraphics[width=\linewidth]{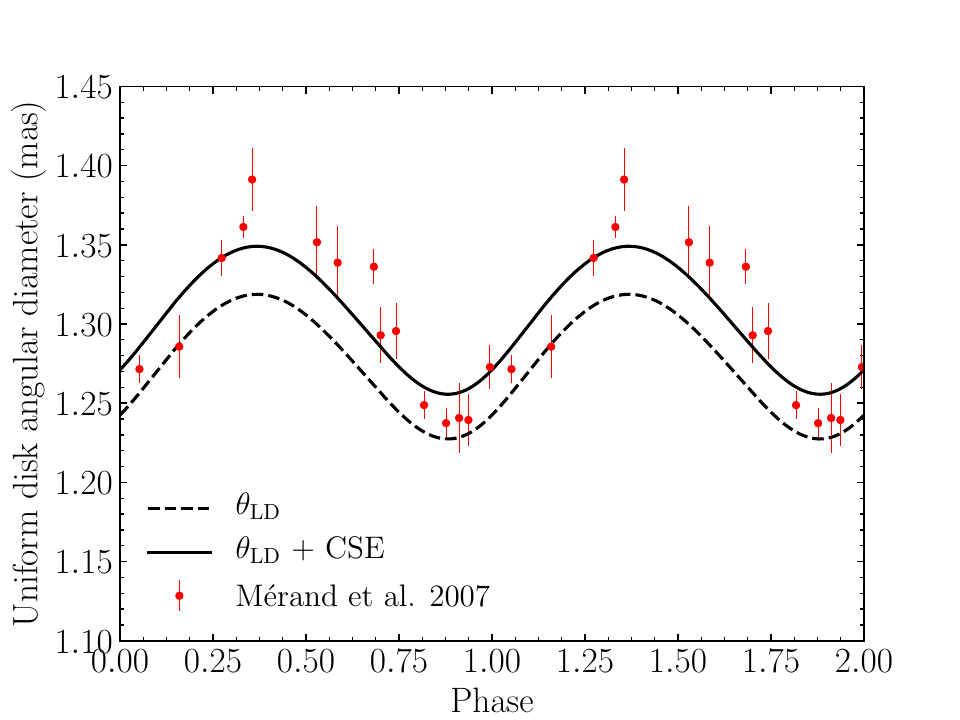}
        \caption{}
        \label{fig:R_4M}
    \end{subfigure}
    \begin{subfigure}{0.33\textwidth}
        \includegraphics[width=\linewidth]{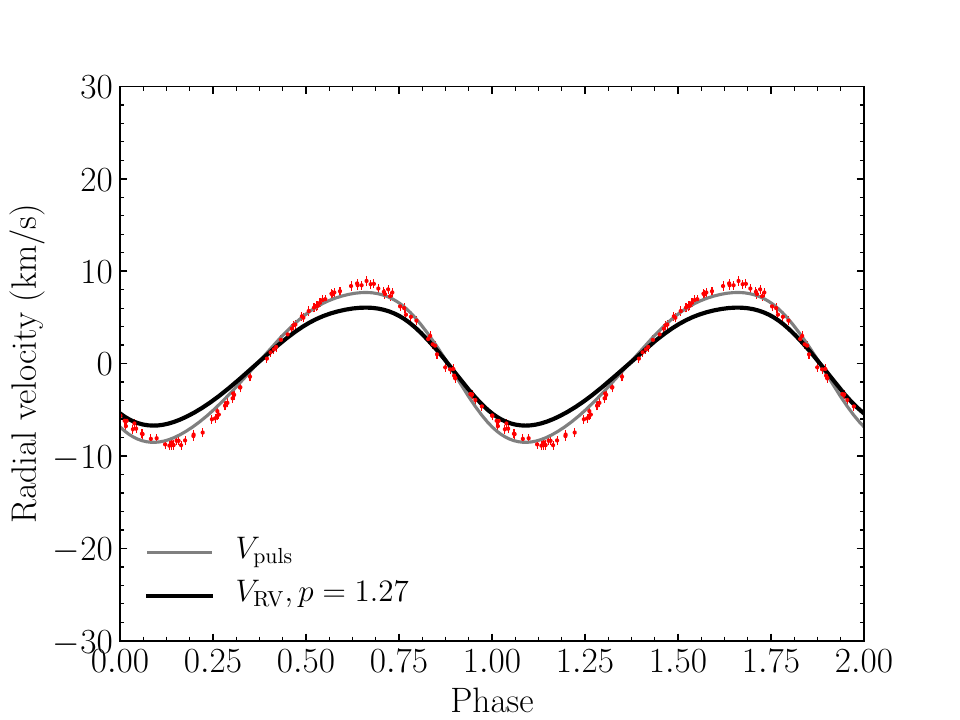}
        \caption{}
        \label{fig:RV_4M}
    \end{subfigure}
    \begin{subfigure}{0.33\textwidth}
        \includegraphics[width=\linewidth]{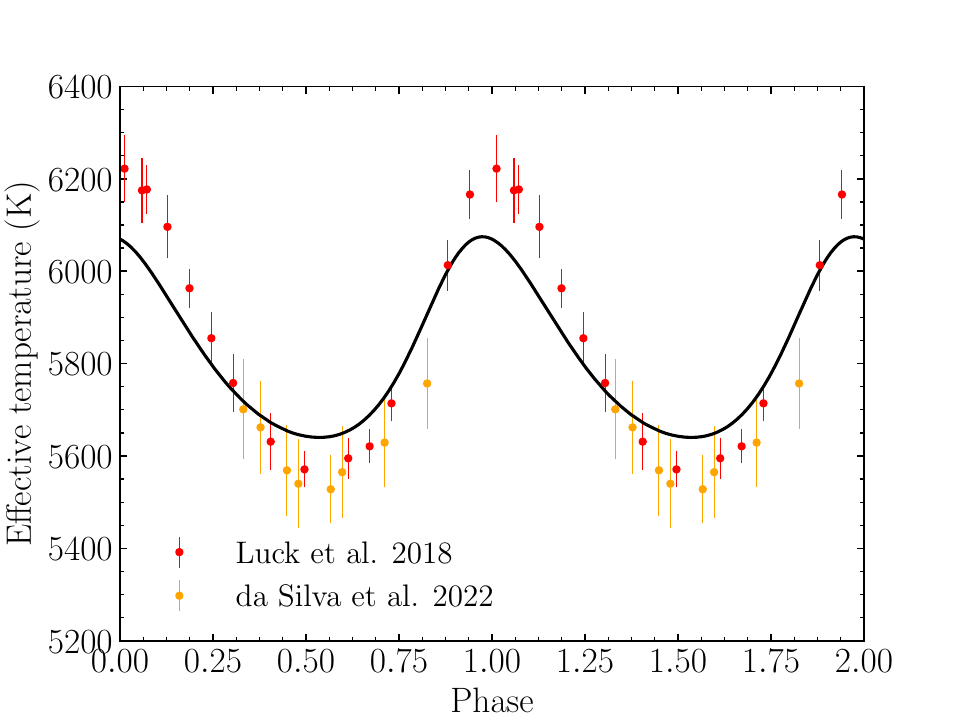}
        \caption{}
        \label{fig:Teff_4M}
    \end{subfigure}

    \medskip
        \begin{subfigure}{0.33\textwidth}
        \includegraphics[width=\linewidth]{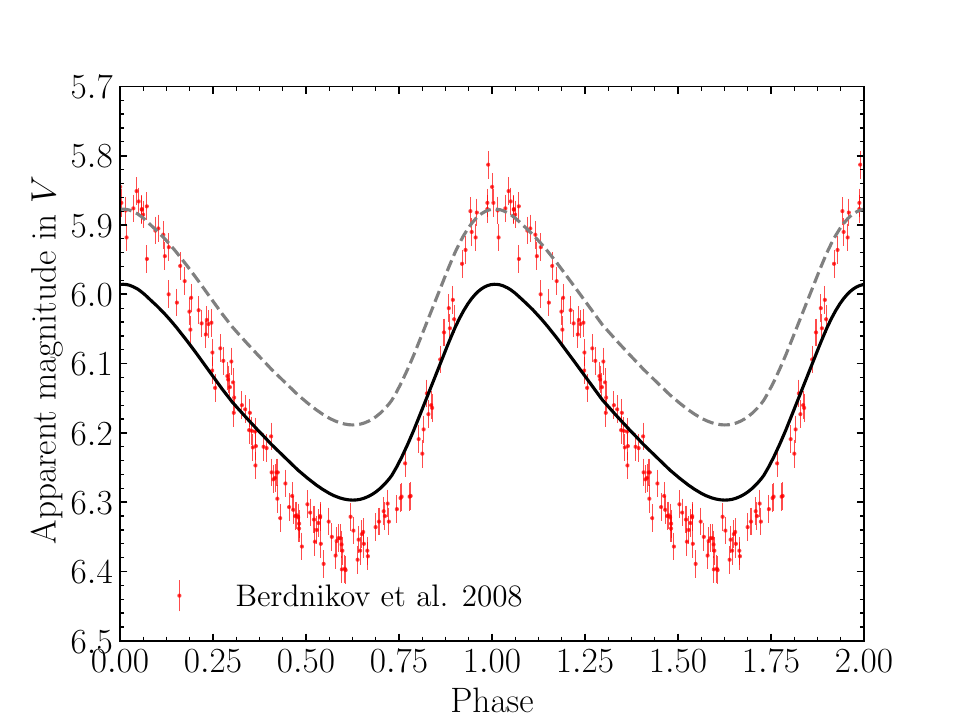}
        \caption{}
        \label{fig:V_4M}
    \end{subfigure}
    \begin{subfigure}{0.33\textwidth}
        \includegraphics[width=\linewidth]{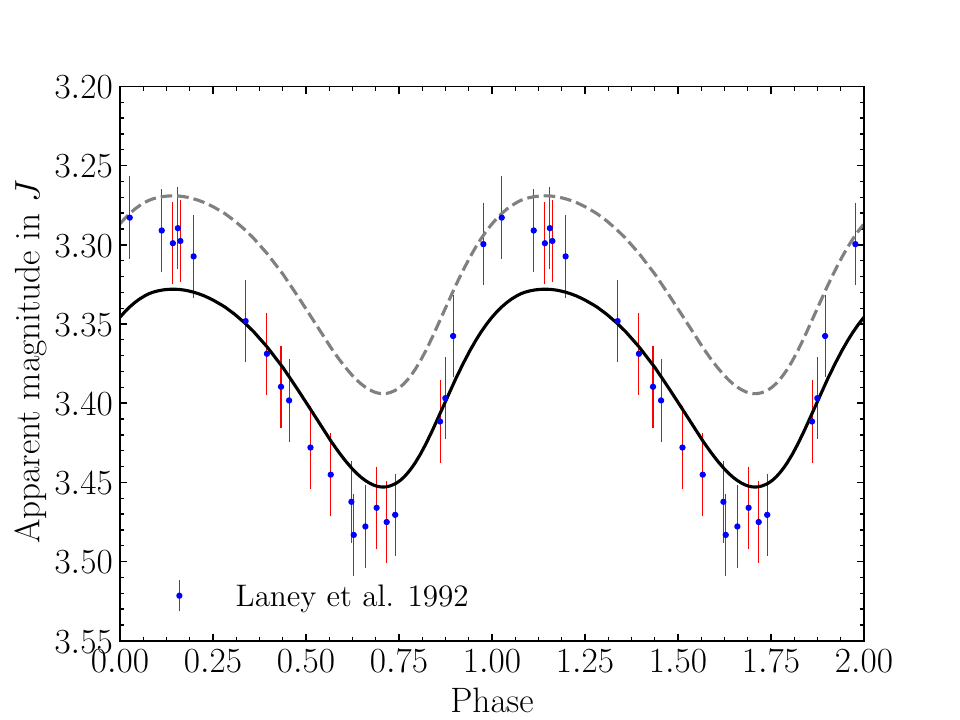}
        \caption{}
        \label{fig:J_4M}
    \end{subfigure}
    \begin{subfigure}{0.33\textwidth}
        \includegraphics[width=\linewidth]{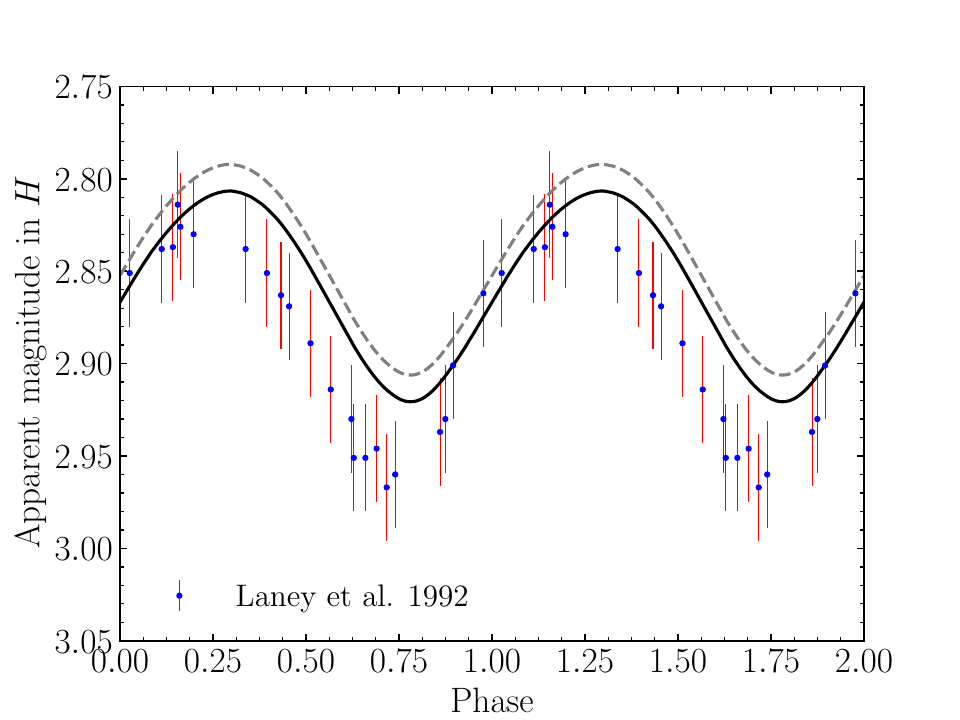}
        \caption{}
        \label{fig:H_4M}
    \end{subfigure}

    \medskip

    \begin{subfigure}{0.33\textwidth}
        \includegraphics[width=\linewidth]{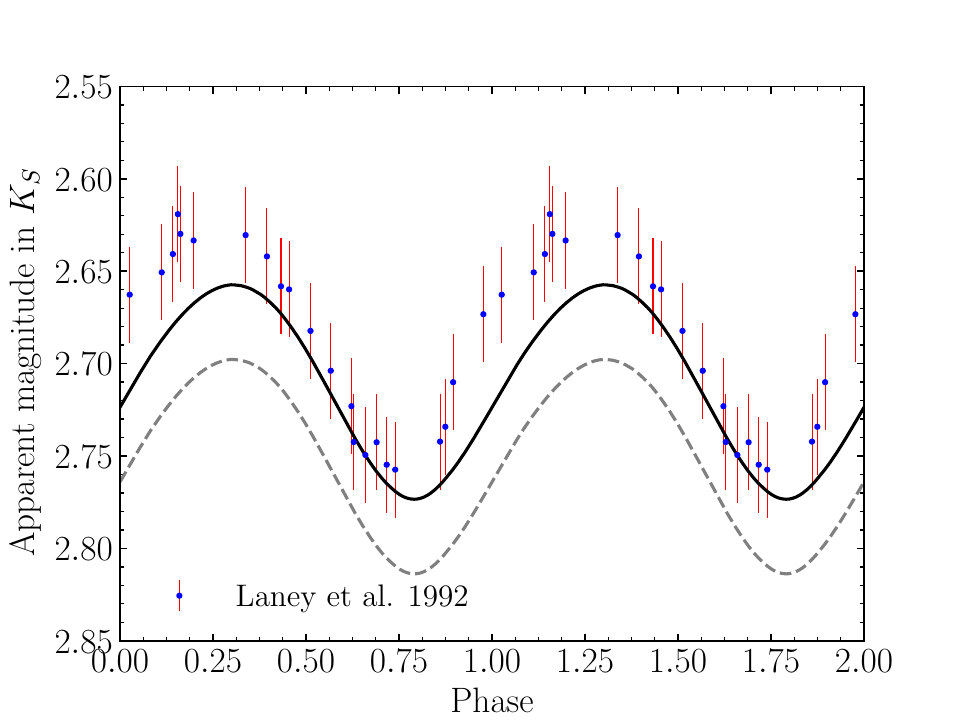}
        \caption{}
        \label{fig:K_4M}
    \end{subfigure}
    \begin{subfigure}{0.33\textwidth}
        \includegraphics[width=\linewidth]{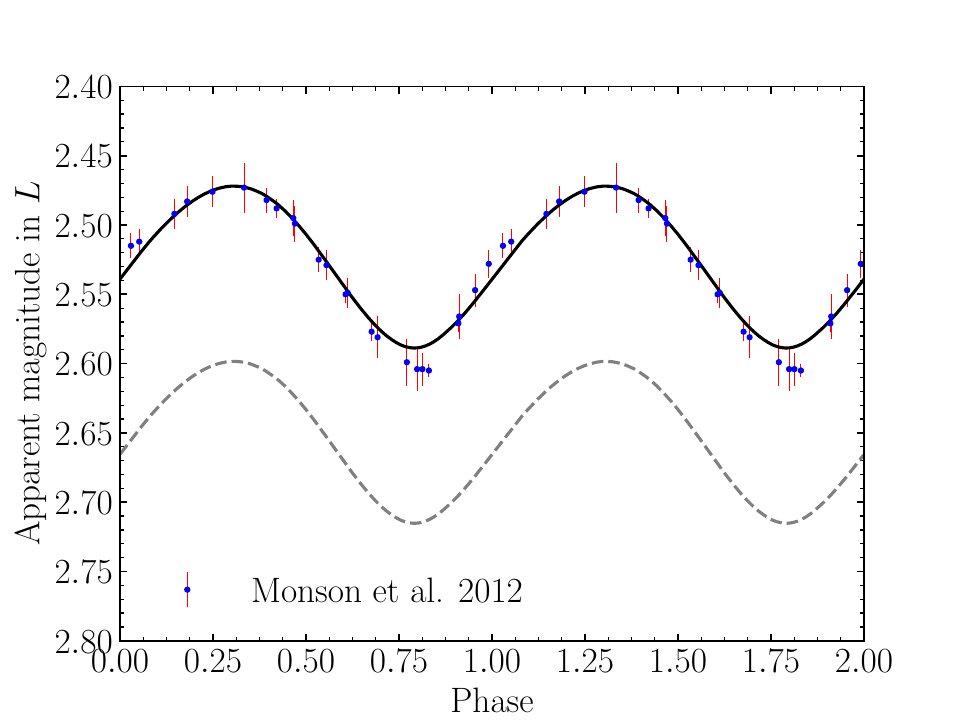}
        \caption{}
        \label{fig:L_4M}
    \end{subfigure}
    \begin{subfigure}{0.33\textwidth}
        \includegraphics[width=\linewidth]{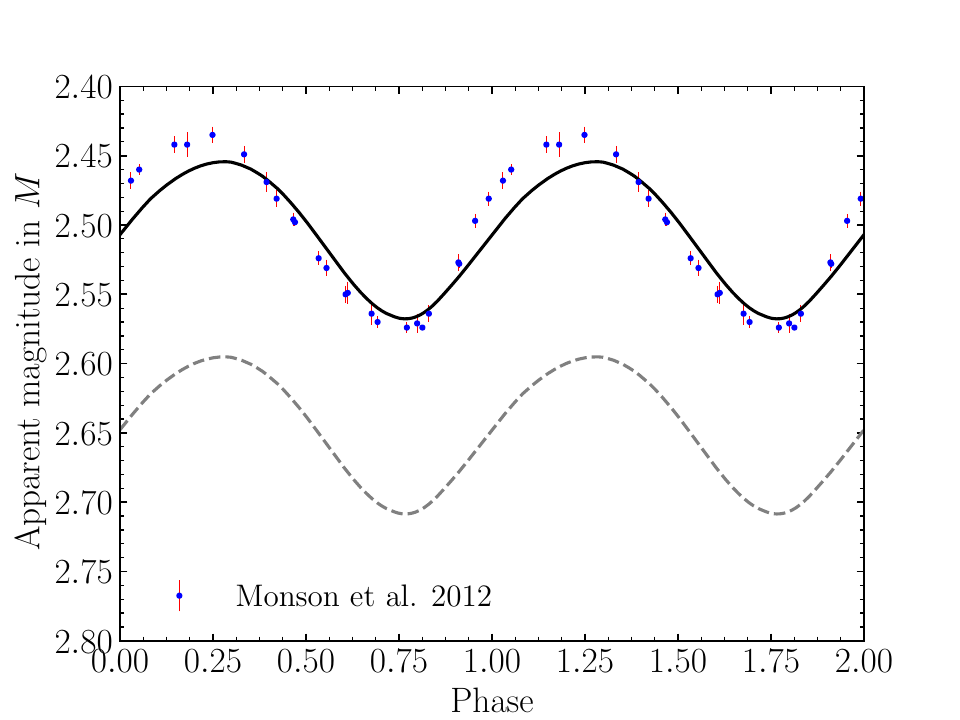}
        \caption{}
        \label{fig:M_4M}
    \end{subfigure}
\caption{Best result of the non-linear analysis with RSP for 4$\,M_\odot$.}\label{fig:non_linear_4M}
\end{figure*}

   \begin{figure*}[]
    \centering
    \begin{subfigure}{0.33\textwidth}
        \includegraphics[width=\linewidth]{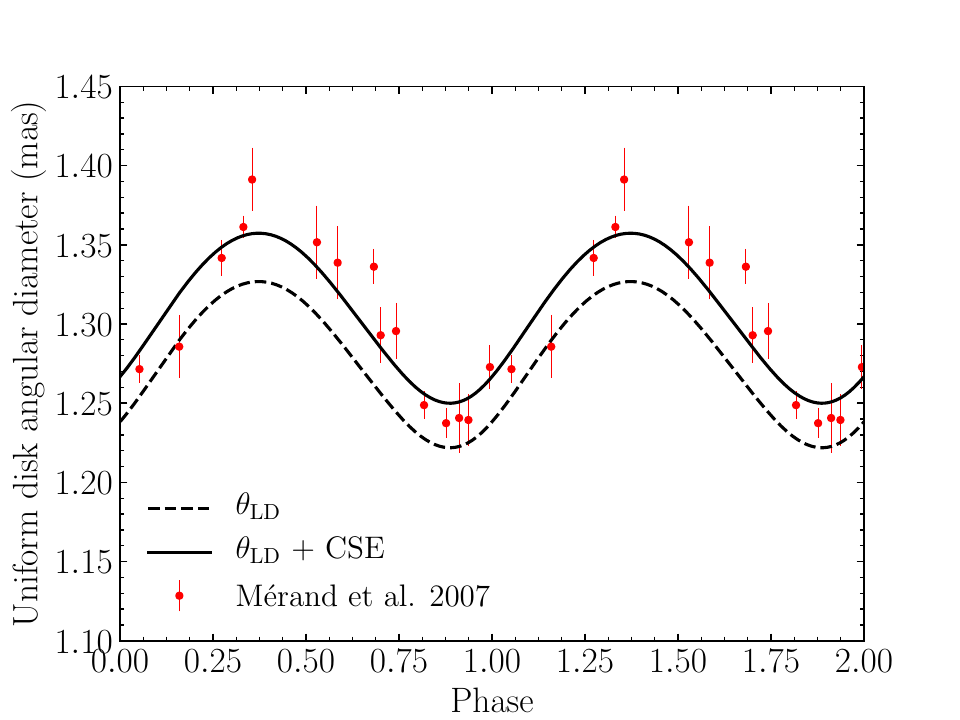}
        \caption{}
        \label{fig:R_5M}
    \end{subfigure}
    \begin{subfigure}{0.33\textwidth}
        \includegraphics[width=\linewidth]{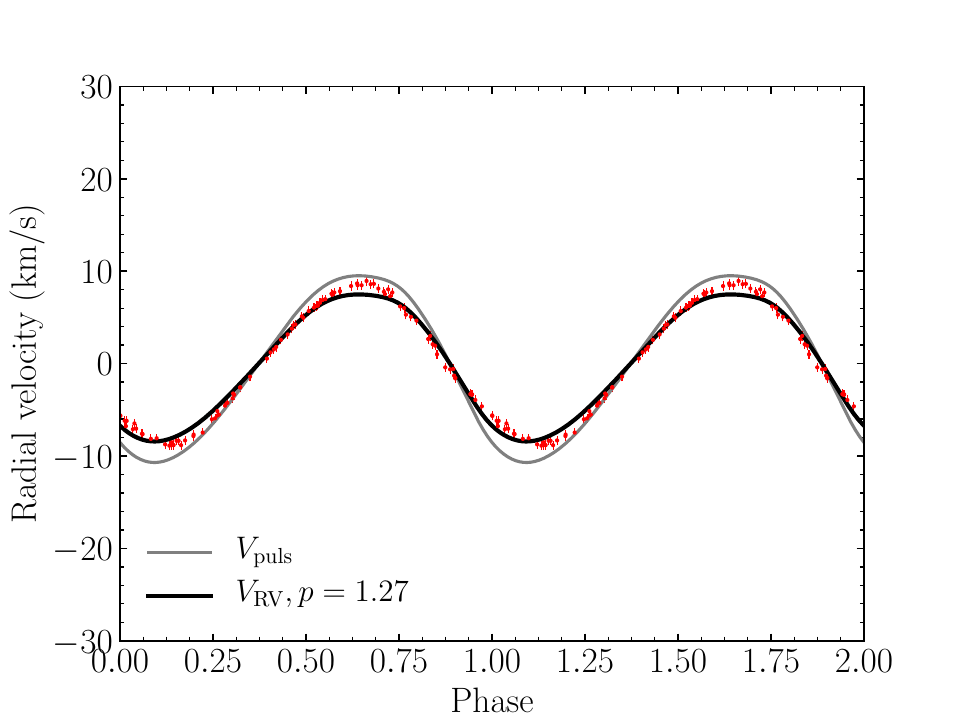}
        \caption{}
        \label{fig:RV_5M}
    \end{subfigure}
    \begin{subfigure}{0.33\textwidth}
        \includegraphics[width=\linewidth]{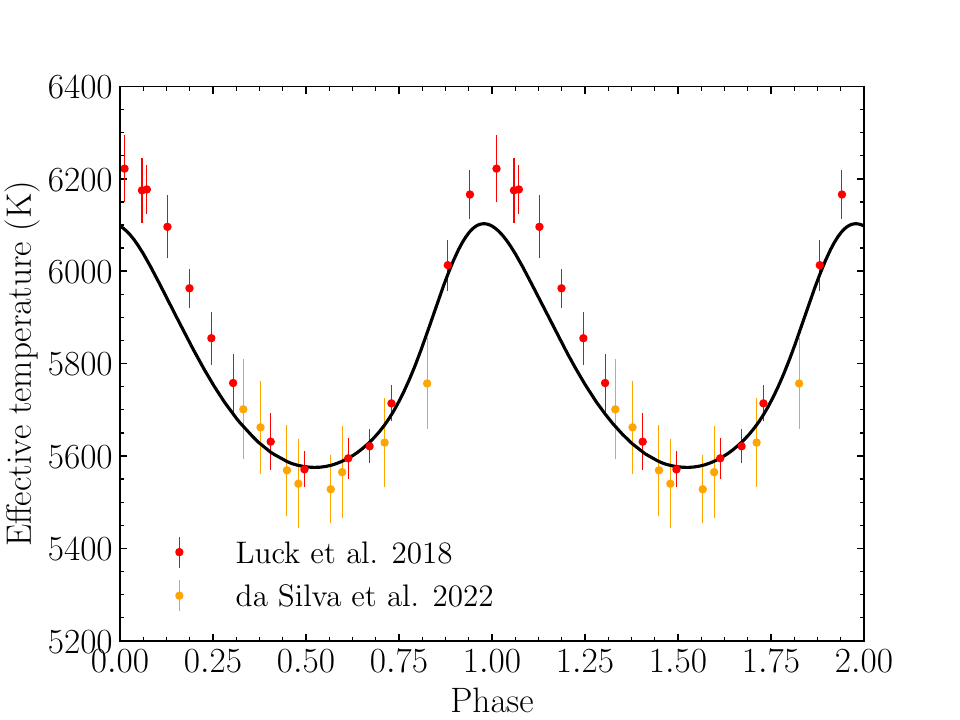}
        \caption{}
        \label{fig:Teff_5M}
    \end{subfigure}

    \medskip
        \begin{subfigure}{0.33\textwidth}
        \includegraphics[width=\linewidth]{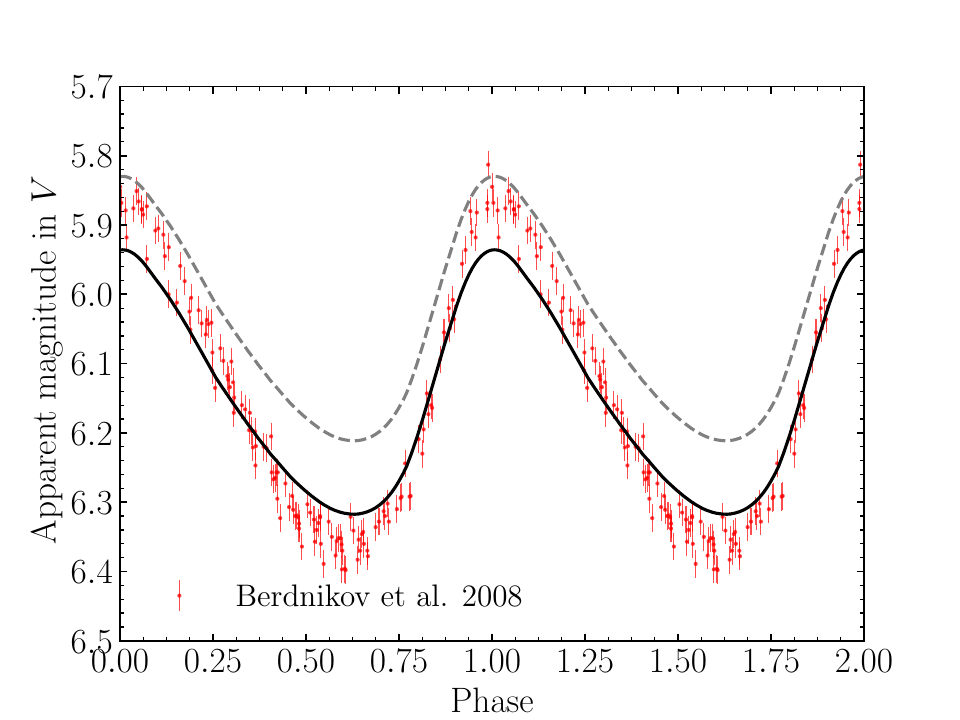}
        \caption{}
        \label{fig:V_5M}
    \end{subfigure}
    \begin{subfigure}{0.33\textwidth}
        \includegraphics[width=\linewidth]{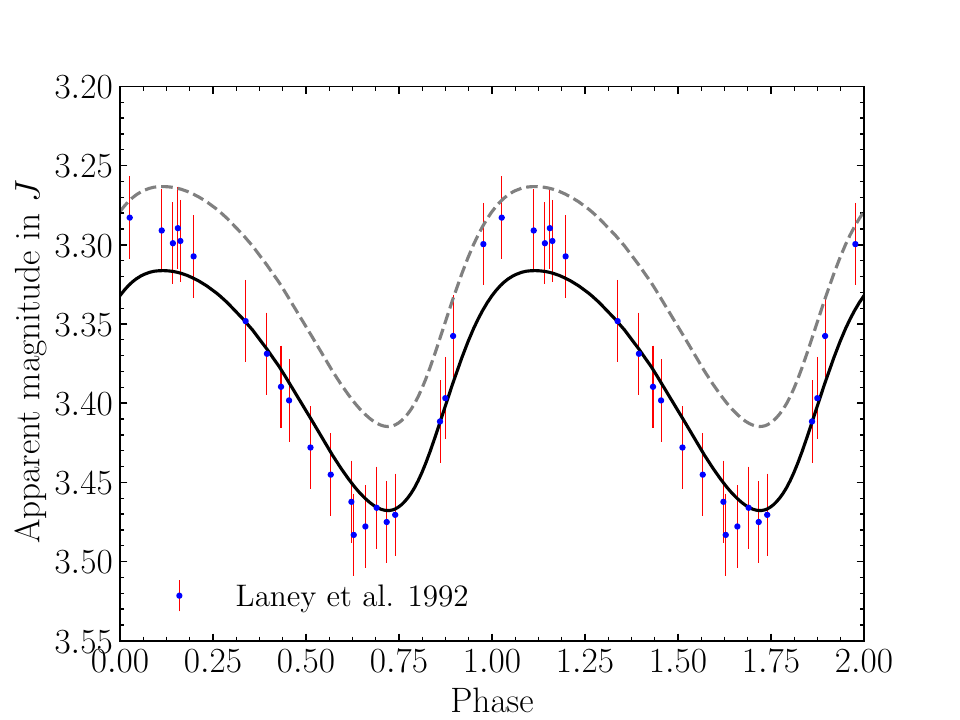}
        \caption{}
        \label{fig:J_5M}
    \end{subfigure}
    \begin{subfigure}{0.33\textwidth}
        \includegraphics[width=\linewidth]{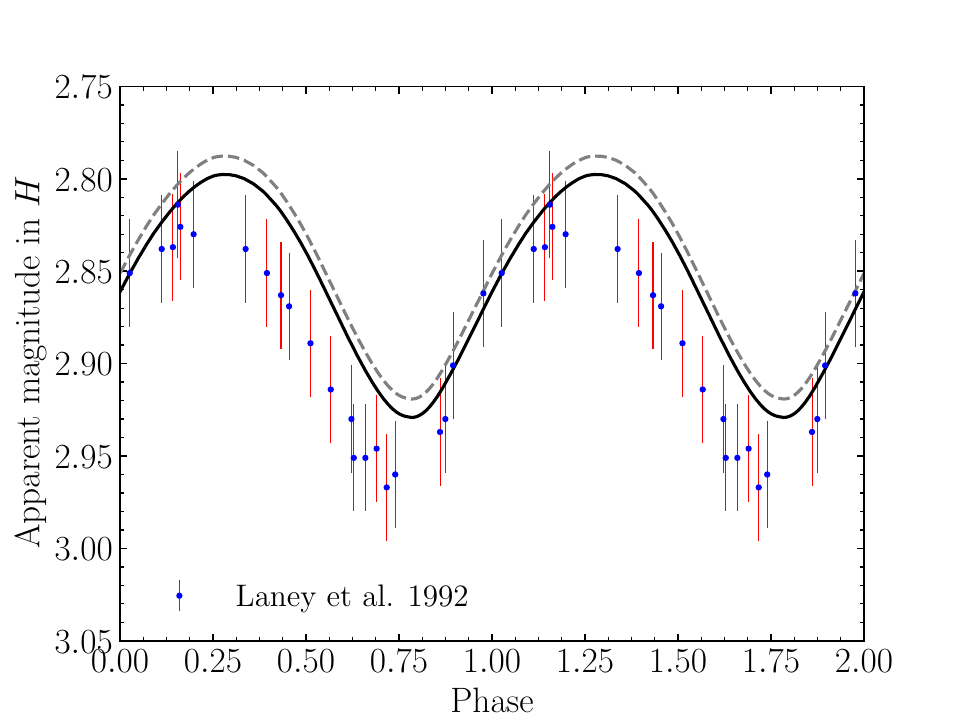}
        \caption{}
        \label{fig:H_5M}
    \end{subfigure}

    \medskip

    \begin{subfigure}{0.33\textwidth}
        \includegraphics[width=\linewidth]{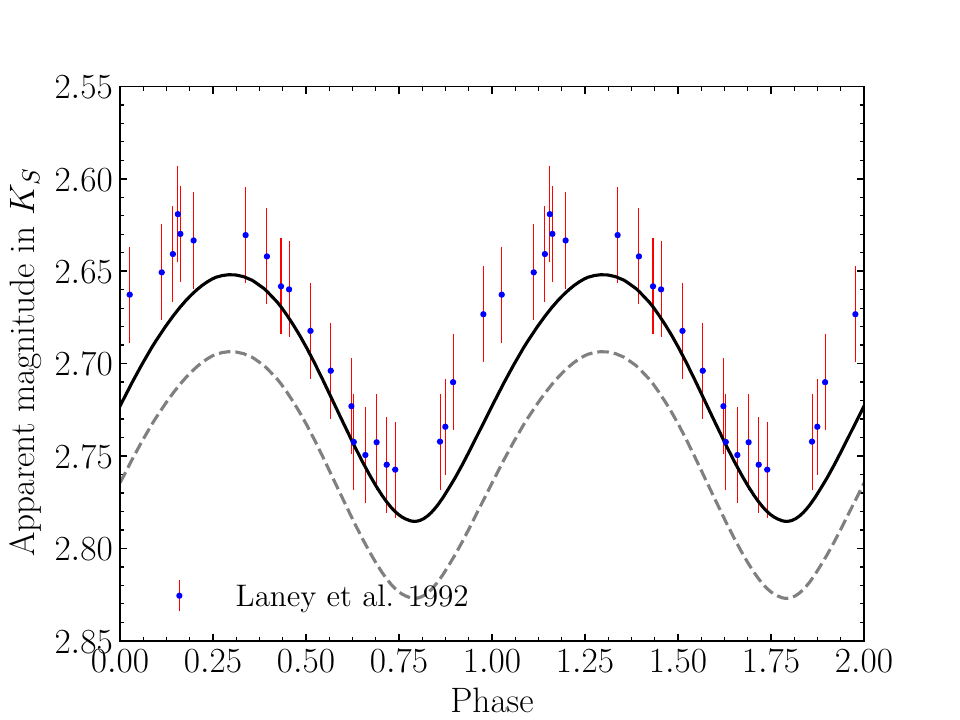}
        \caption{}
        \label{fig:K_5M}
    \end{subfigure}
    \begin{subfigure}{0.33\textwidth}
        \includegraphics[width=\linewidth]{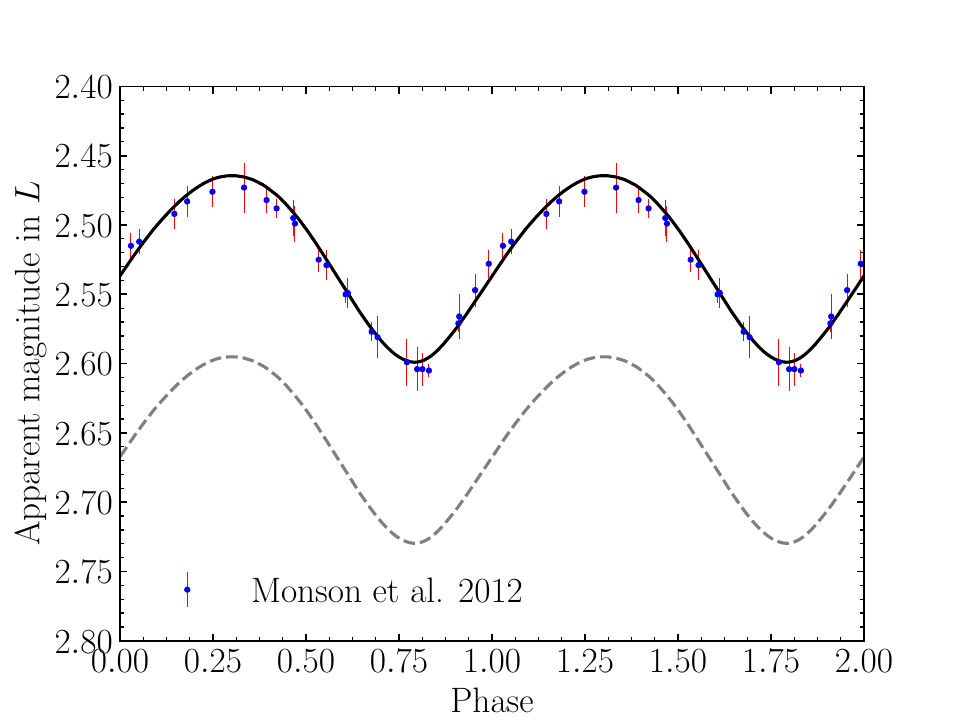}
        \caption{}
        \label{fig:L_5M}
    \end{subfigure}
    \begin{subfigure}{0.33\textwidth}
        \includegraphics[width=\linewidth]{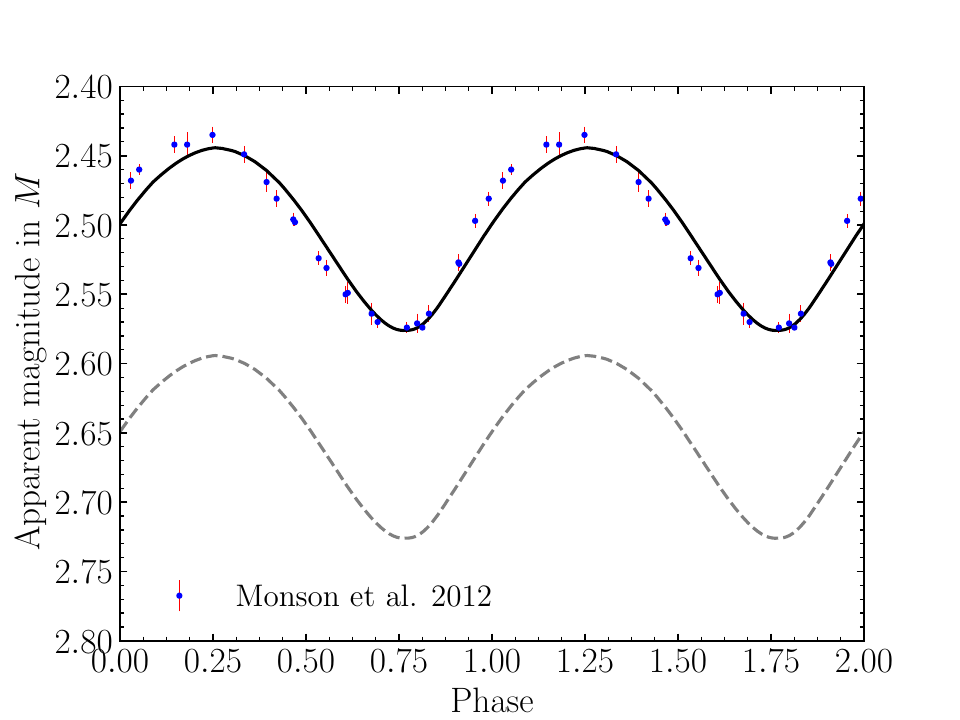}
        \caption{}
        \label{fig:M_5M}
    \end{subfigure}
\caption{Best result of the non-linear analysis with RSP for 5$\,M_\odot$.}\label{fig:non_linear_5M}
\end{figure*}

   \begin{figure*}[]
    \centering
    \begin{subfigure}{0.33\textwidth}
        \includegraphics[width=\linewidth]{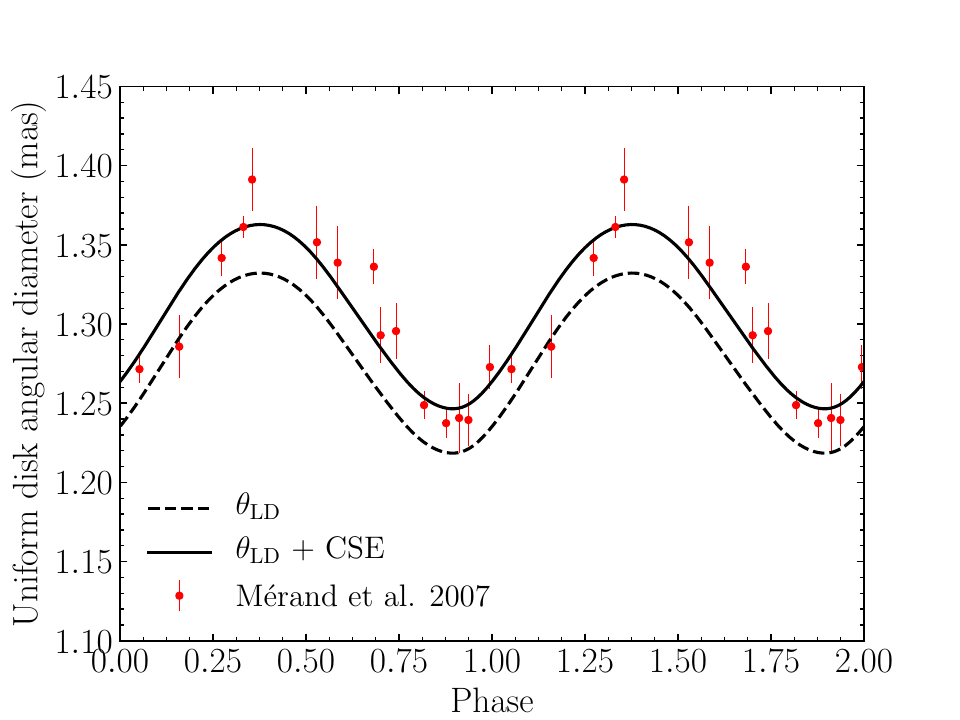}
        \caption{}
        \label{fig:R_6M}
    \end{subfigure}
    \begin{subfigure}{0.33\textwidth}
        \includegraphics[width=\linewidth]{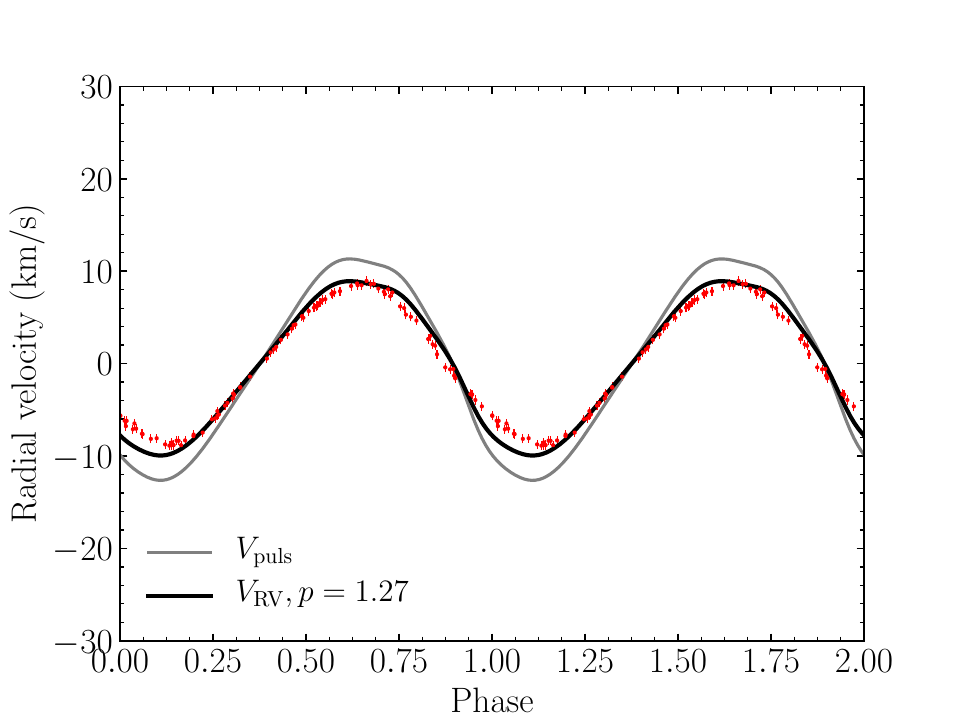}
        \caption{}
        \label{fig:RV_6M}
    \end{subfigure}
    \begin{subfigure}{0.33\textwidth}
        \includegraphics[width=\linewidth]{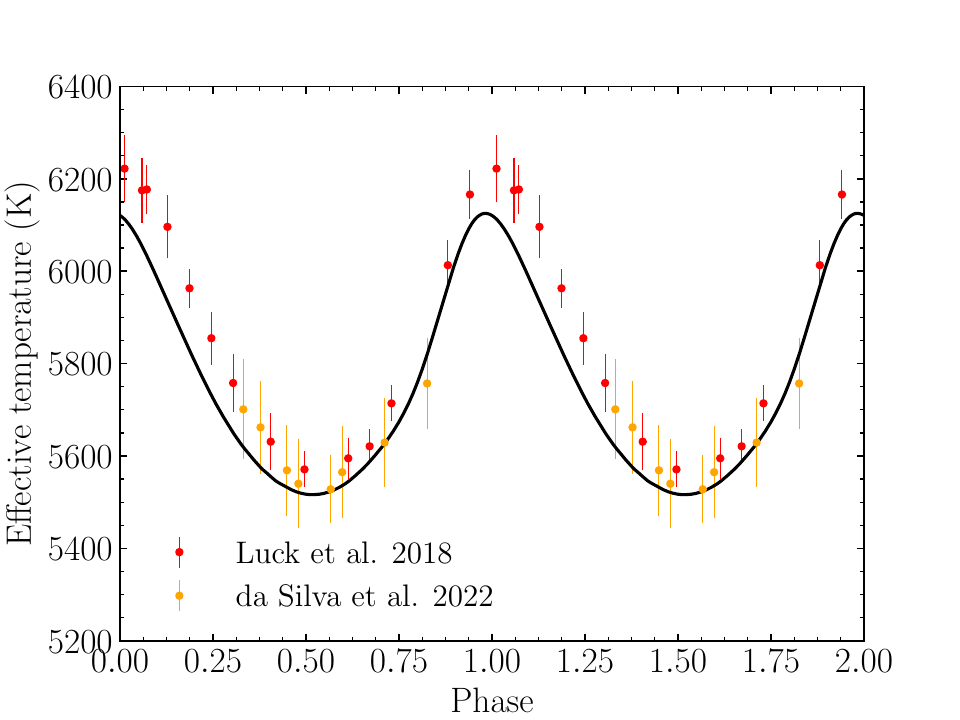}
        \caption{}
        \label{fig:Teff_6M}
    \end{subfigure}

    \medskip
        \begin{subfigure}{0.33\textwidth}
        \includegraphics[width=\linewidth]{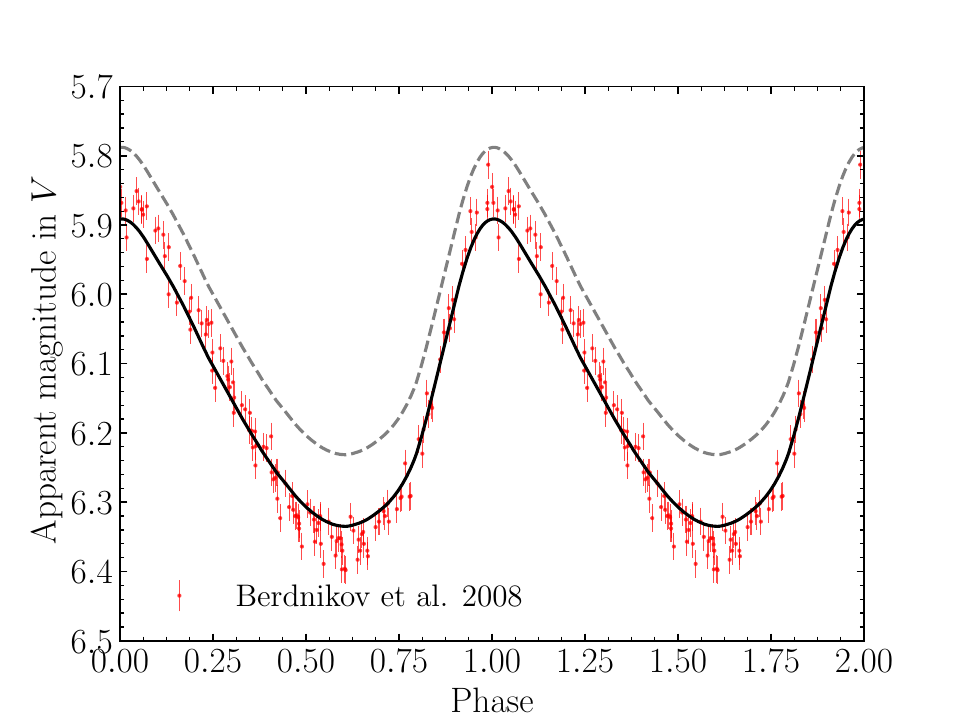}
        \caption{}
        \label{fig:V_6M}
    \end{subfigure}
    \begin{subfigure}{0.33\textwidth}
        \includegraphics[width=\linewidth]{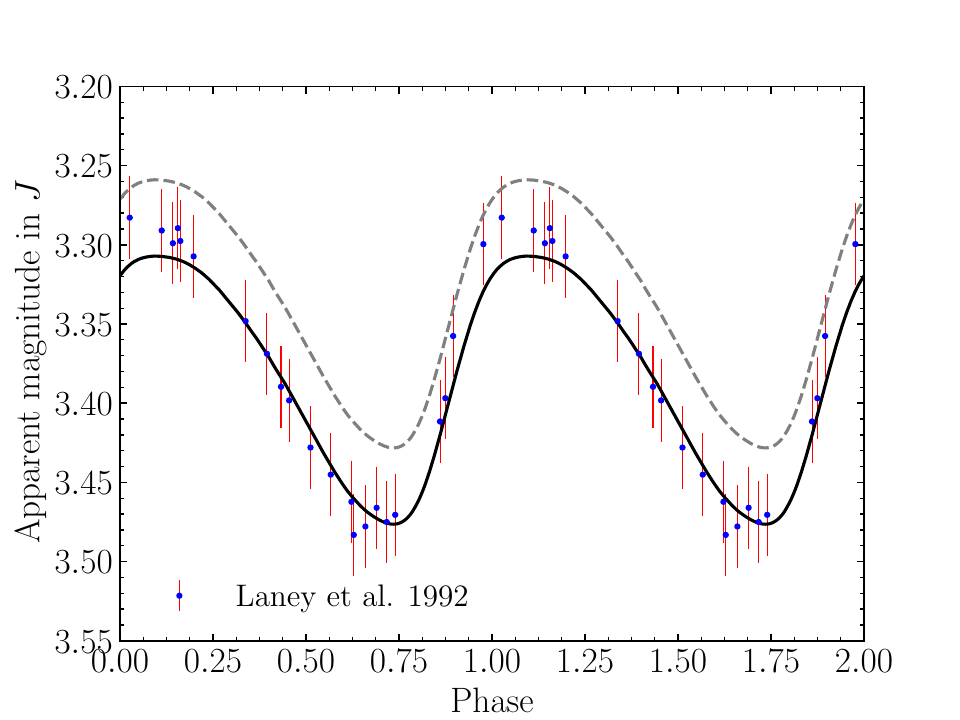}
        \caption{}
        \label{fig:J_6M}
    \end{subfigure}
    \begin{subfigure}{0.33\textwidth}
        \includegraphics[width=\linewidth]{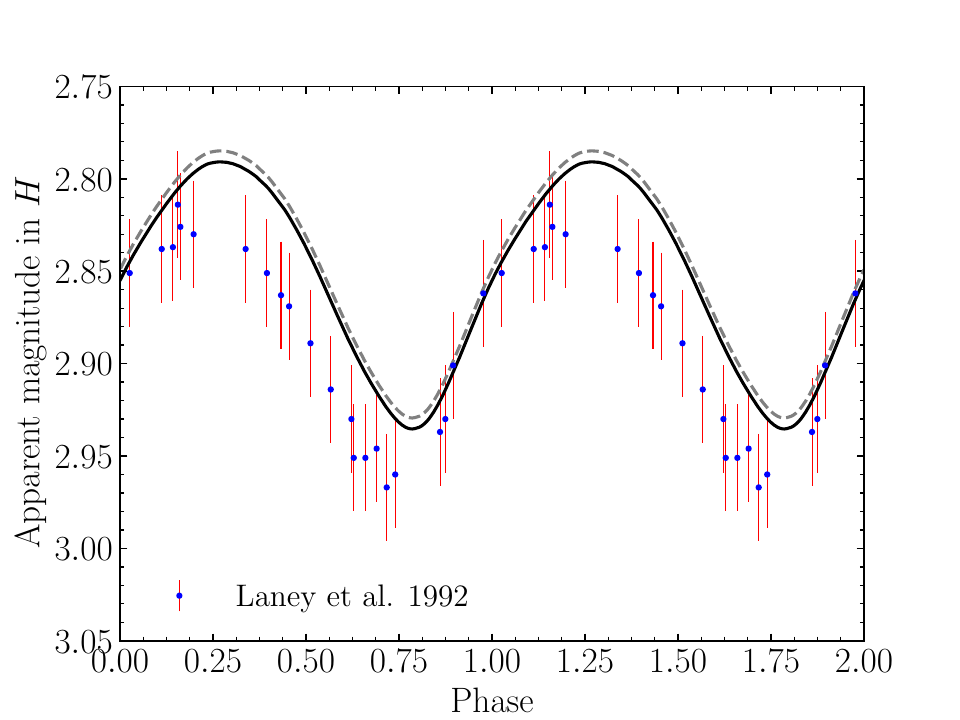}
        \caption{}
        \label{fig:H_6M}
    \end{subfigure}

    \medskip

    \begin{subfigure}{0.33\textwidth}
        \includegraphics[width=\linewidth]{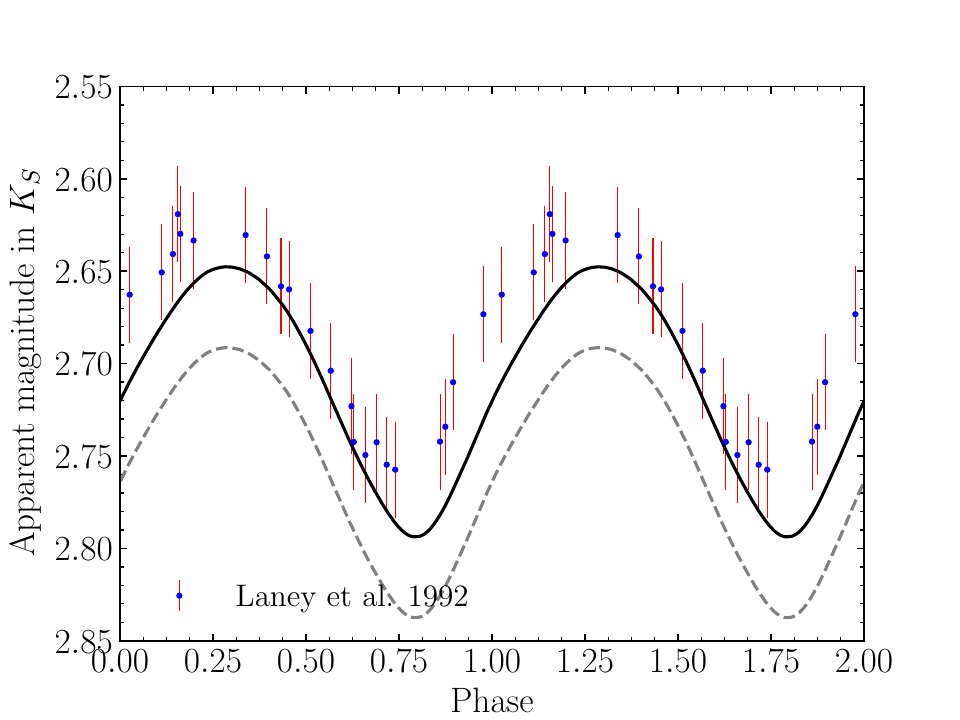}
        \caption{}
        \label{fig:K_6M}
    \end{subfigure}
    \begin{subfigure}{0.33\textwidth}
        \includegraphics[width=\linewidth]{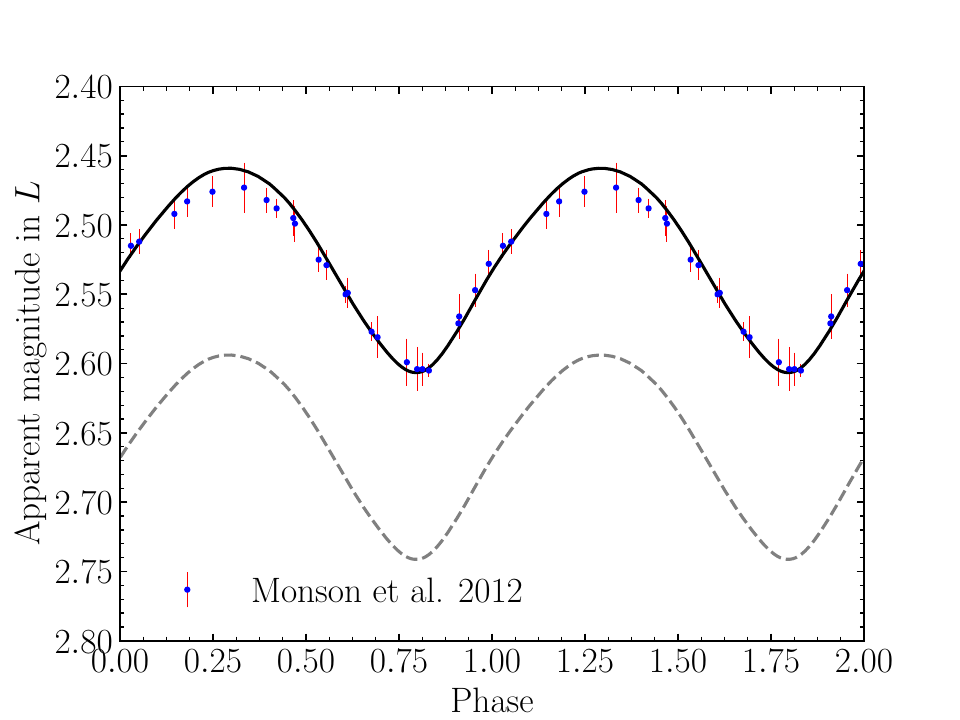}
        \caption{}
        \label{fig:L_6M}
    \end{subfigure}
    \begin{subfigure}{0.33\textwidth}
        \includegraphics[width=\linewidth]{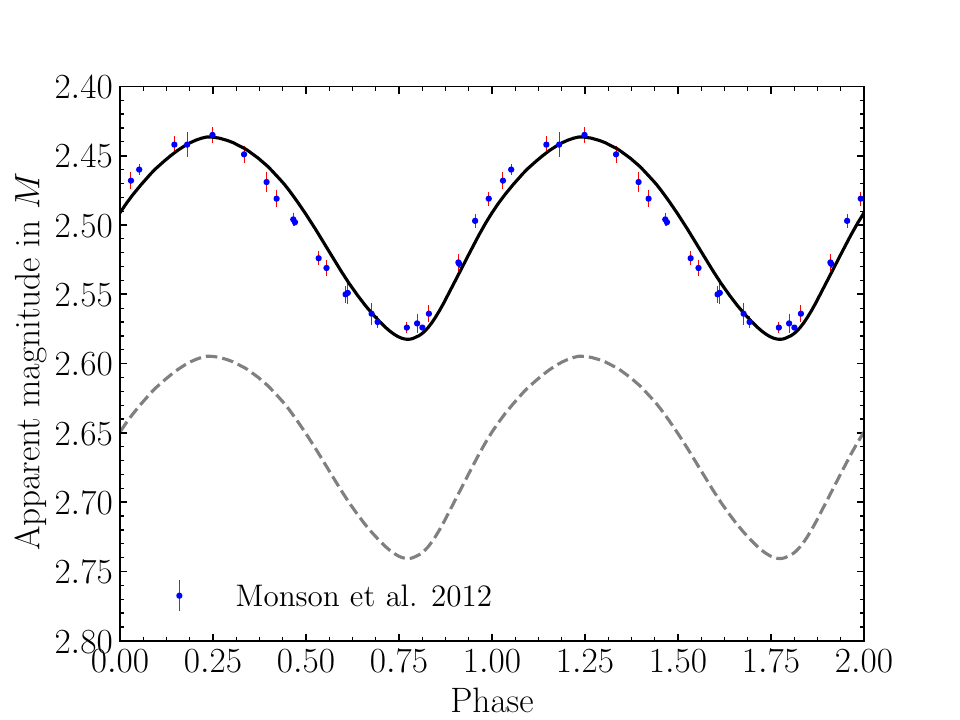}
        \caption{}
        \label{fig:M_6M}
    \end{subfigure}
\caption{Best result of the non-linear analysis with RSP for 6$\,M_\odot$.}\label{fig:non_linear_6M}
\end{figure*}

   \begin{figure*}[]
    \centering
    \begin{subfigure}{0.33\textwidth}
        \includegraphics[width=\linewidth]{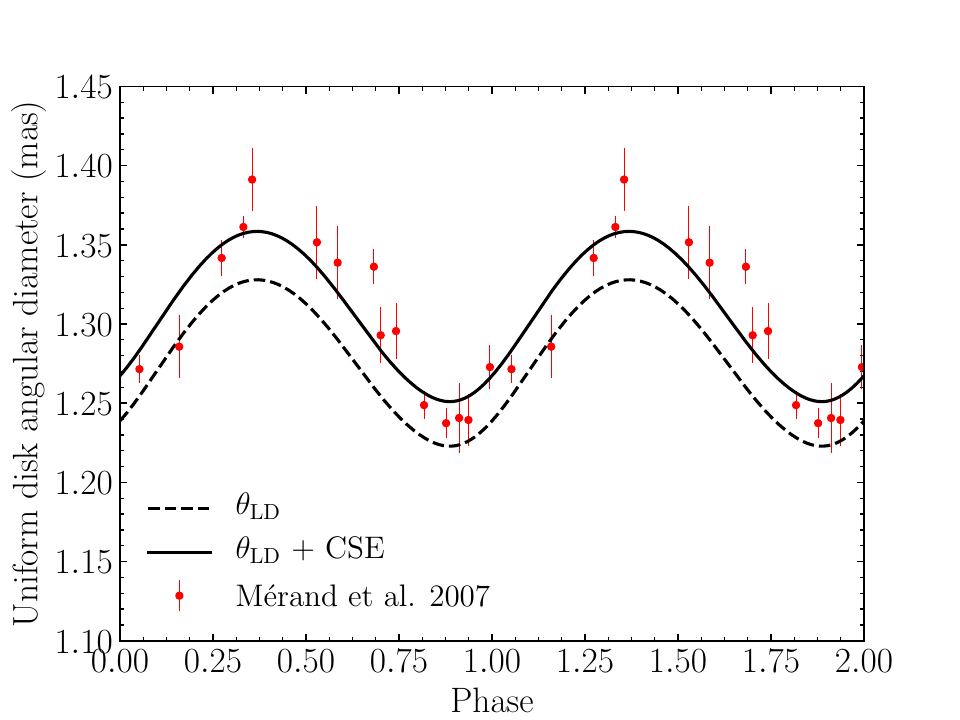}
        \caption{}
        \label{fig:R_7M}
    \end{subfigure}
    \begin{subfigure}{0.33\textwidth}
        \includegraphics[width=\linewidth]{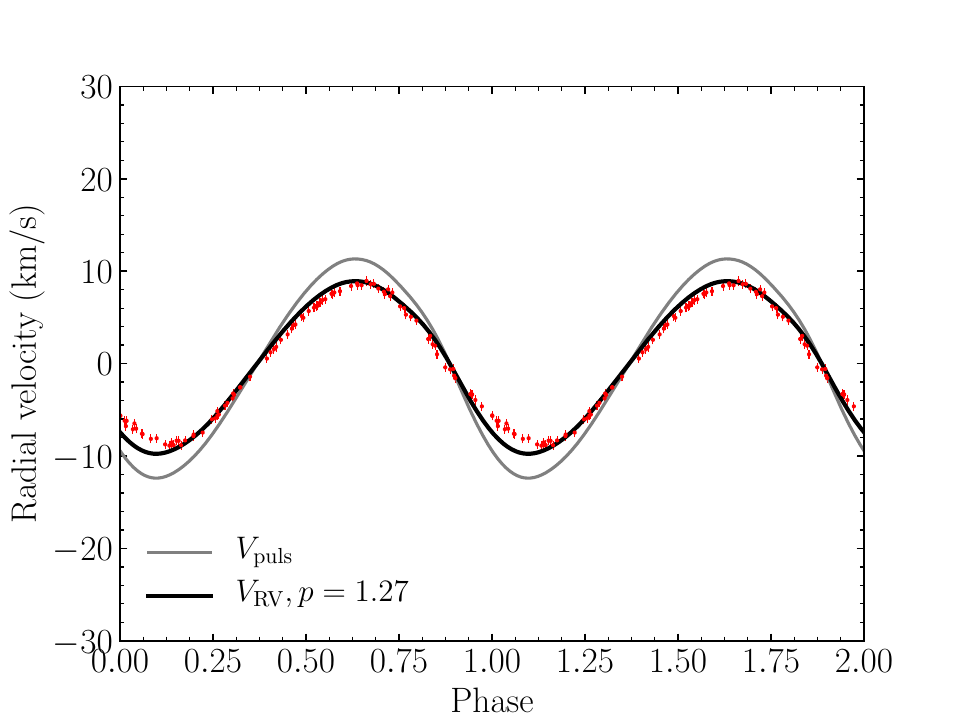}
        \caption{}
        \label{fig:RV_7M}
    \end{subfigure}
    \begin{subfigure}{0.33\textwidth}
        \includegraphics[width=\linewidth]{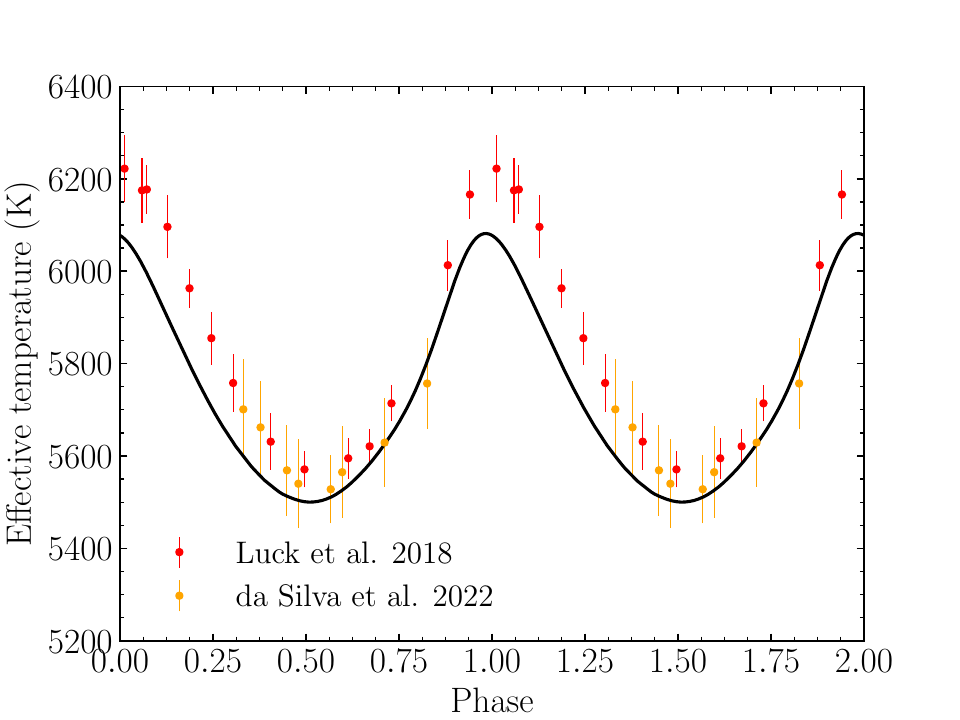}
        \caption{}
        \label{fig:Teff_7M}
    \end{subfigure}

    \medskip
        \begin{subfigure}{0.33\textwidth}
        \includegraphics[width=\linewidth]{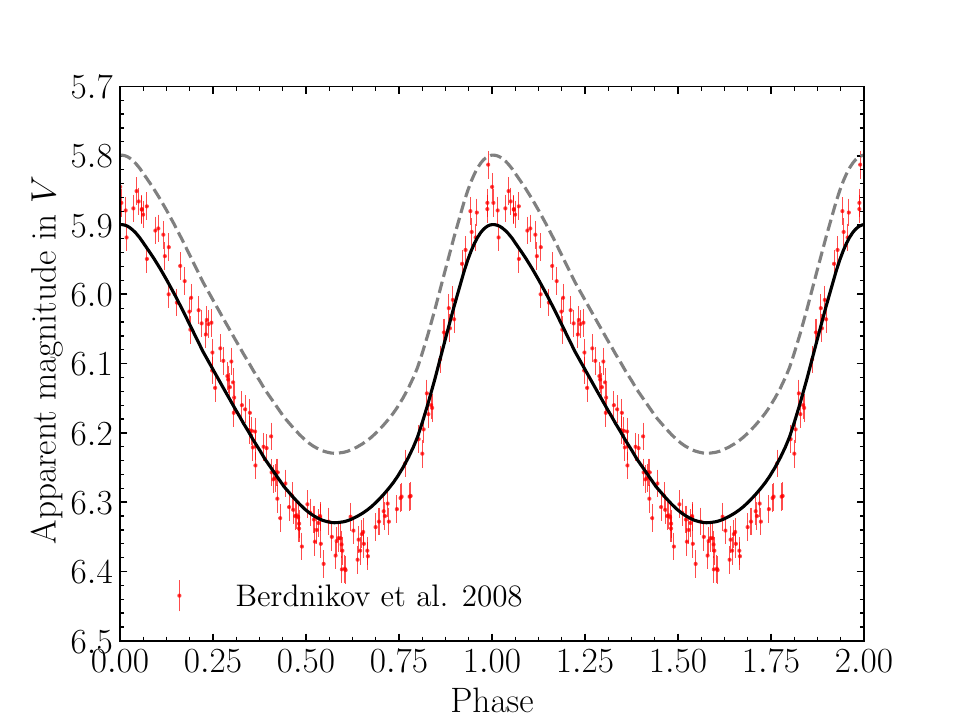}
        \caption{}
        \label{fig:V_7M}
    \end{subfigure}
    \begin{subfigure}{0.33\textwidth}
        \includegraphics[width=\linewidth]{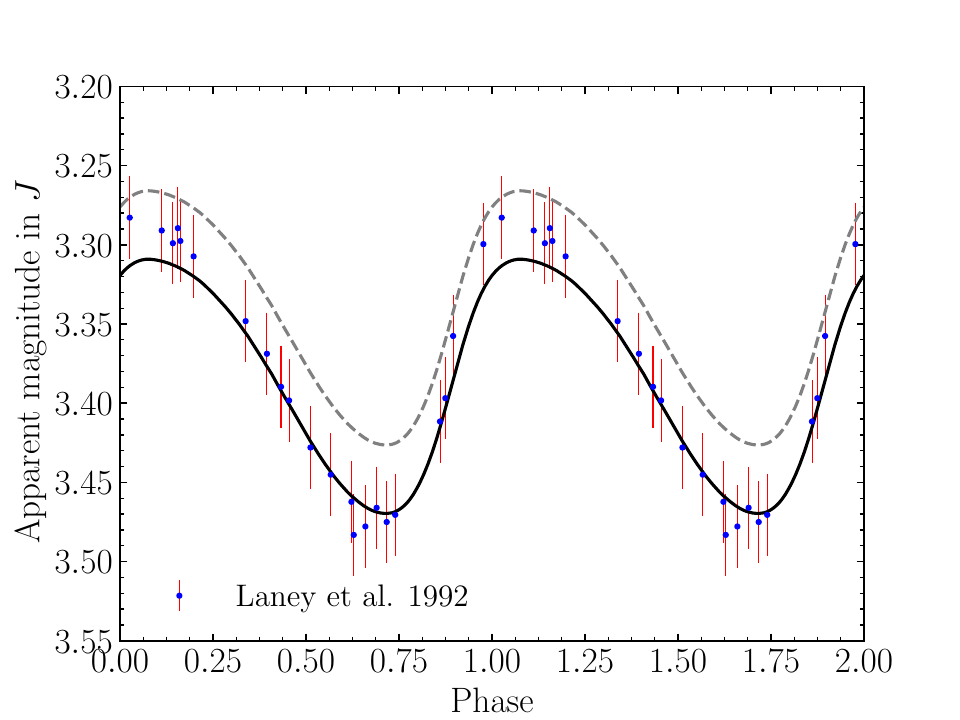}
        \caption{}
        \label{fig:J_7M}
    \end{subfigure}
    \begin{subfigure}{0.33\textwidth}
        \includegraphics[width=\linewidth]{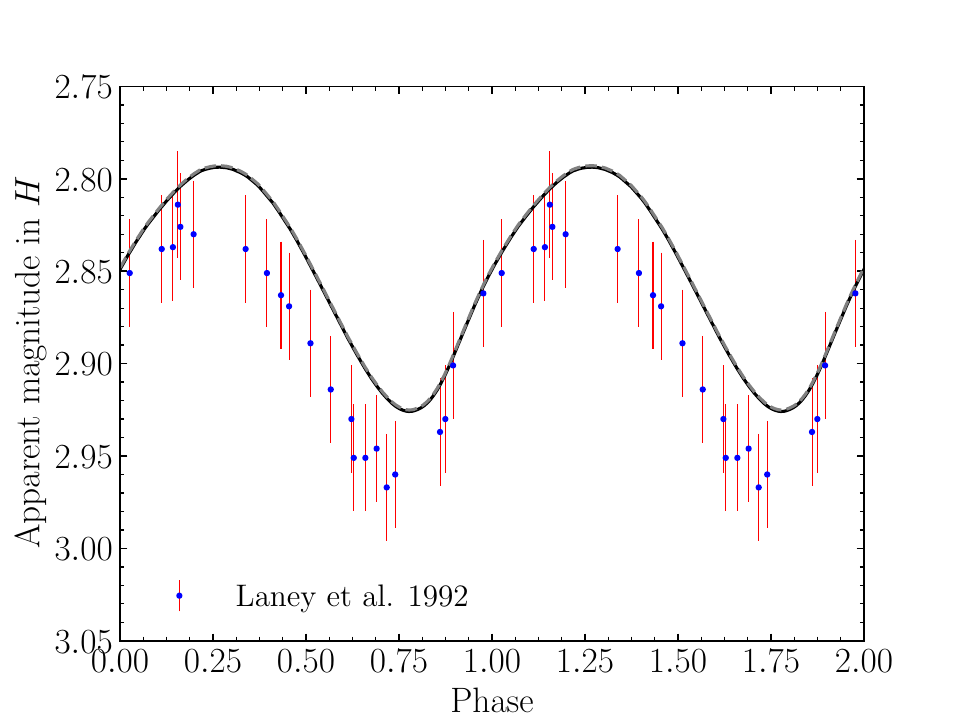}
        \caption{}
        \label{fig:H_7M}
    \end{subfigure}

    \medskip

    \begin{subfigure}{0.33\textwidth}
        \includegraphics[width=\linewidth]{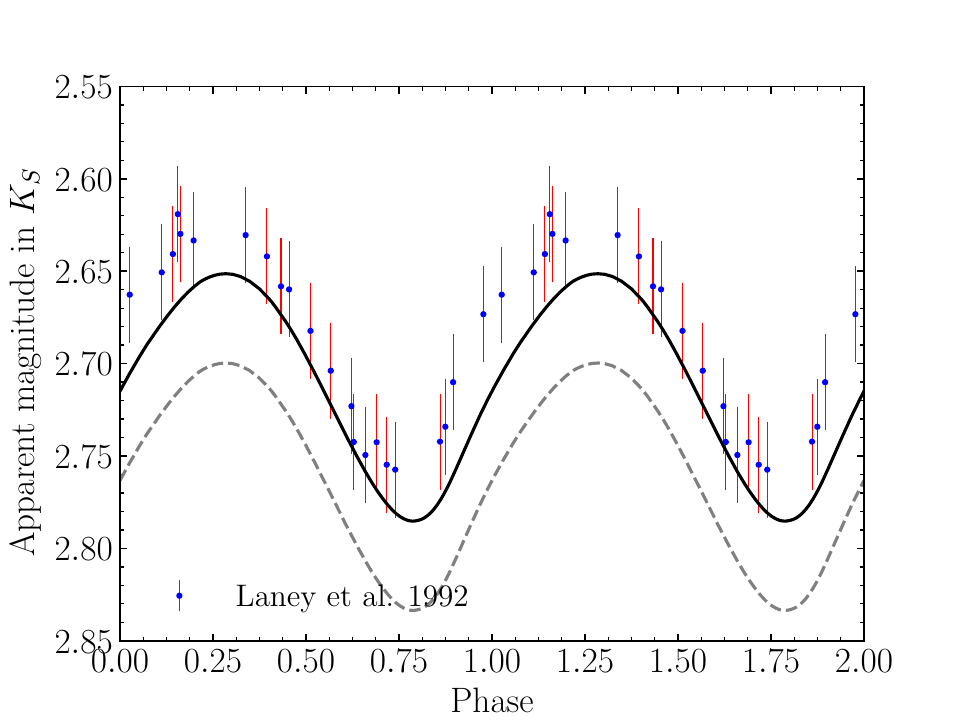}
        \caption{}
        \label{fig:K_7M}
    \end{subfigure}
    \begin{subfigure}{0.33\textwidth}
        \includegraphics[width=\linewidth]{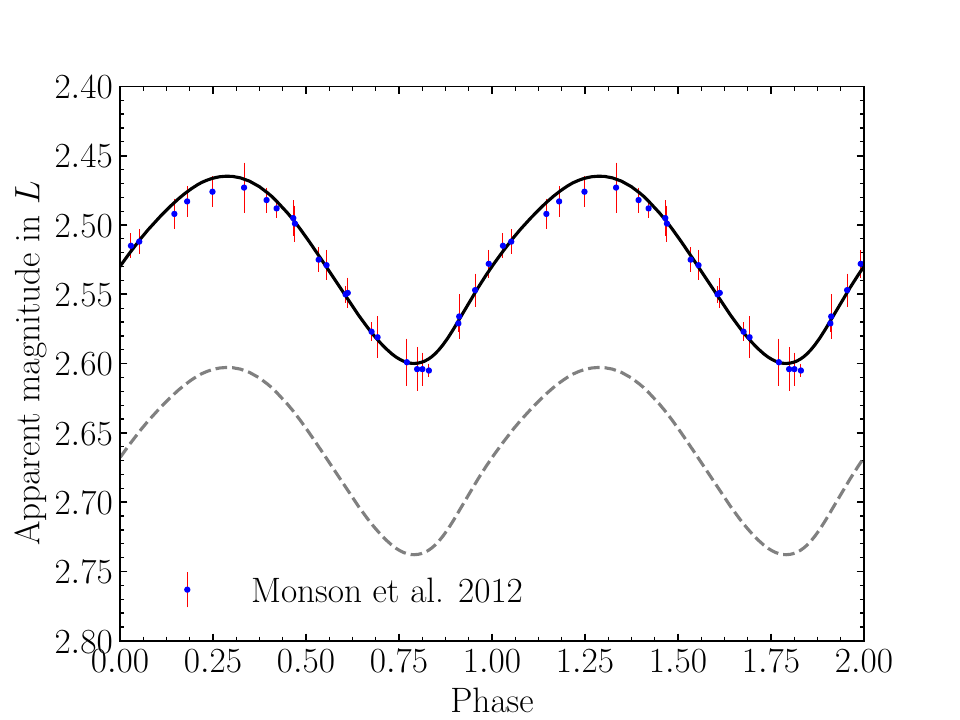}
        \caption{}
        \label{fig:L_7M}
    \end{subfigure}
    \begin{subfigure}{0.33\textwidth}
        \includegraphics[width=\linewidth]{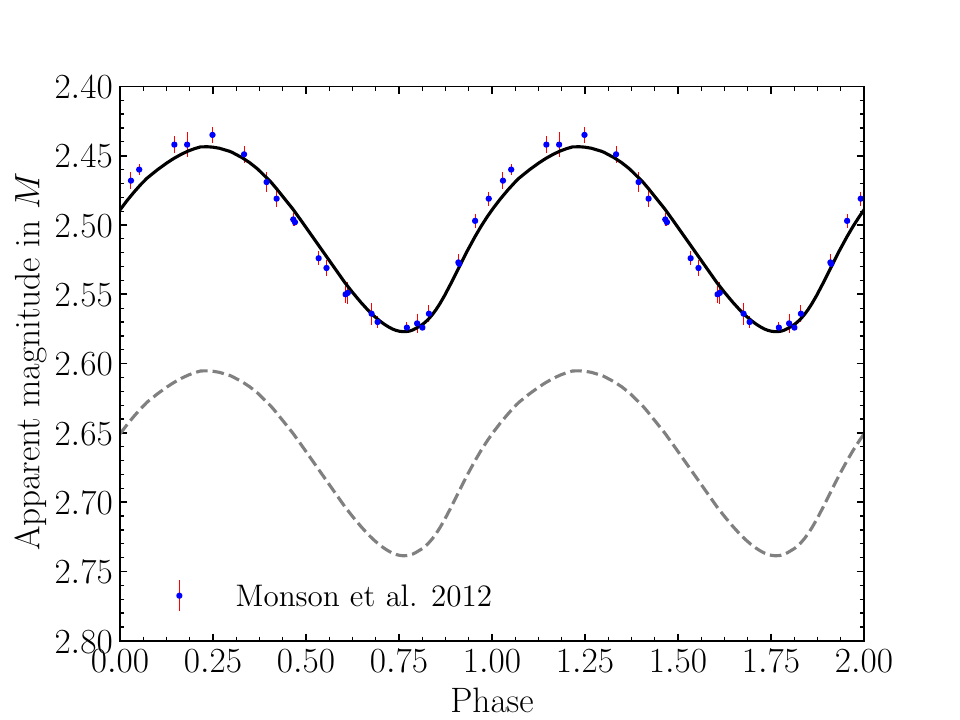}
        \caption{}
        \label{fig:M_7M}
    \end{subfigure}
\caption{Best result of the non-linear analysis with RSP for 7$\,M_\odot$.}\label{fig:non_linear_7M}
\end{figure*}

\end{appendix}

\end{document}